\documentclass[iop]{emulateapj}
\usepackage{apjfonts}
\newcommand\jcap{J. Cosmology Astropart. Phys.}%

\shorttitle{PROPERTIES OF KEPLER TARGETS FROM AMP}
\shortauthors{METCALFE ET AL.}

\begin{document}

\title{Properties of 42 Solar-type \textit{Kepler} Targets from the 
Asteroseismic Modeling Portal}

\author{T.~S.~Metcalfe\altaffilmark{1,2}, 
O.~L.~Creevey\altaffilmark{3},
G.~Do{\u g}an\altaffilmark{2,4},
S.~Mathur\altaffilmark{1,4},
H.~Xu\altaffilmark{5},
T.~R.~Bedding\altaffilmark{6},
W.~J.~Chaplin\altaffilmark{7},
J.~Christensen-Dalsgaard\altaffilmark{2},
C.~Karoff\altaffilmark{2},
R.~Trampedach\altaffilmark{2,8},
O.~Benomar\altaffilmark{6,9},
B.~P.~Brown\altaffilmark{10,11},
D.~L.~Buzasi\altaffilmark{12},
T.~L.~Campante\altaffilmark{7},
Z.~\c{C}elik\altaffilmark{13},
M.~S.~Cunha\altaffilmark{14},
G.~R.~Davies\altaffilmark{7},
S.~Deheuvels\altaffilmark{15,16},
A.~Derekas\altaffilmark{17,18},
M.~P.~Di~Mauro\altaffilmark{19},
R.~A.~Garc{\'\i}a\altaffilmark{20},
J.~A.~Guzik\altaffilmark{21},
R.~Howe\altaffilmark{7},
K.~B.~MacGregor\altaffilmark{4},
A.~Mazumdar\altaffilmark{22}, 
J.~Montalb{\'a}n\altaffilmark{23},
M.~J.~P.~F.~G.~Monteiro\altaffilmark{14},
D.~Salabert\altaffilmark{20},
A.~Serenelli\altaffilmark{24},
D.~Stello\altaffilmark{6},
M.~St\c{e}{\'s}licki\altaffilmark{25},
M.~D.~Suran\altaffilmark{26},
M.~Y{\i}ld{\i}z\altaffilmark{13},
C.~Aksoy\altaffilmark{13},
Y.~Elsworth\altaffilmark{7},
M.~Gruberbauer\altaffilmark{27},
D.~B.~Guenther\altaffilmark{27},
Y.~Lebreton\altaffilmark{28,29},
K.~Molaverdikhani\altaffilmark{30},
D.~Pricopi\altaffilmark{26},
R.~Simoniello\altaffilmark{31},
T.~R.~White\altaffilmark{6,32}}

\altaffiltext{1}{Space Science Institute, 4750 Walnut St. Suite 205, Boulder CO 80301 USA}
\altaffiltext{2}{Stellar Astrophysics Centre, Department of Physics and Astronomy, Aarhus University, Ny Munkegade 120, DK-8000 Aarhus C, Denmark}
\altaffiltext{3}{Institut d'Astrophysique Spatiale, Universit\'e Paris XI, UMR 8617, CNRS, Batiment 121, 91405 Orsay Cedex, France}
\altaffiltext{4}{High Altitude Observatory, NCAR, PO Box 3000, Boulder CO 80307 USA}
\altaffiltext{5}{Computational \& Information Systems Laboratory, NCAR, PO Box 3000, Boulder CO 80307 USA}
\altaffiltext{6}{Sydney Institute for Astronomy (SIfA), School of Physics, University of Sydney, NSW 2006, Australia}
\altaffiltext{7}{School of Physics and Astronomy, University of Birmingham, Birmingham B15 2TT, UK}
\altaffiltext{8}{JILA, University of Colorado and National Institute of Standards and Technology, 440 UCB, Boulder CO 80309 USA}
\altaffiltext{9}{Department of Astronomy, The University of Tokyo, Tokyo 113-0033, Japan}
\altaffiltext{10}{Department of Astronomy and Center for Magnetic Self Organization in Laboratory and Astrophysical Plasmas, University of Wisconsin, Madison, WI 53706 USA}
\altaffiltext{11}{Kavli Institute for Theoretical Physics, University of California, Santa Barbara CA 93106 USA}
\altaffiltext{12}{Department of Chemistry and Physics, Florida Gulf Coast University, Fort Myers, FL 33965 USA}
\altaffiltext{13}{Ege University, Department of Astronomy and Space Sciences, Bornova, 35100, Izmir, Turkey}
\altaffiltext{14}{Centro de Astrof{\'\i}sica e Faculdade de Ci{\^e}ncias, Universidade do Porto, Rua das Estrelas, 4150-762, Porto, Portugal}
\altaffiltext{15}{Universit\'e de Toulouse; UPS-OMP; IRAP; Toulouse, France}
\altaffiltext{16}{CNRS; IRAP; 14, avenue Edouard Belin, F-31400 Toulouse, France}
\altaffiltext{17}{Remaining affiliations removed due to arXiv processing error}

\submitted{The Astrophysical Journal Supplement Series, ACCEPTED}

\begin{abstract}

Recently the number of main-sequence and subgiant stars exhibiting 
solar-like oscillations that are resolved into individual mode frequencies 
has increased dramatically. While only a few such data sets were available 
for detailed modeling just a decade ago, the {\it Kepler} mission has 
produced suitable observations for hundreds of new targets. This rapid 
expansion in observational capacity has been accompanied by a shift in 
analysis and modeling strategies to yield uniform sets of derived stellar 
properties more quickly and easily. We use previously published 
asteroseismic and spectroscopic data sets to provide a uniform analysis of 
42 solar-type {\it Kepler} targets from the Asteroseismic Modeling Portal 
(AMP). We find that fitting the individual frequencies typically doubles 
the precision of the asteroseismic radius, mass and age compared to 
grid-based modeling of the global oscillation properties, and improves the 
precision of the radius and mass by about a factor of three over empirical 
scaling relations. We demonstrate the utility of the derived properties 
with several applications.

\end{abstract}


\section{BACKGROUND}\label{SEC1}

It is difficult to overstate the impact of the {\it Kepler} mission on the 
observation and analysis of solar-like oscillations in main-sequence and 
subgiant stars. In a review from just a decade ago, \cite{bk03} 
highlighted the tentative detections of individual oscillation frequencies 
in just a few such stars from ground-based observations, and {\it Kepler} 
was not even mentioned. Despite funding issues that delayed the mission 
from an original deployment date in 2006, {\it Kepler} finally launched in 
March 2009 and operated almost flawlessly for more than 4 years, slightly 
exceeding its design lifetime \citep{bor10}. The archive of public data 
now includes nearly uninterrupted observations for many thousands of 
solar-type stars, including short-cadence data \citep[58.85~s 
sampling,][]{gil10} for hundreds of these targets. In the span of a 
decade, the study of solar-like oscillations has been transformed 
dramatically \citep{cm13}.

During the first 10 months of science operations, {\it Kepler} performed a 
survey for solar-like oscillations in more than 2000 main-sequence and 
subgiant stars, yielding detections in more than 500 targets from the 
1-month data sets. The initial analysis of this ensemble, using empirical 
scaling relations \citep{kb95} to generate estimates of radius and mass, 
suggested a significant departure from the mass distribution expected from 
Galactic population synthesis models \citep{cha11}. Subsequent analysis of 
the sample, using updated effective temperatures \citep{pin12} and a 
substantial grid-based modeling effort, led to more precise estimates of 
the radii and masses as well as information about the stellar ages 
\citep{cha14}. Some of the brightest stars from the survey were subjected 
to a more detailed analysis, including spectroscopic follow-up to 
determine more precise atmospheric properties \citep{bru12} plus the 
identification and detailed modeling of dozens of oscillation frequencies 
in each star \citep{mat12}. These studies gave us a preview of what to 
expect from the subsequent phase of the mission.

Starting with Quarter 5 (Q5), {\it Kepler's} short-cadence study of 
solar-like oscillations transitioned to a specific target phase, where 
extended observations began for a fixed sample of stars identified during 
the survey. The target list during this phase gave priority to stars 
showing oscillations with the highest signal-to-noise ratio (S/N), but it 
also retained the brightest main-sequence stars cooler than the Sun, where 
the lower intrinsic oscillation amplitudes yielded relatively weak 
detections from the survey. From Q5 through the end of the mission (Q17), 
about 200 of the 512 available short-cadence targets were typically 
specified by the {\it Kepler} Asteroseismic Science Consortium 
\citep[KASC,][]{kje10} and about half of those were intended for the study 
of solar-like oscillations.

Just like the exoplanet side of the mission, the KASC team gradually 
improved the data reduction and analysis methods while additional data 
swelled the archive. Never in the history of the field had such extended 
monitoring been possible at all, let alone for such a large sample of 
stars. As a consequence, the availability of reliable sets of input data 
for stellar modeling lagged well behind the continually expanding 
time-series for each star in the archive. This delay was primarily due to 
the challenge of coordinating the efforts of multiple teams, first to 
produce optimized light curves from the raw {\it Kepler} data 
\citep{gar11}, then to fit the global oscillation properties and remove 
the stellar granulation background from the power spectra 
\citep{ver11,mat11}, and finally to extract and identify the individual 
oscillation frequencies using so-called ``peak-bagging'' techniques 
\citep{app12}. Also like the exoplanet program, ground-based follow up 
observations were difficult to obtain for the faintest targets, further 
limiting the number of stars for which detailed modeling was feasible.

Even after reliable sets of observational constraints became available, an 
analogous effort was required to consolidate the results from many stellar 
modeling teams. Initially this effort sought to define objective metrics 
of model quality, and to use the ensemble of results from different codes 
and methods to estimate the systematic uncertainties for a few specific 
targets \citep{met10,met12,cre12,dog13,sil13}. The first large sample to 
emerge from the survey made this ``boutique'' modeling approach 
impractical, and motivated the initial large-scale application of the 
Asteroseismic Modeling Portal \citep[AMP,][]{met09,woi09}. \cite{mat12} 
presented a uniform analysis of 22 {\it Kepler} targets observed for 1 
month each during the survey phase, and compared detailed modeling from 
AMP with empirical scaling relations and with results from several 
grid-based modeling methods that matched the global oscillation properties 
($\Delta\nu$ and $\nu_{\rm max}$, see below) rather than the individual 
frequencies from peak-bagging. The results clearly demonstrated the 
improved level of precision that was possible from detailed modeling of 
the individual oscillation frequencies, particularly for stellar ages.

In this paper we present stellar modeling results from AMP for the first 
large sample of {\it Kepler} targets with extended observations during the 
specific target phase of the mission. In section~\ref{SEC2} we describe 
the sample, which was drawn from the most recently published observations. 
We outline our stellar modeling approach in section~\ref{SEC3}, including 
several improvements to the previous version of AMP and using slightly 
customized procedures for different types of stars. We present the main 
results and initial applications in section~\ref{SEC4}, and we discuss 
conclusions and future prospects in section~\ref{SEC5}.

\section{OBSERVATIONAL CONSTRAINTS}\label{SEC2}

Solar-like oscillations are stochastically excited and intrinsically 
damped by turbulent convection near the stellar surface \citep{gk77,gk88, 
hou99,sg01}. Each oscillation mode is characterized by its radial order 
$n$ and spherical degree $\ell$, and only the low-degree ($\ell\le3$) 
modes are generally detectable without spatial resolution across the 
surface. The consecutive radial orders define the average large separation 
$\langle\Delta\nu\rangle$, which reflects the mean stellar density 
\citep{tas80}. The power in each mode is governed by a roughly Gaussian 
envelope with a maximum at frequency $\nu_{\rm max}$, which approximately 
scales with the acoustic cutoff frequency \citep{bro91,bel11}. These 
global oscillation properties are well-constrained, even in the relatively 
short time-series obtained during the {\it Kepler} survey phase. Longer 
observations improve the frequency precision and also reveal additional 
oscillation modes, with lower and higher radial orders, as the S/N 
improves in the wings of the Gaussian envelope. This is the primary 
motivation for gathering extended observations: to maximize both the 
number and quality of asteroseismic constraints that are available for 
stellar modeling.

\cite{app12} published asteroseismic data sets for 61 main-sequence and 
subgiant stars observed by {\it Kepler}, based on an analysis of 9-month 
time-series. The data were collected during Q5--Q7 (2010 Mar 20 through 
2010 Dec 22), archived on 2011 Apr 23, and the final sets of identified 
frequencies were published about one year later. Most of these frequency 
sets came from maximum likelihood estimators \citep{app08} with errors 
determined from the inverse of the Hessian matrix, but frequencies for the 
F-like stars were obtained from a Bayesian analysis with MultiNest 
\citep{gru09} and errors were estimated from the 68\% credible interval. 
When asymmetric errors were provided, we adopted the mean of the two 
quoted values. No similar analysis of more extended data sets has yet been 
published, so we adopted this sample of 61 stars as our uniform source of 
asteroseismic constraints.

In addition to the asteroseismic data, precise spectroscopic constraints 
on the effective temperature $T_{\rm eff}$ and metallicity [M/H] are also 
required for detailed stellar modeling. A uniform spectroscopic analysis 
of 93 solar-type {\it Kepler} targets was published by \cite{bru12}, 
including 46 of the brightest stars in the \citeauthor{app12}\ sample with 
magnitudes in the {\it Kepler} bandpass $K\!p=7.4$--9.8. Six of the 15 
missing targets are also in this magnitude range: KIC\,3735871, 11772920, 
12317678 and 12508433, as well as the two bright spectroscopic binaries 
KIC\,8379927 and 9025370. The remaining 9 stars fall in the magnitude 
range $K\!p=9.9$--11.4, and are difficult targets for high-resolution 
spectroscopy on all but the largest telescopes.

  \begin{figure}[t] 
  \centerline{\includegraphics[angle=270,width=\columnwidth]{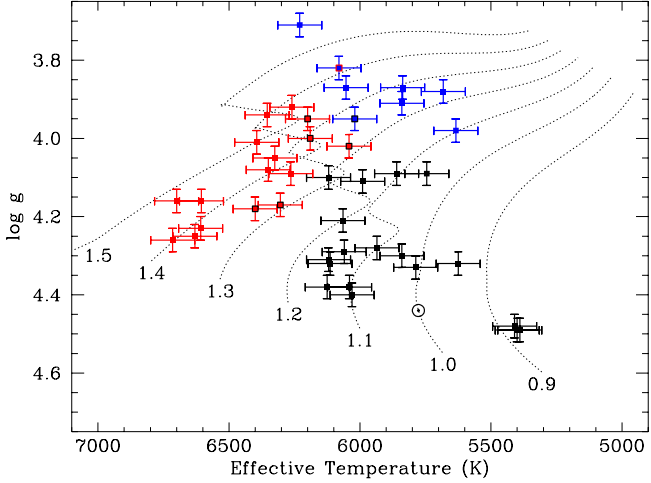}} 
  \caption{Spectroscopic H-R diagram for our final sample of 42
  asteroseismic targets, including simple (black), F-like (red) and 
  mixed-mode stars (blue). Points outlined in a different color indicate 
  the original classification by \cite{app12}. Error bars were adopted 
  from \cite{cha14}. Solar-composition evolution tracks from ASTEC for 
  masses between 0.9 and 1.5 $M_\odot$ are shown as dotted lines, with 
  the position of the Sun indicated by the $\odot$ symbol. Stars 
  similar to the Sun are missing from the KASC sample because the data 
  were sequestered by the {\it Kepler} exoplanet team.\label{fig1}}
  \vspace*{6pt}
  \end{figure} 

Recently, \cite{mol13} published spectroscopic constraints for a larger 
sample of 169 stars in the {\it Kepler} field. The overlap with the 
\citeauthor{app12} sample only includes two additional stars 
(KIC\,11772920 and 12508433) for the low-precision ROTFIT results 
\citep{fra03,fra06}, while the high-precision ARES+MOOG results 
\citep{sou07,sne73} contain fewer asteroseismic targets than are available 
in the \citeauthor{bru12}\ sample. Considering that our goal is to produce 
a uniform analysis, we adopted the spectroscopic constraints from 
\cite{bru12}, limiting the available sample to 46 stars. Four of these 
targets are evolved subgiants with too many mixed modes for successful 
automated modeling, so our final sample includes 42 stars. Following 
\cite{cha14}, we adopted larger uncertainties on $T_{\rm eff}$ ($\pm84$~K, 
see Figure~\ref{fig1}) that fold in a systematic error of 59~K, as 
suggested by \cite{tor12}. For 12 stars with {\it Hipparcos} parallaxes 
\citep{van07}, we used the spectroscopic $T_{\rm eff}$ to obtain 
bolometric corrections from \cite{flo96}. Following \cite{tor10}, we 
adopted $M_{\rm bol,\odot}=4.73\pm0.03$ and used the extinction estimates 
from \cite{amm06} to derive luminosity constraints.

\section{STELLAR MODELING APPROACH}\label{SEC3}

The Asteroseismic Modeling Portal \citep[AMP,][]{woi09} is a web-based 
interface to the stellar model-fitting pipeline described in detail by 
\cite{met09}. The underlying science code uses a parallel genetic 
algorithm \citep[GA,][]{mc03} on XSEDE supercomputing resources to 
optimize the match between asteroseismic models produced by the Aarhus 
stellar evolution code \citep[ASTEC,][]{jcd08a} and adiabatic pulsation 
code \citep[ADIPLS,][]{jcd08b} and a given set of observational 
constraints. \cite{mat12} were the first to apply AMP to a large sample of 
{\it Kepler} targets, motivating several improvements to the physical 
inputs and fitting procedures that are described below.

\subsection{Updated physics}\label{SEC3.1}

The version of AMP that was used for the models presented by \cite{mat12} 
was configured to use the \cite{gn93} solar mixture with the OPAL 2005 
equation of state \citep[see][]{rn02} and the most recent OPAL opacities 
\citep[see][]{ir96}, supplemented by \cite{af94} opacities at low 
temperatures. The updated version of AMP uses the low-T opacities from 
\cite{fer05}. We have also updated the default nuclear reaction rates, 
replacing the \cite{bp95} rates with those from the NACRE collaboration 
\citep{ang99}. Convection is still described by the mixing-length theory 
from \cite{bv58} without overshoot, and we continue to include the effects 
of helium diffusion and settling as described by \cite{mp93}. To correct 
the model frequencies for so-called ``surface effects'' due to incomplete 
modeling of the near-surface layers, we use the empirical prescription of 
\cite{kje08}.

We originally performed our analysis shortly after the publication of 
asteroseismic constraints by \cite{app12}, using the updated physics 
described above but without modifying the fitting procedures. The approach 
used by \cite{mat12} simultaneously optimized the match between the models 
and two sets of constraints: [1] the individual oscillation frequencies 
and [2] the atmospheric parameters from spectroscopy. This procedure 
generally yielded stellar radii, masses and ages that were consistent with 
empirical scaling relations and grid-based modeling of the global 
oscillation properties ($\Delta\nu$ and $\nu_{\rm max}$)---but with 
significantly improved precision. However, the optimal models for the 22 
targets included six stars with an initial helium mass fraction $Y_{\rm 
i}$ significantly below the primordial value from standard Big Bang 
nucleosynthesis \citep[$Y_{\rm P}=0.2482\pm0.0007$,][]{ste10}, and four 
additional stars that were marginally below $Y_{\rm P}$. The original 
motivation for including these sub-primordial values in the search was a 
recognition that there could be systematic errors in the determination of 
$Y_{\rm i}$, but the source of the potential bias was not identified. Our 
first attempts to fit the data described in section~\ref{SEC2} using the 
same methods as \cite{mat12} were plagued by an even higher fraction of 
models with low initial helium, so we revised our procedures.

\subsection{Updated fitting procedures}\label{SEC3.2}

As part of a study of convective cores in two {\it Kepler} targets, AMP 
was compared to several other fitting methods by \cite{sil13}. In addition 
to the individual frequencies and spectroscopic constraints, some of these 
methods also used sets of frequency ratios that eliminated the need to 
correct the model frequencies for surface effects \citep{rv03}. A 
comparison of the AMP results with models that used the frequency ratios 
as additional constraints revealed systematic differences in the interior 
structure that were correlated with the initial helium abundance. We 
subsequently modified the AMP optimization procedure to try to avoid this 
bias, by adopting the frequency ratios as additional constraints and by 
reducing the weight at higher frequency, where the surface correction is 
larger (see details below).

The ratios proposed by \cite{rv03} are constructed from individual 
frequency separations, including the large separations defined by 
$\Delta\nu_\ell(n) = \nu_{n,\ell} - \nu_{n-1,\ell},$ and the small 
separations defined by $d_{\ell,\ell+2}(n) = \nu_{n,\ell} - 
\nu_{n-1,\ell+2}.$ These can be used to define one set of ratios that 
relates the small separation between modes of degree 0 and 2 to the large 
separation of $\ell=1$ modes at the same radial order:
 \begin{equation}
 r_{02}(n) = \frac{d_{0,2}(n)}{\Delta\nu_1(n)}.\label{eq:r02}
 \end{equation}
Note that these ratios involve modes with all three degrees. Another set 
of ratios only involves the small and large separations between modes of 
degree 0 and 1:
 \begin{equation}
 r_{01}(n) = \frac{d_{01}(n)}{\Delta\nu_1(n)},\ 
 r_{10}(n) = \frac{d_{10}(n)}{\Delta\nu_0(n+1)},\label{eq:r010}
 \end{equation} 
where $d_{01}(n)$ and $d_{10}(n)$ are 5-frequency smoothed small 
separations defined by equations (4) and (5) in \cite{rv03}. This 
smoothing introduces correlations between the individual ratios that more 
than double the effective uncertainties\footnote{We examined a specific 
case from \cite{sil13} and compared the quadratic sum of all terms in the 
covariance matrix to the diagonal element for each ratio. We determined 
that the off-diagonal elements inflate the effective uncertainty by a 
factor of 2--4 with the largest boost near the center of the observed 
frequency range.}, but it also shifts some weight from the center of the 
frequency range toward the edges where the S/N of the modes is 
lower\footnote{Ratios near the edges of the observed frequency range are 
directly correlated with fewer than 4 other ratios, so the correlated 
errors are not inflated as much relative to the diagonal elements of the 
covariance matrix and these less certain ratios are assigned higher 
relative weights.}. To account for these correlations without shifting 
weight toward the edges of the frequency range, we adopted 3$\sigma$ 
uncorrelated uncertainties on all ratios from Eq.(\ref{eq:r010}). In 
addition, \cite{sil13} noted that the ratios formed from the highest 
radial orders were typically unreliable due to large line-widths, and 
recommended that they be excluded from the set of constraints. We excluded 
all ratios that are centered on frequencies from the highest three radial 
orders. Hereafter, we refer to the set of ratios $r_{02}(n)$ as $r_{02}$ 
and the set of ratios $r_{01}(n)$ and $r_{10}(n)$ as $r_{010}$.

Although the frequency ratios help to discriminate between families of 
models that provide comparable matches to the other sets of constraints, 
the individual frequencies contain additional information that we would 
like to exploit. The primary difficulty is that the model frequencies need 
to be corrected for surface effects, and the commonly-used empirical 
correction \citep{kje08} appears to inject a bias in the determination of 
some stellar properties. What is the source of this bias, and how can we 
mitigate it? Essentially, \citeauthor{kje08}\ assumed that the differences 
between the observed and optimal model frequencies can be described by
 \begin{equation}
 \nu_{\rm obs} - \nu_{\rm mod} \approx a_0 \left(\frac{\nu}{\nu_0}\right)^b,
 \label{eq:surf}
 \end{equation}
where $a_0$ is the size of the correction at a reference frequency $\nu_0$ 
(typically chosen to be $\nu_{\rm max}$), and the exponent $b$ is fixed at 
a solar-calibrated value near 4.9. They demonstrated that this simple 
parametrization\footnote{Note that \cite{kje08} also scaled the model 
frequencies by a homology factor $r$ to provide a better match to the 
observations. By definition, the best model should have $r=1$. The net 
effect of applying homology scaling to every model sampled by the GA is to 
decrease the dynamic range of the frequency $\chi^2$-space, so we omitted 
this term from our surface correction.} can adequately describe the 
frequency differences between the observations and models of several 
solar-type stars, including $\beta$~Hyi and $\alpha$~Cen A and B. They 
cautioned that the value of the exponent depends on the number of radial 
orders considered for the solar calibration, varying from 4.4--5.25 when 
including 7--13 orders. To facilitate comparisons with previous work, we 
adopted the solar-calibrated value $b=4.82$ determined by \cite{met09}. 
For mixed modes we scaled the surface correction by the mode inertia 
ratio, as described in \cite{bra11}.

\begin{figure}[t] 
  \centerline{\includegraphics[angle=270,width=\columnwidth]{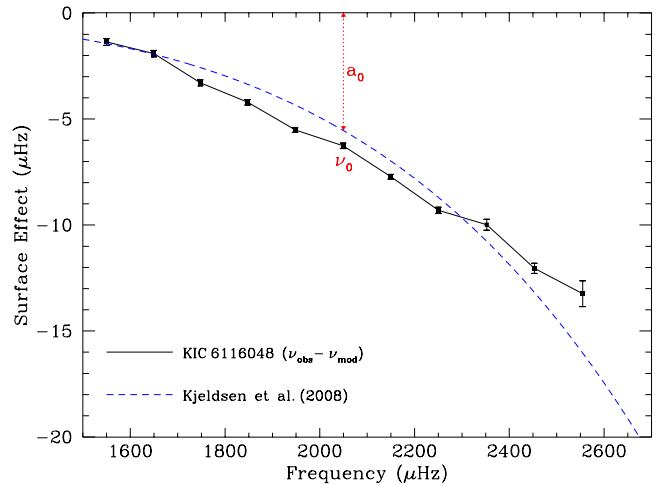}} 
  \caption{Comparison of the actual surface effect and the empirical 
  correction of \cite{kje08} for the AMP model of KIC\,6116048. 
  Differences between the observed $\ell=0$ frequencies and those of the 
  AMP model (connected points) are reasonably well represented by the 
  empirical correction (dashed line) with amplitude $a_0$ at the 
  reference frequency $\nu_0$, but it substantially overestimates the 
  correction at high frequencies.\label{fig2}} 
  \vspace*{6pt}
  \end{figure}

The actual solar surface effect appears more linear at high frequencies 
\citep{jcd96}, so assuming any fixed exponent will tend to over-correct 
the highest-order modes (see Figure~\ref{fig2}). This tendency appears to 
interact with intrinsic parameter correlations---in particular, the 
well-known correlation between mass and initial helium abundance in 
stellar models---to favor higher-mass low-helium models that fit the 
frequencies better while getting the interior structure wrong. Including 
the frequency ratios as constraints favors the lower-mass higher-helium 
models, but it does not eliminate the bias caused by the high-frequency 
modes. To mitigate this bias, we adopted an uncertainty for each frequency 
that is the quadratic sum of the statistical error and half the surface 
correction \citep{br92}. In doing so, we are acknowledging that surface 
effects represent a systematic error in the models \citep{gb04}. We also 
imposed a penalty on models with $Y_{\rm i}<Y_{\rm P}$, such that the 
spectroscopic quality metric (see section~\ref{SEC3.3}) was inflated by 1 
for every 0.01 that $Y_{\rm i}$ fell below $Y_{\rm P}$. Although this does 
not explicitly rule out low-helium solutions, it does require that they 
provide a substantially better match to the other sets of constraints to 
be considered superior. Without a precise constraint on the luminosity 
and/or radius, this approach is required even to recover accurate solar 
properties from Sun-as-a-star helioseismic data \citep{met09}.

  \begin{figure*}[t] 
  \centerline{\includegraphics[width=2.25in]{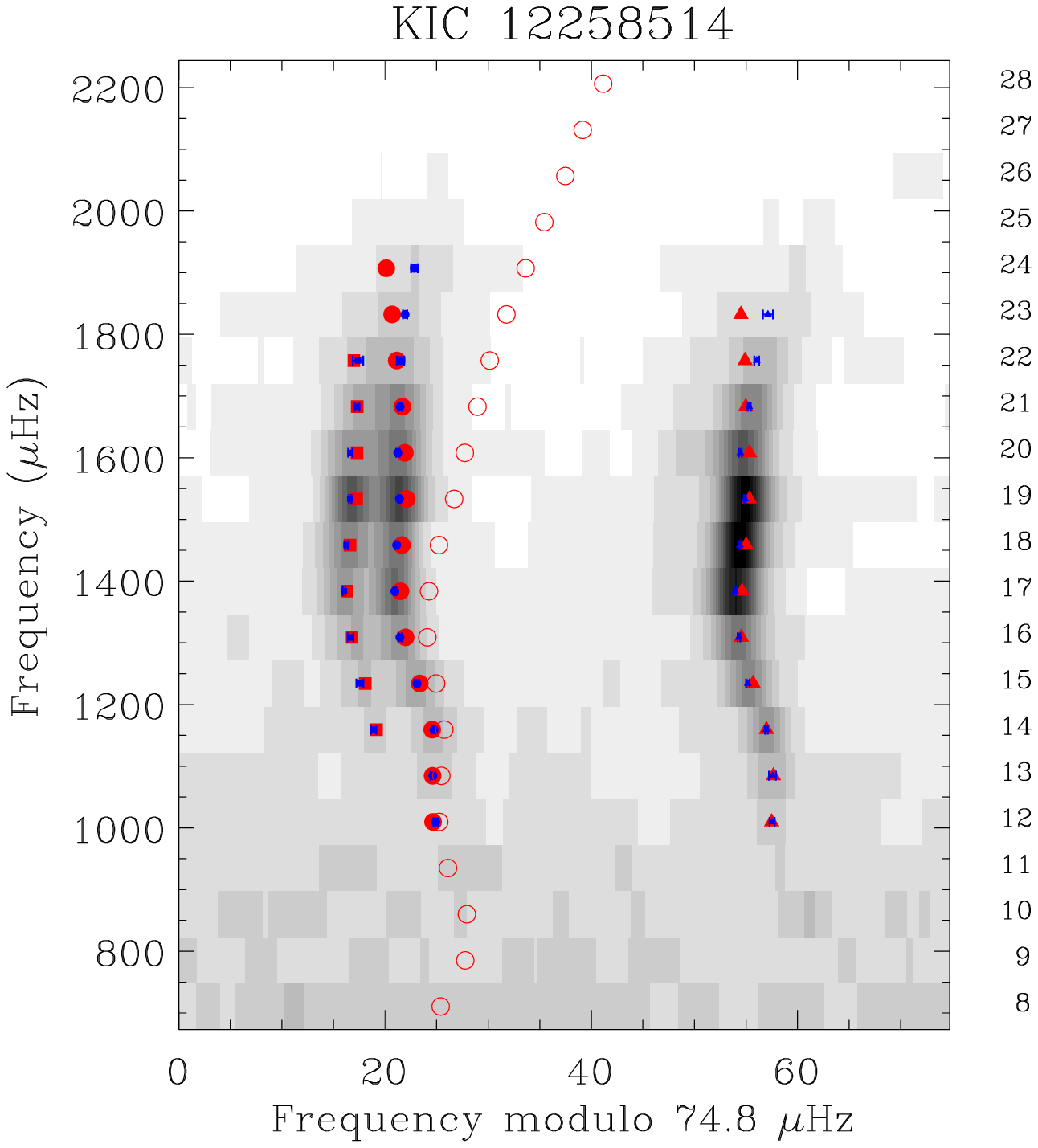}\hspace*{0.125in}\includegraphics[width=2.25in]{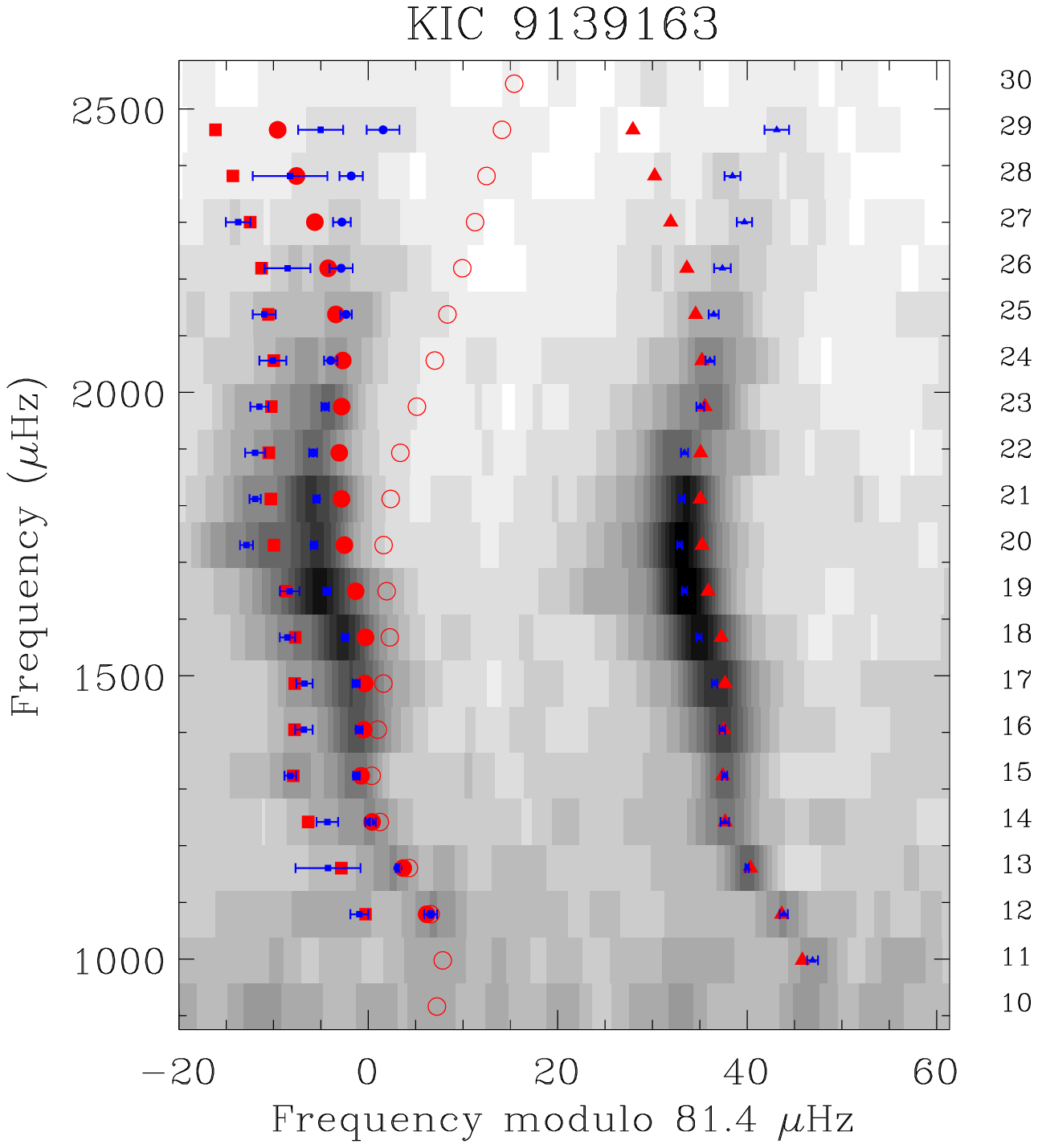}\hspace*{0.125in}\includegraphics[width=2.25in]{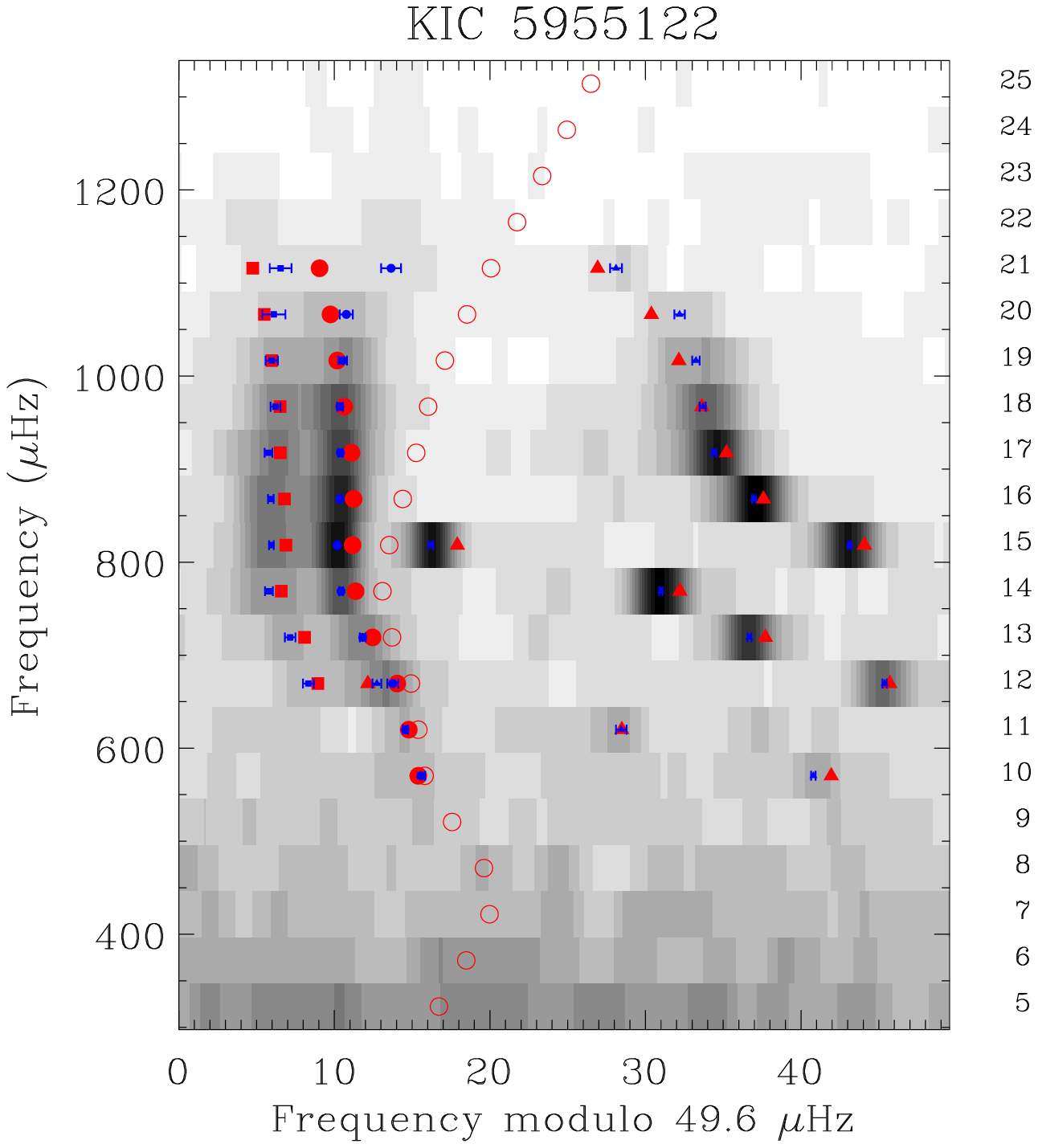}}
  \caption{{\'E}chelle diagrams for typical examples of the three 
  different star types, including the simple star KIC\,12258514 (left), 
  the F-like star KIC\,9139163 (center), and the mixed-mode star 
  KIC\,5955122 (right). Each plot shows a smoothed greyscale representation 
  of the power spectrum overplotted with the frequencies and errors from
  \citeauthor{app12}\ (2012, blue points) and from the surface-corrected 
  AMP model (solid red points). The uncorrected $\ell=0$ model 
  frequencies (open red circles) are shown to illustrate the size of the 
  surface effect, and the $\ell=0$ radial order is given on the right 
  axis (Figures 3.1--3.42 are available in the online version of the
  Journal).\label{fig3}}
  \vspace*{6pt}
  \end{figure*} 

\subsection{Customization by star type}\label{SEC3.3}

\cite{app12} categorized their sample of 61 asteroseismic targets into 
three classes, based on the appearance of the oscillation modes in an 
{\'e}chelle diagram \citep{gre80}. Dividing the frequency spectrum into 
segments having width equal to the large separation and then stacking them 
vertically, modes with the same spherical degree form approximately 
vertical ridges for {\it simple stars} like the Sun (see 
Figure~\ref{fig3}). Significantly hotter main-sequence stars have larger 
intrinsic line-widths, blurring the individual modes and complicating the 
identification of mode geometry ({\it F-like stars}). Finally, the 
$\ell=1$ ridge in subgiants can be disrupted when buoyancy modes in the 
evolved stellar core couple with pressure modes in the envelope, creating 
an avoided crossing \citep{osa75,aiz77} that leads to deviations from the 
regular frequency spacing ({\it mixed-mode stars}). This final category is 
usually unambiguous, but there is no clear dividing line between the first 
two.

\cite{app12} suggested a boundary between the simple and F-like stars at 
an effective temperature near 6400~K or a line-width at maximum mode 
height around 4~$\mu$Hz \citep[see][]{whi12}. We adopted a slightly 
different convention based on whether or not the $\ell=0$ and $\ell=2$ 
ridges in the {\'e}chelle diagram are cleanly separated. This led us to 
treat 5 stars as F-like that were identified as simple by 
\citeauthor{app12}: KIC\,3632418, 7206837, 8228742, 9139163 and 10162436. 
In addition, KIC\,3424541 (originally classified as F-like) and 
KIC\,10018963 (classified as simple by \citeauthor{app12}) both show 
evidence of avoided crossings, so we treated them as mixed-mode stars.

The use of frequency ratios as additional constraints for asteroseismic 
model-fitting can improve the uniqueness of the solution, but the ratios 
cannot all be used for certain types of stars. For simple stars, the large 
and small spacings can be measured cleanly, and the underlying assumption 
that the frequencies are all pure p-modes is justified. In this case, AMP 
attempts to match four sets of observational constraints simultaneously: 
[1] the individual oscillation frequencies, with uncertainties inflated in 
proportion to the surface correction, [2] the ratios $r_{010}$ with 
3$\sigma$ uncorrelated statistical uncertainties, [3] the ratios $r_{02}$ 
with errors propagated from the quoted frequency uncertainties, and [4] 
the spectroscopic and other constraints, such as a luminosity or 
interferometric radius. A normalized $\chi^2$ is calculated for each of 
these sets of constraints:
 \begin{equation}
 \chi^2 = \frac{1}{N} \sum_{i=1}^{N} \left( \frac{O_i - C_i}{\sigma_i} \right)^2,
 \end{equation} 
where $O_i$ and $C_i$ are the sets of $N$ observed and calculated 
quantities, and $\sigma_i$ are the associated uncertainties\footnote{We 
did not consider covariances for any of our quality metrics. \cite{lg14} 
recently demonstrated that including a full treatment of correlations has 
a negligible impact on the derived stellar properties.}. The GA then 
attempts to minimize the average of the $\chi^2$ values for the adopted 
sets of constraints\footnote{This decision compromises the statistical 
purity of our metric, as discussed in section~\ref{SEC4.1}, so future 
updates to AMP will preserve the individual $\chi^2$ values for subsequent 
error analysis.}. This approach recognizes that each oscillation frequency 
is not completely independent, but it uses the information content in 
several different ways to create metrics that can be traded off against 
each other and/or against the spectroscopic constraints.

These procedures must be modified slightly for F-like and mixed-mode 
stars. In F-like stars, large line-widths make the measurement of $\ell=0$ 
and $\ell=2$ frequencies more difficult. Consequently, the small spacings 
$d_{0,2}(n)$ are compromised and the ratios $r_{02}$ are unreliable. In 
this case, set [3] above is excluded from consideration, and the GA uses 
the average $\chi^2$ for the remaining three sets of constraints. For 
mixed-mode stars, the $\ell=1$ frequencies are not pure p-modes, so the 
theoretical insensitivity of the ratios $r_{010}$ to the near-surface 
layers is no longer valid and they lose their utility. This is not limited 
to the modes that are immediately adjacent to an avoided 
crossing---several radial orders on either side are generally perturbed, 
depending on the strength of the mode coupling \citep{deh11,ben13}. In 
addition, note that the ratios $r_{02}$ in Eq.(\ref{eq:r02}) depend on the 
large separations $\Delta\nu_1(n)$, which are also contaminated by mixed 
modes. So in this case we must exclude both sets [2] and [3] above, 
leaving only the individual frequencies (assigned double 
weight\footnote{For mixed-mode stars, the information content of the 
individual frequencies is not redistributed among any sets of frequency 
ratios, so we compensate by assigning double weight to the frequencies 
relative to the spectroscopic constraints.}) and spectroscopic 
constraints. The modified treatment of errors on the individual 
frequencies is still in effect for the mixed-mode stars, so the new 
procedure is different from the approach taken by \cite{mat12}.

  \begin{deluxetable*}{lccrcccrrc}
  \tablewidth{\textwidth}
  \tablecaption{Properties of the optimal models and surface correction from AMP\label{tab1}}
  \tablehead{\colhead{KIC} & \colhead{$R/R_\odot$\tablenotemark{a}} & \colhead{$M/M_\odot$\tablenotemark{a}} & \colhead{$t$/Gyr\tablenotemark{a}} & 
  \colhead{$Z$} & \colhead{$Y_{\rm i}$} & \colhead{$\alpha$} & \colhead{$a_0$} & \colhead{$f$} & \colhead{AMP\tablenotemark{b}}}
  \startdata
  \sidehead{Simple stars}
   3427720       & $1.125\pm0.014$ & $1.13\pm0.04$ & $ 2.23\pm0.17$ & $0.0168\pm0.0016$ & $0.248\pm0.020$ & $1.98\pm0.11$ &  $-$3.65 &   1.2 & 641 \\
   6116048       & $1.219\pm0.009$ & $1.01\pm0.03$ & $ 6.23\pm0.37$ & $0.0118\pm0.0011$ & $0.255\pm0.014$ & $1.80\pm0.10$ &  $-$5.55 &   6.2 & 642 \\
   6603624       & $1.181\pm0.015$ & $1.09\pm0.03$ & $ 8.11\pm0.46$ & $0.0309\pm0.0029$ & $0.250\pm0.021$ & $2.12\pm0.09$ &  $-$1.65 &  10.3 & 643 \\
   6933899       & $1.599\pm0.018$ & $1.14\pm0.03$ & $ 6.87\pm0.34$ & $0.0203\pm0.0022$ & $0.257\pm0.014$ & $2.10\pm0.07$ &  $-$1.98 &   2.8 & 497 \\
   7871531       & $0.874\pm0.008$ & $0.84\pm0.02$ & $ 9.15\pm0.47$ & $0.0125\pm0.0014$ & $0.263\pm0.018$ & $2.02\pm0.12$ &  $-$4.07 &   4.2 & 523 \\
   8006161       & $0.947\pm0.007$ & $1.04\pm0.02$ & $ 5.04\pm0.17$ & $0.0427\pm0.0052$ & $0.259\pm0.015$ & $2.22\pm0.08$ &  $-$1.99 &   2.0 & 494 \\
   8394589       & $1.116\pm0.019$ & $0.94\pm0.04$ & $ 2.92\pm0.18$ & $0.0082\pm0.0005$ & $0.308\pm0.028$ & $1.62\pm0.07$ &  $-$9.68 &   1.5 & 526 \\
   8694723       & $1.436\pm0.024$ & $0.96\pm0.03$ & $ 4.90\pm0.54$ & $0.0058\pm0.0006$ & $0.298\pm0.025$ & $1.52\pm0.10$ &  $-$7.70 &  36.8 & 516 \\
   8760414       & $1.010\pm0.004$ & $0.78\pm0.01$ & $ 3.69\pm0.74$ & $0.0032\pm0.0003$ & $0.238\pm0.006$ & $1.90\pm0.12$ &  $-$6.01 &  20.3 & 644 \\
   9098294       & $1.154\pm0.009$ & $1.00\pm0.03$ & $ 7.28\pm0.51$ & $0.0143\pm0.0018$ & $0.252\pm0.017$ & $2.00\pm0.12$ &  $-$4.76 &   1.6 & 509 \\
   9139151       & $1.146\pm0.011$ & $1.14\pm0.03$ & $ 1.71\pm0.19$ & $0.0224\pm0.0014$ & $0.289\pm0.018$ & $2.04\pm0.07$ &  $-$4.67 &   0.7 & 508 \\
   9955598       & $0.883\pm0.008$ & $0.89\pm0.02$ & $ 6.72\pm0.20$ & $0.0231\pm0.0017$ & $0.291\pm0.017$ & $2.06\pm0.09$ &  $-$2.69 &   3.6 & 524 \\
   10454113      & $1.250\pm0.015$ & $1.19\pm0.04$ & $ 2.03\pm0.29$ & $0.0168\pm0.0012$ & $0.252\pm0.018$ & $1.86\pm0.08$ &  $-$4.67 &   2.6 & 645 \\
   10644253      & $1.108\pm0.016$ & $1.13\pm0.05$ & $ 1.07\pm0.25$ & $0.0239\pm0.0024$ & $0.290\pm0.025$ & $1.96\pm0.12$ &  $-$7.48 &   1.5 & 527 \\
   10963065      & $1.213\pm0.008$ & $1.05\pm0.02$ & $ 4.30\pm0.23$ & $0.0118\pm0.0010$ & $0.262\pm0.012$ & $1.84\pm0.08$ &  $-$5.34 &   8.4 & 518 \\
   11244118      & $1.589\pm0.026$ & $1.10\pm0.05$ & $ 6.43\pm0.58$ & $0.0272\pm0.0034$ & $0.310\pm0.031$ & $2.16\pm0.19$ &  $-$2.08 &  10.0 & 499 \\
   12009504      & $1.375\pm0.015$ & $1.12\pm0.03$ & $ 3.64\pm0.26$ & $0.0152\pm0.0011$ & $0.282\pm0.023$ & $1.76\pm0.06$ &  $-$5.80 &   7.0 & 498 \\
   12258514      & $1.573\pm0.010$ & $1.19\pm0.03$ & $ 4.03\pm0.32$ & $0.0197\pm0.0015$ & $0.290\pm0.017$ & $1.88\pm0.08$ &  $-$3.57 &   3.9 & 490 \\
  \sidehead{F-like stars}
   1435467       & $1.641\pm0.027$ & $1.27\pm0.05$ & $ 1.87\pm0.14$ & $0.0203\pm0.0016$ & $0.317\pm0.023$ & $1.74\pm0.05$ &  $-$5.41 &   8.9 & 552 \\
   2837475       & $1.592\pm0.027$ & $1.39\pm0.06$ & $ 0.83\pm0.12$ & $0.0224\pm0.0018$ & $0.315\pm0.029$ & $2.12\pm0.12$ &  $-$4.08 &   1.9 & 553 \\
   3632418       & $1.835\pm0.034$ & $1.27\pm0.03$ & $ 2.88\pm0.38$ & $0.0143\pm0.0015$ & $0.288\pm0.027$ & $1.88\pm0.08$ &  $-$3.51 &  14.2 & 646 \\
   3733735       & $1.367\pm0.023$ & $1.32\pm0.04$ & $ 0.12\pm0.06$ & $0.0179\pm0.0011$ & $0.300\pm0.021$ & $1.72\pm0.11$ &  $-$5.33 &   1.4 & 554 \\
   6508366       & $2.081\pm0.021$ & $1.36\pm0.04$ & $ 2.25\pm0.15$ & $0.0168\pm0.0011$ & $0.305\pm0.022$ & $1.84\pm0.09$ &  $-$2.77 &   2.9 & 536 \\
   6679371       & $2.186\pm0.015$ & $1.56\pm0.03$ & $ 1.81\pm0.12$ & $0.0191\pm0.0019$ & $0.245\pm0.014$ & $1.56\pm0.09$ &  $-$2.70 &   6.4 & 647 \\
   7103006       & $1.898\pm0.026$ & $1.43\pm0.05$ & $ 1.49\pm0.14$ & $0.0247\pm0.0025$ & $0.313\pm0.026$ & $1.88\pm0.10$ &  $-$3.78 &  12.1 & 530 \\
   7206837       & $1.555\pm0.016$ & $1.46\pm0.05$ & $ 0.22\pm0.04$ & $0.0299\pm0.0012$ & $0.271\pm0.022$ & $1.34\pm0.08$ & $-$10.35 &  13.1 & 556 \\
   8228742       & $1.809\pm0.014$ & $1.27\pm0.02$ & $ 3.84\pm0.29$ & $0.0147\pm0.0015$ & $0.247\pm0.017$ & $1.64\pm0.06$ &  $-$5.02 &  24.9 & 648 \\
   9139163       & $1.532\pm0.021$ & $1.36\pm0.03$ & $ 1.07\pm0.16$ & $0.0272\pm0.0014$ & $0.302\pm0.017$ & $1.84\pm0.07$ &  $-$4.33 &   7.3 & 557 \\
   9206432       & $1.479\pm0.014$ & $1.40\pm0.03$ & $ 0.19\pm0.07$ & $0.0290\pm0.0015$ & $0.312\pm0.011$ & $1.68\pm0.10$ &  $-$5.94 &   5.5 & 532 \\
   9812850       & $1.745\pm0.024$ & $1.39\pm0.05$ & $ 1.68\pm0.13$ & $0.0162\pm0.0011$ & $0.252\pm0.022$ & $1.26\pm0.09$ &  $-$7.82 &   6.1 & 649 \\
   10162436      & $1.903\pm0.020$ & $1.23\pm0.02$ & $ 2.86\pm0.33$ & $0.0162\pm0.0009$ & $0.316\pm0.012$ & $1.58\pm0.08$ &  $-$5.46 &  24.2 & 539 \\
   10355856      & $1.667\pm0.015$ & $1.32\pm0.03$ & $ 1.54\pm0.13$ & $0.0152\pm0.0009$ & $0.280\pm0.018$ & $1.18\pm0.15$ &  $-$6.84 &   7.1 & 650 \\
   11081729      & $1.382\pm0.021$ & $1.26\pm0.03$ & $ 0.86\pm0.21$ & $0.0162\pm0.0006$ & $0.303\pm0.017$ & $2.00\pm0.11$ &  $-$6.72 &   7.1 & 551 \\
   11253226      & $1.551\pm0.019$ & $1.41\pm0.05$ & $ 0.56\pm0.20$ & $0.0197\pm0.0012$ & $0.279\pm0.025$ & $1.22\pm0.26$ &  $-$7.64 &   4.7 & 555 \\
  \sidehead{Mixed-mode stars}
   3424541       & $2.526\pm0.065$ & $1.64\pm0.04$ & $ 2.28\pm0.13$ & $0.0309\pm0.0036$ & $0.257\pm0.018$ & $1.74\pm0.12$ &  $-$5.44 &   2.3 & 636 \\
   5955122       & $2.042\pm0.025$ & $1.12\pm0.05$ & $ 5.26\pm0.58$ & $0.0143\pm0.0023$ & $0.290\pm0.022$ & $1.76\pm0.15$ &  $-$2.69 &  15.6 & 537 \\
   7747078       & $1.889\pm0.023$ & $1.06\pm0.05$ & $ 6.26\pm0.92$ & $0.0103\pm0.0019$ & $0.271\pm0.020$ & $1.76\pm0.20$ &  $-$5.63 &  38.6 & 541 \\
   7976303       & $1.961\pm0.041$ & $1.10\pm0.05$ & $ 4.78\pm0.58$ & $0.0077\pm0.0015$ & $0.268\pm0.016$ & $1.92\pm0.19$ &  $-$7.36 & 150.0 & 538 \\
   8026226       & $2.753\pm0.041$ & $1.50\pm0.03$ & $ 2.23\pm0.13$ & $0.0125\pm0.0010$ & $0.247\pm0.007$ & $1.28\pm0.13$ &  $-$4.99 &  32.5 & 651 \\
   8524425       & $1.733\pm0.015$ & $1.00\pm0.07$ & $ 7.98\pm0.46$ & $0.0197\pm0.0022$ & $0.313\pm0.041$ & $1.72\pm0.11$ &  $-$3.27 &  15.2 & 564 \\
   10018963      & $1.915\pm0.020$ & $1.18\pm0.03$ & $ 4.36\pm0.34$ & $0.0097\pm0.0011$ & $0.255\pm0.010$ & $1.96\pm0.11$ &  $-$2.66 & 113.2 & 544 \\
   11026764      & $2.106\pm0.025$ & $1.27\pm0.06$ & $ 5.00\pm0.53$ & $0.0197\pm0.0029$ & $0.254\pm0.016$ & $2.10\pm0.37$ &  $-$1.62 &   4.6 & 567 \\
  \enddata
  \tablenotetext{a}{Formal uncertainties do not include a typical systematic component of 1.3\% for radius, 3.7\% for mass, and 12\% for age.}
  \tablenotetext{b}{Comprehensive model output is available at http://amp.phys.au.dk/browse/simulation/\#\#\#}
  \end{deluxetable*}
  \vspace*{12pt}

\section{RESULTS}\label{SEC4}

The properties derived by AMP for our sample of 42 main-sequence and 
subgiant {\it Kepler} targets are listed in Table~\ref{tab1}. For each 
star in the three categories, we give the {\it Kepler} Input Catalog 
\citep[KIC,][]{bro11} number, the asteroseismic radius $R$, mass $M$, age 
$t$, heavy element mass fraction $Z$, initial helium mass fraction $Y_{\rm 
i}$, mixing-length $\alpha$, and amplitude of the surface correction $a_0$ 
at the reference frequency $\nu_{\rm max}$ \cite[where $\nu_{\rm max}$ was 
taken from][]{app12}. As an indication of the relative quality of each 
model, we list an average of the normalized $\chi^2$ values discussed in 
section~\ref{SEC3.3}---similar to the metric that was used by the GA to 
identify the optimal model, but using the original frequency errors from 
\cite{app12} rather than inflating them in proportion to the surface 
correction. This modified metric $f$ is more useful for comparing the 
quality of the models for different stars, and can be used in conjunction 
with the {\'e}chelle diagrams in Figure~\ref{fig3} to judge the 
reliability of each result. Some of the models are clearly better 
representations of the observations than others (see discussion below), so 
we caution readers not to treat them all as equivalent. In the final 
column of Table~\ref{tab1} we list the AMP run number so that interested 
readers can access a comprehensive archive with the evolution and 
structure of each stellar model.

  \begin{figure*}[t] 
  \centerline{\includegraphics[width=2.25in]{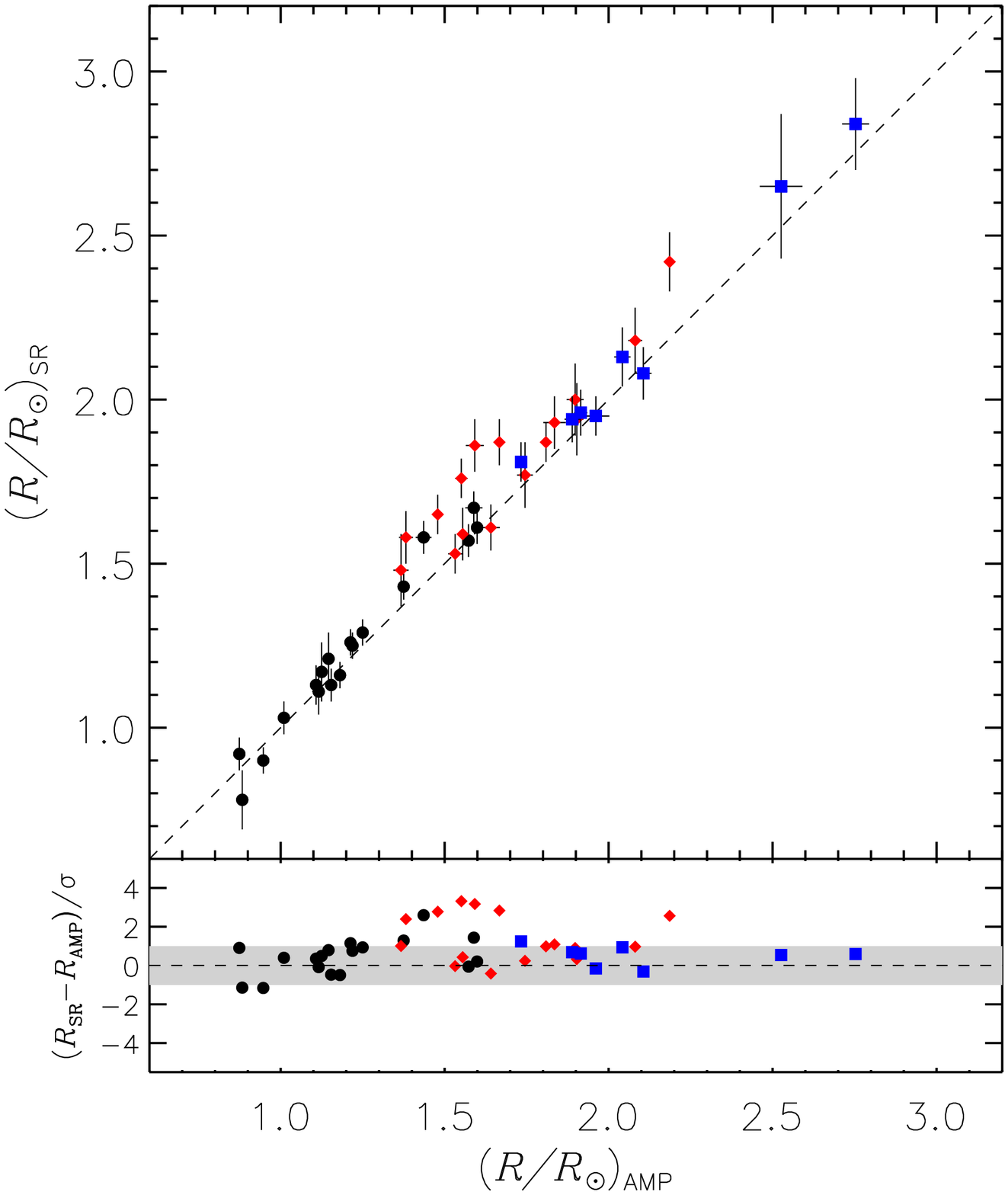}\hspace*{0.125in}\includegraphics[width=2.25in]{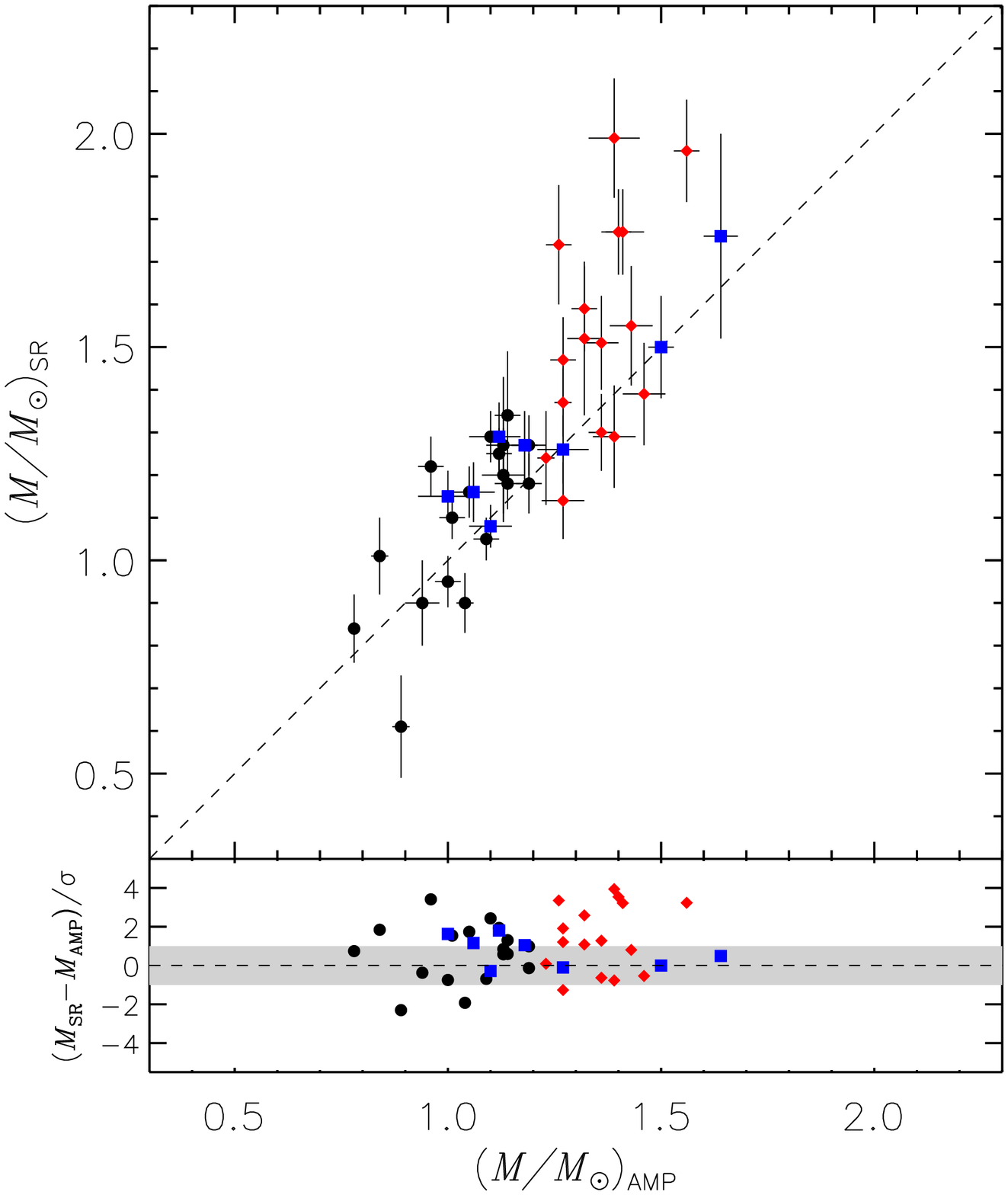}}
  \caption{Comparison of the asteroseismic radii (left) and masses 
  (right) derived from scaling relations (SR) with the AMP estimates 
  from Table~\ref{tab1}, including the simple (black circles), F-like 
  (red diamonds), and mixed-mode stars (blue squares). The top panels 
  compare the values and uncertainties, while the bottom panels show 
  the differences between the estimates normalized by the uncertainty on 
  the difference.\label{fig4}}
  \vspace*{6pt}
  \end{figure*} 

Figure~\ref{fig3} shows {\'e}chelle diagrams for typical examples of each 
of the three different star types, with similar plots for all 42 stars 
included in the online version of the Journal. A common feature in many of 
these diagrams is a divergence between the observations (blue points) and 
the surface corrected model frequencies (solid red points), which appear 
to curve off towards the left at higher frequencies. This feature is most 
obvious in the F-like star shown in the center panel, which has a larger 
number of observed radial orders. As discussed in section~\ref{SEC3.2}, 
this is an indication that the empirical surface correction of 
\cite{kje08} tends to over-correct the highest order modes, motivating our 
decision to decrease the weight of these frequencies in proportion to the 
size of the surface correction (i.e.\ the difference between the solid and 
open red circles in each plot). Thus, the primary constraints from the 
individual frequencies are concentrated at low frequency, while the ratios 
capture the information content of the higher frequencies (excluding the 
highest three radial orders, see section~\ref{SEC3.2}). Forcing a better 
fit to the high frequency modes, under the current approach to the surface 
correction, leads to the systematic bias towards low-helium solutions that 
affected the results of \cite{mat12}.

\subsection{Statistical and systematic uncertainties}\label{SEC4.1}

The assessment of uncertainties on the adjustable parameters and other 
model properties requires some degree of pragmatism. Our previous 
approach, based on the local shape of the $\chi^2$ surface using singular 
value decomposition \citep[SVD,][]{cre09,met09}, generally fails to 
capture uncertainties due to non-uniqueness of the solution---a common 
outcome when we combine several different metrics of model quality. Rather 
than accept the underestimated uncertainties from SVD, we opted for a more 
conservative approach using an ensemble of the best models sampled by the 
GA on its way to finding the optimal solution. First, we ranked the 
50,000--80,000 unique models for each star by the average $\chi^2$ value 
and assigned each model a likelihood:
 \begin{equation}
 \mathcal{L} = \exp\left(-\chi^2/2\right). \label{eq:L}
 \end{equation}
Next, we calculated a likelihood-weighted mean value and standard 
deviation for each parameter, including additional models in the mean 
until the uncertainty on $[Z/X]_{\rm i}$ was comparable to the 
observational error on [M/H]. This is a first approximation, since helium 
diffusion and settling gradually changes the value of $X$ as the model 
evolves to its final age. Using these uncertainties for the five 
adjustable model parameters ($M, t, Z, Y_{\rm i}, \alpha$), we rescaled 
the covariance matrix from the optimal solution and calculated a set of 
models to define the 1$\sigma$ error ellipse. The uncertainties on other 
model properties, such as $R$ and $T_{\rm eff}$, were determined from the 
range of values represented in these 1$\sigma$ models. Finally, we refined 
the number of models that were included in the mean so that the output 
uncertainty on the model $T_{\rm eff}$ was equal to the input error on the 
spectroscopic $T_{\rm eff}$. This procedure leads to an inherently 
conservative estimate of the uncertainties, because it implicitly assumes 
that the asteroseismic data do not contribute to the determination of 
$T_{\rm eff}$ in the final solution. With the ensemble of best models 
defined in this way for each star, we used them to determine the 
likelihood-weighted standard deviation (taken to be the uncertainty) on 
each parameter of the best solution identified by the GA (for a detailed 
example, see Appendix~\ref{APPA}).

  \begin{figure*}[t] 
  \centerline{\includegraphics[width=2.25in]{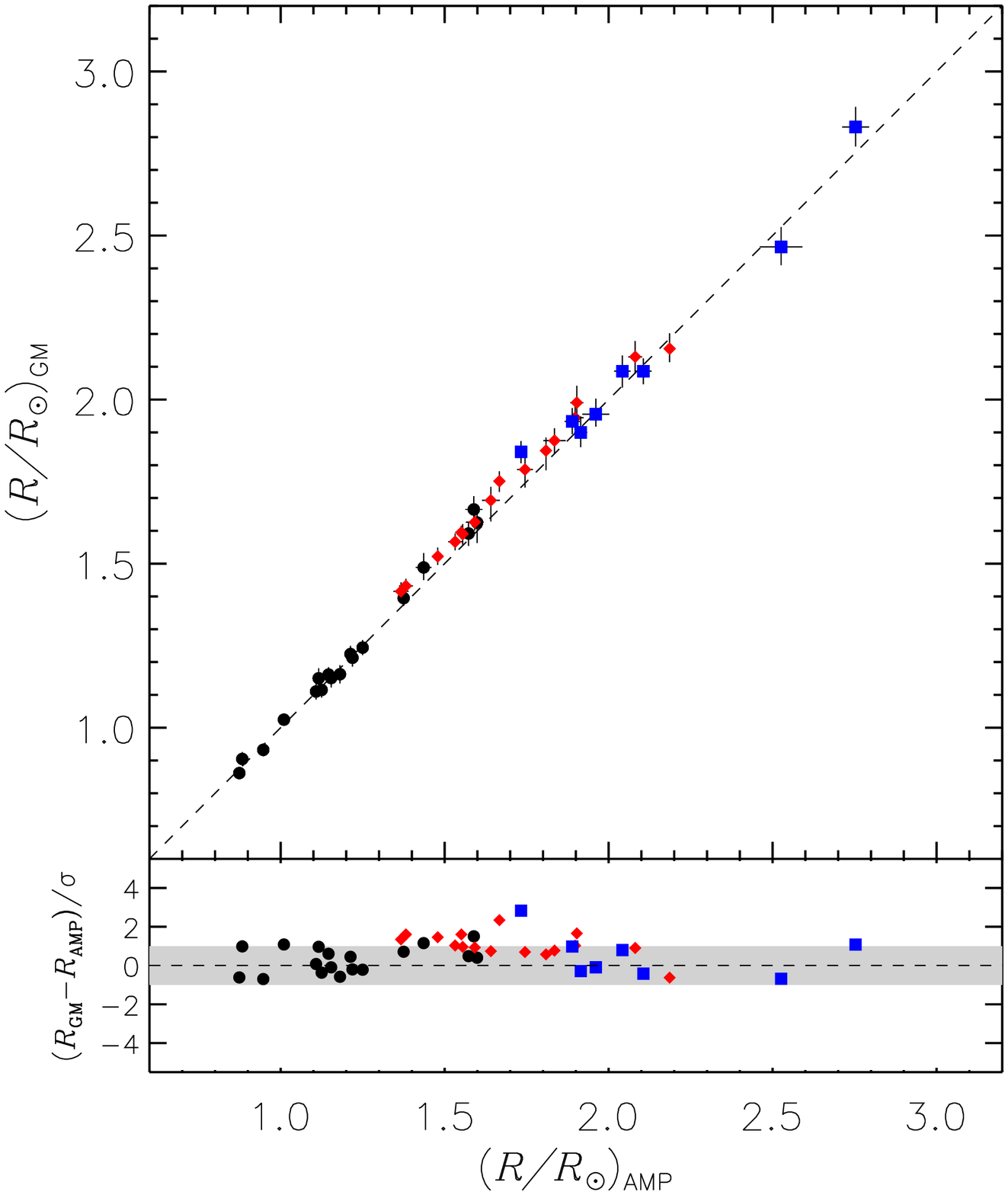}\hspace*{0.125in}\includegraphics[width=2.25in]{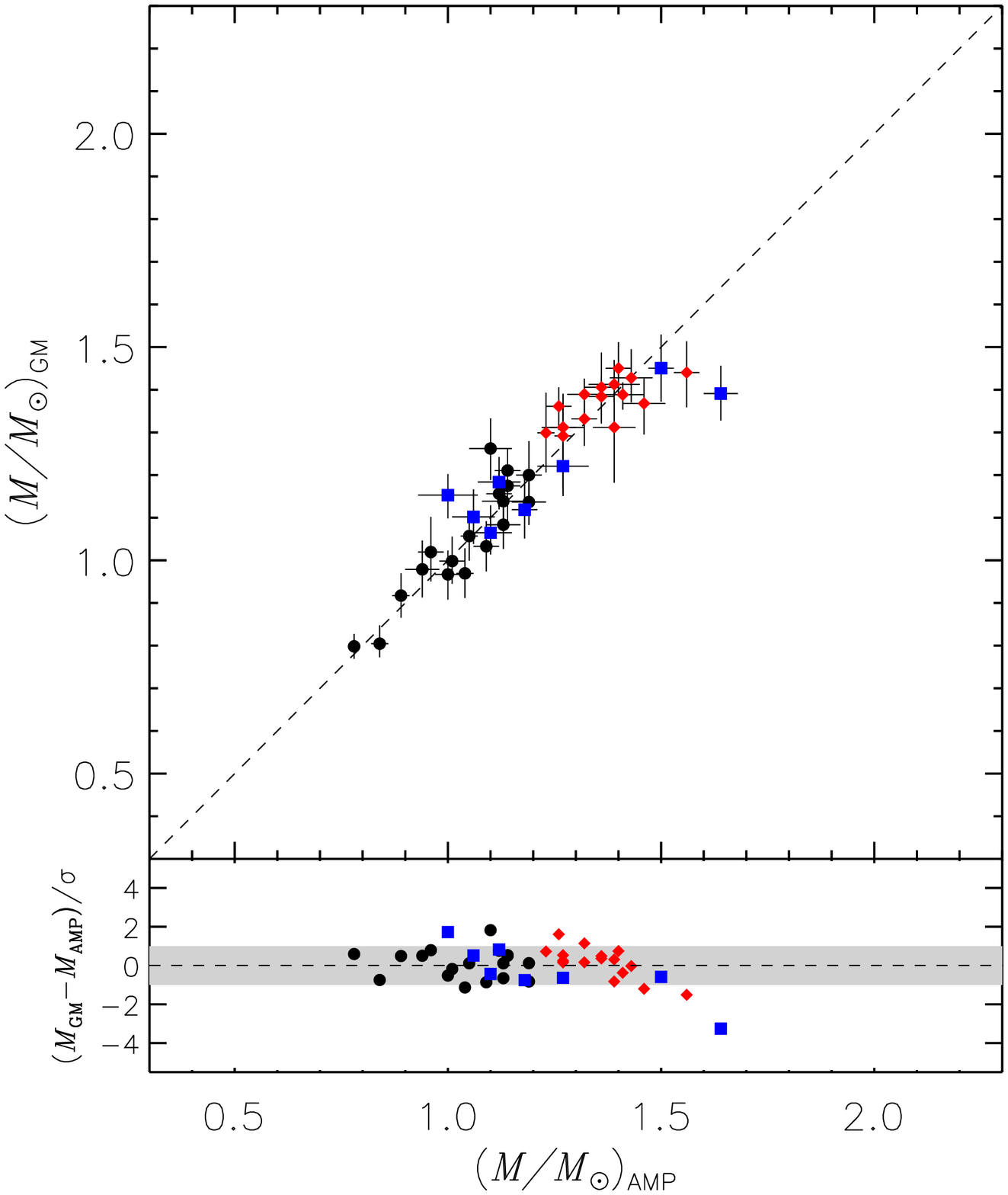}\hspace*{0.125in}\includegraphics[width=2.25in]{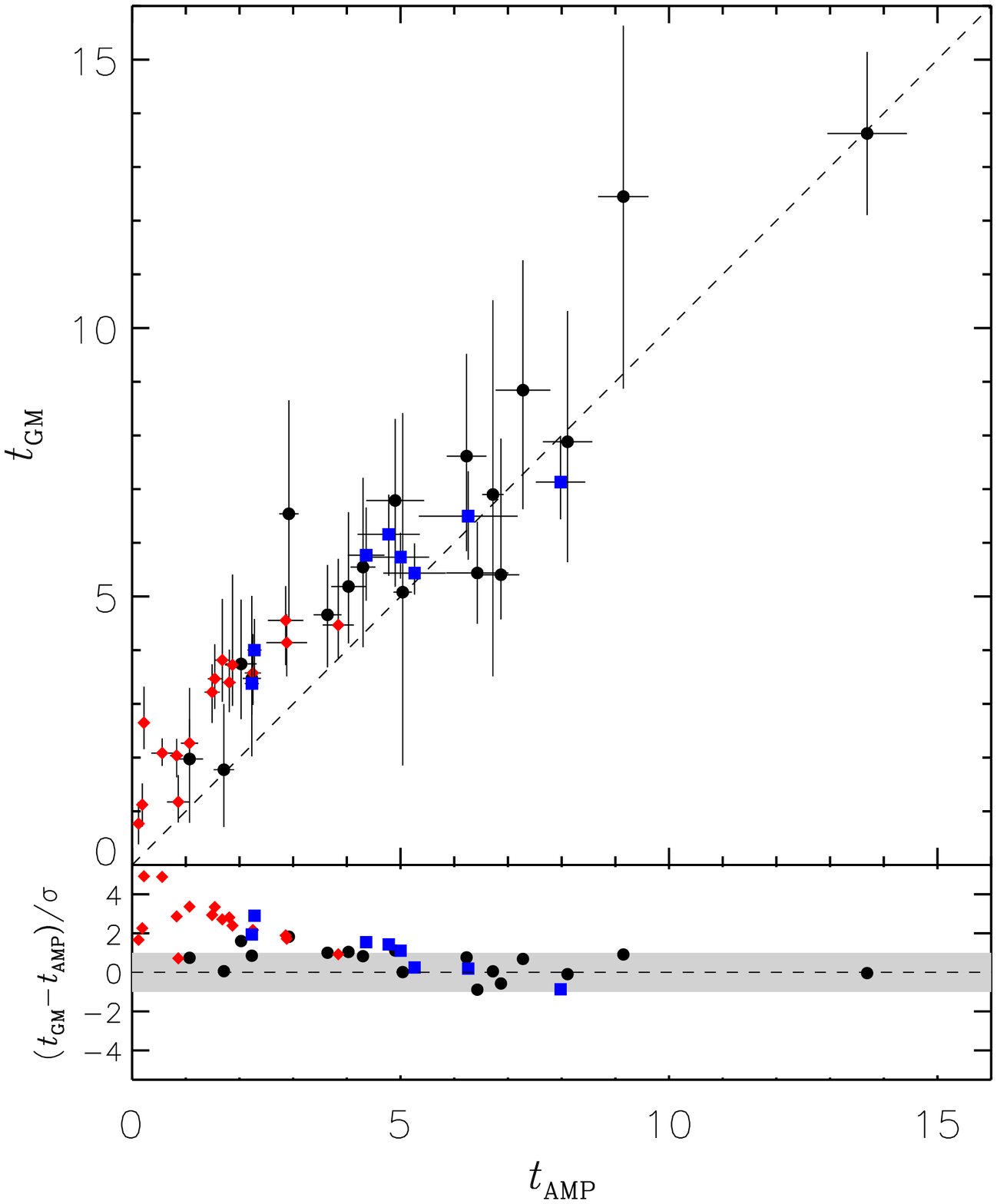}}
  \caption{Comparison of the asteroseismic radii (left), masses 
  (center), and ages (right) derived from grid-modeling 
  \citep[GM,][]{cha14} with the AMP estimates from Table~\ref{tab1}, 
  including the simple (black circles), F-like (red diamonds), and 
  mixed-mode stars (blue squares). The top panels compare the values and 
  uncertainties, while the bottom panels show the differences between the 
  estimates normalized by the uncertainty on the difference.\label{fig5}}
  \vspace*{6pt}
  \end{figure*} 

It is important to emphasize that the resulting uncertainties, listed in 
Table~\ref{tab1}, do not include systematic errors from our particular 
choice of model physics or fitting strategy. \cite{cha14} used results 
from six pipeline methodologies coupled to 11 different model grids to 
quantify the associated systematic uncertainties. They found that, in 
addition to the statistical uncertainties from a given model grid and 
fitting methodology, the systematic errors for stars with a spectroscopic 
$T_{\rm eff}$ and [M/H] were typically 1.3\% on the radius, 3.7\% on the 
mass, and 12\% on the age. Although these additional systematic errors may 
have an impact on the {\it absolute} values of the stellar properties 
derived by AMP, the uniformity of the data sources and modeling approach 
means that the {\it relative} values can be considered as reliable as the 
uncertainties listed in Table~\ref{tab1}. Readers who wish to combine our 
results with those from other sources should consider the additional 
systematic errors noted above.

\subsection{Comparison with other methods}\label{SEC4.2}

Setting aside the question of absolute accuracy for the derived stellar 
properties, we can compare the internal precision of the AMP results to 
other common methods of inferring asteroseismic radii, masses and ages. 
Such a comparison can quantify the benefits of modeling the individual 
frequencies, relative to using only the global oscillation properties such 
as $\Delta\nu$ and $\nu_{\rm max}$. Empirical scaling relations can 
provide model-independent estimates of the asteroseismic radius and mass 
using only the observed values of $\Delta\nu$, $\nu_{\rm max}$ and $T_{\rm 
eff}$ \citep{kb95}. In Figure~\ref{fig4} we compare the radii (left) and 
masses (right) derived from the scaling relations (SR) to those found by 
AMP. The top panels compare the actual values and associated 
uncertainties, while the bottom panels show differences between the 
estimates normalized by the uncertainty on the difference. For clarity, 
simple (black circles), F-like (red diamonds), and mixed-mode stars (blue 
squares) are shown in different colors. The median uncertainties from the 
scaling relations are 4.2\% on the radius, and 6.8\% on the mass. The 
scaling relation values for F-like stars are systematically higher on 
average than the estimates from AMP, but the agreement is generally better 
than 2$\sigma$.

In Figure~\ref{fig5} we show a similar comparison of AMP results with 
grid-modeling from \cite{cha14}, including the asteroseismic radii (left), 
masses (center) and ages (right). These grid-modeling (GM) results use the 
same spectroscopic constraints from \cite{bru12}, but the adopted model 
grid (yielding values closest to the median over all grids and methods) 
came from the GARSTEC code \citep{ws08}. The median uncertainties from 
grid-modeling of this sample are 2.1\% on the radius, 5.5\% on the mass, 
and 20\% on the age---an improvement of a factor of two for the radius and 
25\% for the mass compared to the typical precision from scaling 
relations. The grid-modeling radii for F-like stars are systematically 
higher compared to AMP, but again the agreement is generally better than 
2$\sigma$. For all three categories of stars, GARSTEC yields ages that are 
systematically older by $\sim$1~Gyr for most targets with AMP ages below 
$\sim$3~Gyr. \cite{cha14} noted this tendency of GARSTEC with respect to 
most of the other model grids they explored and attributed the offset to 
differences in the treatment of convective core overshoot, which is not 
included in the AMP models.

Modeling the individual frequencies with AMP led to a significant 
improvement in the internal precision of the derived stellar properties 
relative to estimates based on scaling relations or grid-modeling. The 
median uncertainties from AMP are 1.2\% on the radius, 2.8\% on the mass, 
and 7.9\% on the age---about a factor of three improvement over the radius 
and mass precision from scaling relations, and more precise than 
grid-modeling by about a factor of two in radius, mass, and age. It is 
more difficult to assess the absolute accuracy of the AMP results, but 12 
stars in our sample (KIC\,3632418, 3733735, 7747078, 8006161, 8228742, 
9139151, 9139163, 9206432, 10162436, 10454113, 11253226, and 12258514) 
have a parallax from {\it Hipparcos} \citep{van07}, allowing us to compare 
the predicted luminosities with those of the AMP models that incorporate 
this constraint. Ten of the AMP luminosities are within 1$\sigma$ of the 
predictions, and only two stars show larger deviations (1.7$\sigma$ for 
10162436, and 1.8$\sigma$ for 10454113)---fewer than expected for Gaussian 
distributed errors. One star (KIC\,8006161) also has a radius from CHARA 
interferometry \citep[$0.952\pm0.021\,R_\odot$,][]{hub12}, which is 
reproduced by the AMP model within 0.2$\sigma$ when the constraint was 
included and 2$\sigma$ when it was not. These subsamples suggest that, in 
addition to being more precise than other methods, the AMP results are 
also reasonably accurate.

\subsection{Applications of the derived stellar properties}\label{SEC4.3}

\begin{figure}[t] 
  \centerline{\includegraphics[width=\columnwidth]{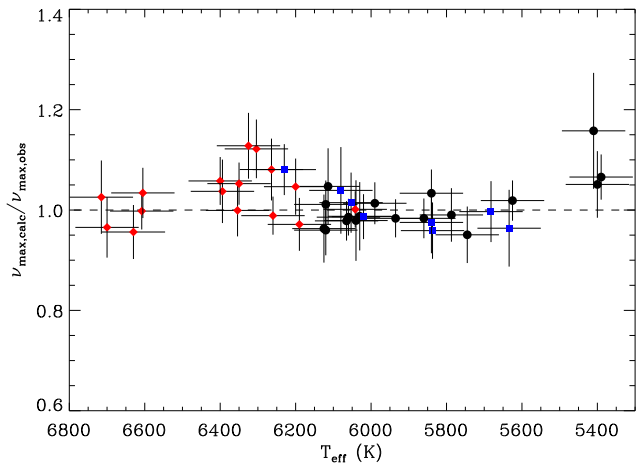}} 
  \caption{Ratio of $\nu_{\rm max}$ calculated from Eq.(\ref{eq:numax}) 
  to that measured from the observed power spectrum \citep{cha14} for 
  simple (black circles), F-like (red diamonds), and mixed-mode stars 
  (blue squares). Calculated values are derived from the radii and 
  masses listed in Table~\ref{tab1} using the $T_{\rm eff}$ values of
  \cite{bru12} with the errors from \cite{cha14}.\label{fig6}}
  \end{figure}

The radii and masses in Table~\ref{tab1} provide a new opportunity to test 
the $\nu_{\rm max}$ scaling relation for main-sequence and subgiant stars 
\citep[for previous discussion, see][]{ste09,bed11,bel11,bel13,hub12}. 
Such a test is meaningful because the AMP results were obtained without 
using the observed values of $\nu_{\rm max}$ as constraints. We consider 
the following form of the scaling relation \citep[Eq.10 of][]{kb95}:
 \begin{equation}
 \frac{\nu_{\rm max}}{\nu_{{\rm max},\odot}} = \frac{M/M_\odot}
 {\left(R/R_\odot\right)^2 \sqrt{T_{\rm eff}/T_{{\rm eff},\odot}}},\label{eq:numax}
 \end{equation}
with $\nu_{{\rm max},\odot} = 3090\,\mu$Hz \citep{cha14}. 
Figure~\ref{fig6} compares the values of $\nu_{\rm max}$ calculated using 
Eq.(\ref{eq:numax}) with those measured for our sample by \cite{cha14}. 
For about 70\% of the stars the agreement is within 1$\sigma$, confirming 
the relation---although there is some evidence of systematic deviations 
that merits further investigation.

\begin{figure}[t] 
  \centerline{\includegraphics[height=2.5in]{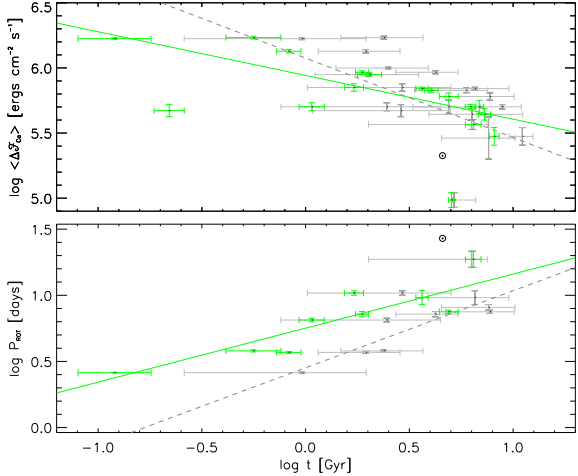}} 
  \caption{Age-activity (top, $\chi^2_R=247$) and age-rotation (bottom, 
  $\chi^2_R=38$) relations from the \cite{kar13} sample, using the 
  original ages from SEEK (grey, dashed line) and the updated ages from 
  AMP (green, solid line). Note that only SEEK results are shown for 
  KIC\,4914923, which is not included in our sample. The position of the 
  Sun is indicated by the $\odot$ symbol.\label{fig7}} 
  \vspace*{6pt}
  \end{figure}

The improved estimates of stellar age in Table~\ref{tab1} allow us to 
refine the age-rotation-activity relations derived by \cite{kar13}. Among 
their sample of 20 stars that have been monitored for chromospheric 
activity since 2009 from the Nordic Optical Telescope, 17 appear in 
Table~\ref{tab1}. The other three targets are: KIC\,4914923, which was not 
analyzed by \cite{app12}; the spectroscopic binary KIC\,8379927, which has 
contaminated atmospheric parameters; and the asteroseismic binary 
KIC\,10124866, which presents difficulties in extracting the overlapping 
oscillation spectra. Adopting the rotation periods $P_{\rm rot}$ and net 
Ca~{\sc ii} fluxes $\Delta\mathcal{F}_{\rm Ca}$ from Table~2 of 
\cite{kar13}, but replacing the grid-modeling ages from SEEK \citep{qui10} 
with the values from Table~\ref{tab1}, we find
 \begin{eqnarray}
 \log \Delta\mathcal{F}_{\rm Ca} &=& (-0.34\pm0.04)\log\ t+(5.94\pm0.02),\\
 \log P_{\rm rot} &=& (0.41\pm0.06) \log\ t + (0.75\pm0.03).
 \end{eqnarray}
The revised exponents on the age-activity and age-rotation relations (see 
Figure~\ref{fig7}) differ significantly from the values ($-0.61\pm0.17$ 
and $0.48\pm0.17$) derived by \cite{kar13}, and from the original values 
of $-$0.54 and 0.51 found by \cite{sku72}. A more detailed analysis of 
this sample has recently been performed by \cite{gar14}.

\begin{figure}[t] 
  \centerline{\includegraphics[width=\columnwidth]{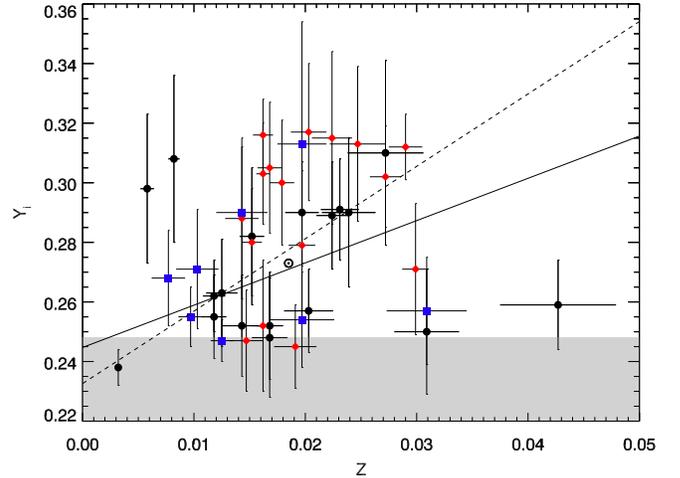}} 
  \caption{Linear least-squares fits for a relation between the 
  metallicity $Z$ and the initial helium mass fraction $Y_{\rm i}$, 
  considering errors only on the latter. Two linear fits are shown, 
  one including all stars (solid line), and another excluding three 
  outliers with $Z>0.03$ (dashed line). The shaded region denotes 
  initial helium below the primordial value \citep{ste10}.\label{fig8}} 
  \vspace*{6pt}
  \end{figure}

The asteroseismic bulk composition can be used to test the predictions of 
Galactic chemical enrichment, from which we expect a linear relation 
between metallicity and initial helium with slope $\Delta Y / \Delta Z$. 
Observational and theoretical techniques have been used to determine this 
slope using H\,{\sc ii} regions \citep{pei07}, eclipsing binaries 
\citep{rib00}, and main-sequence broadening \citep{cas07}, with results 
ranging from 1.0 to 2.5. The determination of stellar helium abundance is 
an inherently difficult problem, so it is important to consider the 
results from a variety of methods, including asteroseismic analysis. The 
least-squares linear fit to all stars is $Y_{\rm i}= (1.42\pm0.27) Z + 
(0.245\pm0.005)$, with $\chi^2_R=2.2$. Excluding a few high-metallicity 
outliers (KIC\,3424541, 6603624 and 8006161), the relation becomes $Y_{\rm 
i}= (2.43\pm0.34) Z + (0.233\pm0.005)$ with $\chi^2_R=1.7$. The slopes of 
these two fits span nearly the full range of other determinations of 
$\Delta Y / \Delta Z$, and the scatter around either relation is rather 
large---for a given $Z$, our values of $Y_{\rm i}$ show deviations of more 
than $\pm 0.03$ (see Figure~\ref{fig8})\footnote{Note that one value of 
$Y_{\rm i}$ is significantly below the primordial helium abundance, while 
three others are marginally below $Y_{\rm P}$. Recall that heavy element 
diffusion and settling was not included in our stellar evolution models.}. 
A more direct determination of helium abundances in stellar envelopes, 
e.g.\ using acoustic glitches \citep{maz14,ver14}, would be very helpful 
as an independent constraint for global model-fitting.

We can compare our estimates of the mixing-length $\alpha$ and the 
amplitude of the surface correction $a_0$ to recent 3D simulations of 
convective stellar atmospheres at solar metallicity \citep{tra13,tra14}. 
The values of $\alpha$ in our sample decrease slightly with $T_{\rm eff}$, 
and a tri-linear regression also reveals small increases with both gravity 
and metallicity. Although the AMP values exhibit a larger range, the 
trends with $T_{\rm eff}$ and gravity qualitatively agree with the latest 
calibration of $\alpha$ \citep{tra14}, while the trend with metallicity 
has not yet been sampled by those simulations. The 3D simulations also 
predict a convective expansion of the atmosphere relative to 
one-dimensional (1D) models---an important contribution to the surface 
term $a_0$. The expansion increases monotonically with higher $T_{\rm 
eff}$ and lower $\log g$ \citep{tra13}, but in 1D models it is strongly 
coupled to the value of $\alpha$, which controls the slope of the 
temperature gradient near the top of the convection zone. Physical 
constraints on $\alpha$ and physical modeling of the surface effect from 
future 3D simulations promise to improve the reliability and robustness of 
asteroseismic fitting.

\section{DISCUSSION}\label{SEC5}

Our uniform asteroseismic analysis for a large sample of {\it Kepler} 
main-sequence and subgiant stars yielded precise determinations of many 
stellar properties (see Table~\ref{tab1}), and some important lessons for 
future work. Most of the simple stars have values of $f\le10$, indicating 
a reasonable match between the models from AMP and the observational 
constraints. The exceptions (KIC\,6603624, 8694723, 8760414 and 11244118) 
are stars with the most extreme metallicities in our sample, which may 
require modifications to the solar-composition mixture in the opacity 
tables employed by AMP for more accurate modeling. More of the F-like 
stars have $f>10$, but these are generally targets where the $\ell=0$ and 
$\ell=2$ modes are most difficult to separate. The {\'e}chelle diagrams in 
Figure~\ref{fig3} show that the extracted frequencies for F-like stars 
from \cite{app12} include variations in the small separations $d_{0,2}(n)$ 
that cannot be reproduced by models---in which the $\ell=0,2$ ridges 
always curve together with only monotonic variations in $d_{0,2}(n)$. This 
suggests that the smoothness criteria in peak-bagging pipelines may need 
to be modified for F-like stars. Automated fitting of the mixed-mode stars 
is limited to those with fewer than $\sim$3 avoided crossings and the 
match to the observed frequencies is generally less precise, with most 
stars showing $f>10$. The greatest difficulties are again encountered for 
extreme metallicity targets (KIC\,7976303 and 10018963), but the 
qualitative agreement with the locations of avoided crossings for most of 
the mixed-mode stars in Figure~\ref{fig3} is remarkable. Ultra-precise 
constraints on the properties of these stars must rely on future dense 
grid-modeling to refine the estimates from Table~\ref{tab1}.

Our modifications to the fitting procedures to try and avoid a bias 
towards low-helium solutions have helped, but the problem was not 
eliminated entirely. As noted in section~\ref{SEC3.1}, of the 22 stars 
considered by \cite{mat12} six of them (27\%) yielded an initial helium 
mass fraction significantly below the primordial value, while four 
additional targets (18\%) were marginally below $Y_{\rm P}$. We attempted 
to address this issue by including the frequency ratios as additional 
constraints, and by adopting larger uncertainties at higher frequencies 
where the surface correction is larger. A smaller fraction of our sample 
of 42 stars appears to be affected by the low-helium bias, but 
Table~\ref{tab1} still includes one star (2.4\%) with $Y_{\rm i}$ 
significantly below $Y_{\rm P}$ and three additional targets (7.1\%) that 
fall marginally below $Y_{\rm P}$. Further improvement may require an 
alternative surface correction \citep[e.g.,][]{jcd12}.

The precision of asteroseismic data sets from extended observations with 
{\it Kepler} now demands that we address the dominant sources of 
systematic error in the stellar models. The largest source arises from 
incomplete modeling of the near-surface layers. Although something like 
the empirical correction of \cite{kje08} will continue to be useful, we 
are now in a position to capture more of the relevant physics. For 
example, \cite{gru13} performed a Bayesian analysis of the 22 {\it Kepler} 
targets from \cite{mat12}, including a simplified non-adiabatic treatment 
of the pulsations. They found that the Bayesian probabilities were higher 
when non-adiabatic rather than adiabatic frequencies were fit to the 
observations, and that for most stars the surface effect was minimized and 
in some cases even eliminated. Their non-adiabatic model accounted for 
radiative losses and gains but neglected perturbations to the convective 
flux and turbulent pressure \citep{gue94}. In the case of the Sun, the 
stability and frequency of the oscillation modes depends substantially on 
turbulent pressure and the inclusion of non-local effects in the treatment 
of convection \citep{bal92a,bal92b,hou10}, and the resulting frequency 
shift is uncertain. The sensitivity of the results to the model of 
convection and the temperature profile in the super-adiabatic layer was 
also emphasized by \cite{ros99} and \cite{li02}. Regardless, the Bayesian 
approach has the advantage of incorporating the unknown sources of 
systematic error directly into the uncertainties on the derived stellar 
properties, and can reveal which approach to the pulsation calculations 
generally improves the model fits.

For future analyses, we intend to augment ADIPLS with the non-adiabatic 
stellar oscillation code GYRE \citep{tt13}, which currently includes a 
limited treatment of non-adiabatic effects but is flexible enough to 
incorporate additional contributions. We would also like to take advantage 
of the numerical stability and modular architecture of the open-source 
MESA code \citep{pax13} to explore different chemical mixtures and to 
include heavy element diffusion and settling in the evolutionary models, 
which is not currently stable for all types of stars with ASTEC. With 
these new modules for stellar evolution and pulsation calculations, we can 
embed a Bayesian formalism into the parallel genetic algorithm to 
complement the simple $\chi^2$ approach. The complete sample of 
asteroseismic targets from extended {\it Kepler} observations spanning up 
to four years will provide a rich data set to validate these new 
ingredients for the next generation of AMP.

The golden age of asteroseismology for main-sequence and subgiant stars 
owes a great debt to the {\it Kepler} mission, but it promises to continue 
with the anticipated launch of NASA's Transiting Exoplanet Survey 
Satellite \citep[TESS,][]{ric14} in 2017. While {\it Kepler} was able to 
provide asteroseismic data for hundreds of targets and could 
simultaneously monitor 512 stars with 1-minute sampling, TESS plans to 
observe $\sim$500,000 of the brightest G- and K-type stars in the sky at a 
cadence sufficient to detect solar-like oscillations. The data sets will 
be nearly continuous for at least 27 days, but in two regions near the 
ecliptic poles the fields will overlap for durations up to a full year. 
These brighter stars will generally be much better characterized than the 
{\it Kepler} targets---with parallaxes from {\it Hipparcos} and ultimately 
{\it Gaia} \citep{per01}, and reliable atmospheric constraints from 
ground-based spectroscopy---making asteroseismic characterization more 
precise and accurate. With several years of development time available, 
AMP promises to be ready to convert this avalanche of data into reliable 
inferences on the properties of our solar system's nearest neighbors.


\acknowledgments We would like to thank Victor Silva Aguirre for helpful 
discussions. This work was supported in part by NASA grants NNX13AC44G and 
NNX13AE91G, and by White Dwarf Research Corporation through the Pale Blue 
Dot project (http://whitedwarf.org/palebluedot/). Computational time on 
Kraken at the National Institute of Computational Sciences was provided 
through XSEDE allocation TG-AST090107. Funding for the Stellar 
Astrophysics Centre is provided by The Danish National Research Foundation 
(Grant DNRF106).
We acknowledge the ASTERISK project (ASTERoseismic Investigations with 
SONG and Kepler) funded by the European Research Council (Grant agreement 
no.: 267864),
the Scientific and Technological Research Council of Turkey 
(T{\"U}B{\.I}TAK:112T989),
and a European Commission grant for the SPACEINN project 
(FP7-SPACE-2012-312844).
BPB was supported in part by NSF Astronomy and Astrophysics postdoctoral
fellowship AST 09-02004. CMSO is supported by NSF grant PHY 08-21899 and
KITP is supported by NSF grant PHY 11-25915.
MSC is supported by an Investigador FCT contract funded by FCT/MCTES 
(Portugal) and POPH/FSE (EC).
AD has been supported by the Hungarian OTKA Grants K83790, KTIA 
URKUT\_10-1-2011-0019 grant, the Lend\"ulet-2009 Young Researchers 
Programme of the Hungarian Academy of Sciences, the J\'anos Bolyai 
Research Scholarship of the Hungarian Academy of Sciences and the City of 
Szombathely under Agreement No. S-11-1027.
AD and RAG acknowledge the support of the European Community Seventh 
Framework Programme ([FP7/2007- 2013]) under the grant agreement no. 
269194 (IRSES/ASK). RAG and D.~Salabert acknowledge the support of the 
CNES grant at CEA-Saclay.
AM acknowledges support from the NIUS programme of HBCSE (TIFR).
AS is supported  by the MICINN grant AYA2011-24704 and by the ESF
EUROCORES Program EuroGENESIS (MICINN grant EUI2009-04170).
D.~Stello is supported by the Australian Research Council.

\appendix\section{A. Uncertainty Estimation Procedure\label{APPA}}

Our previous approach to uncertainty estimation using singular value 
decomposition (SVD) is no longer appropriate for several reasons. First, 
SVD assumes that the observables are independent. This was reasonable when 
we were only fitting the individual frequencies and spectroscopic 
constraints. Now that we also fit the frequency ratios $r_{02}$ and 
$r_{010}$ (see section~\ref{SEC3.2}), some of the observables are no 
longer independent and SVD cannot be used. Second, even if we do not 
include the frequency ratios as constraints, the SVD method determines the 
uncertainties from the local shape of the $\chi^2$ surface. Such an 
analysis fails to capture the uncertainties due to non-uniqueness of the 
optimal solution, yielding errors that are too optimistic for practical 
purposes. Finally, the metric used by the GA for optimization, which 
defines the local shape of the $\chi^2$ surface, is a composite of 
normalized $\chi^2$ values from several distinct sets of observables, 
making it difficult to interpret in a statistically robust manner.

We can use the ensemble of models sampled by the GA to provide a more 
conservative and global estimate of the uncertainties. During an AMP run, 
the parameter values and average $\chi^2$ metric are recorded for each 
trial model that is compared to the observations. The nature of the 
fitting process ensures that each trial model is a much better match to 
the observations than a random model in the search space. For each stellar 
evolution track generated by AMP, the age is optimized internally using a 
binary decision tree to match the observed large separation. In addition, 
the final age is interpolated between time steps on the track to match the 
lowest observed radial mode, and the empirical surface correction of 
\cite{kje08} is applied to improve the match to higher frequency modes. In 
effect, the GA is producing the best possible match to the observations 
given the four fixed parameters $(M, Z, Y_{\rm i}, \alpha)$ for each trial 
model. As a consequence, we need to use a subset of the full ensemble of 
GA models if we want a reasonable estimate of the uncertainties from the 
limited information that is available.

  \begin{figure*}[h] 
  \centerline{\includegraphics[width=\columnwidth]{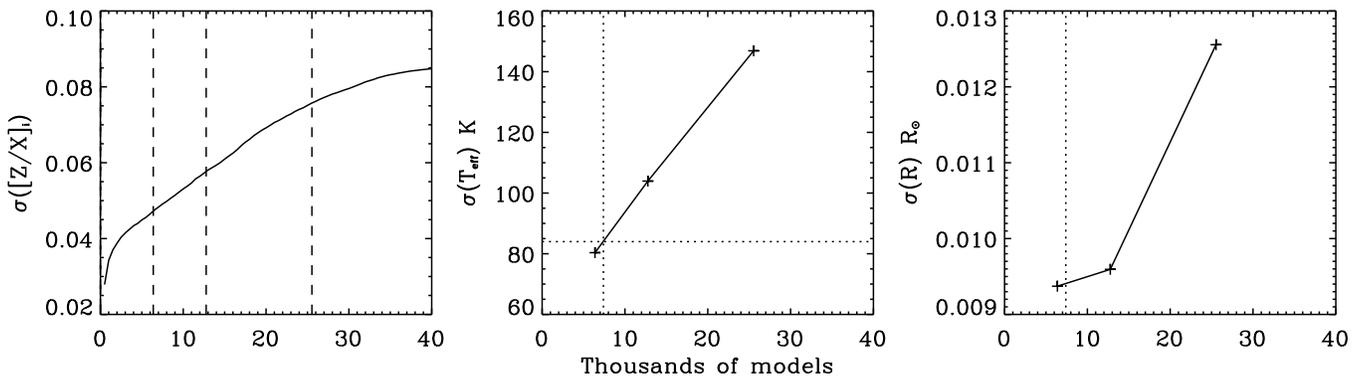}}
  \caption{Uncertainties in $[Z/X]_{\rm i}$ (left), $T_{\rm eff}$ 
  (center), and $R$ (right) for KIC\,6116048 as the number of models 
  included in the likelihood-weighted mean is increased. Several cuts 
  are defined from the uncertainty in $[Z/X]_{\rm i}$ (left panel, 
  vertical dashed lines), which yield corresponding uncertainties in 
  $T_{\rm eff}$ and $R$ from the 1$\sigma$ models (center and right 
  panels, + points). Interpolating to yield an uncertainty of 84~K in 
  $T_{\rm eff}$ (center panel, horizontal dotted line) provides the 
  final cut (center and right panels, vertical dotted line) that is used 
  to define the uncertainties on all model properties.\label{figA1}}
  \end{figure*} 

To determine the number of GA trial models that we should include in our 
uncertainty estimation procedure, we initially attempted to match the 
observational error on [M/H] \citep[0.09~dex,][]{cha14,bru12}. As outlined 
briefly in section~\ref{SEC4.1}, we ranked the unique trial models by 
their average $\chi^2$ value and assigned each one a relative likelihood 
using Eq.(\ref{eq:L}). This allowed us to calculate a likelihood-weighted 
mean and standard deviation for each adjustable model parameter, including 
$Z$ and $Y_{\rm i}$ to generate an uncertainty on the initial composition 
$[Z/X]_{\rm i}$ as we gradually included more models in the mean. We only 
had access to the initial value of $X$, which increases at the surface 
over time as helium diffusion and settling operates in the models, so this 
only provided a first estimate of the appropriate number of models to 
include. The resulting uncertainty on $[Z/X]_{\rm i}$ for our example star 
KIC\,6116048 is shown in the left panel of Figure~\ref{figA1} as a 
function of the number of models included in the likelihood-weighted mean.

Uncertainties on the other properties of the optimal models can only be 
determined after a cut on the number of models has been adopted. The cut 
establishes the uncertainties on the adjustable parameters $(M, t, Z, 
Y_{\rm i}, \alpha)$ as well as the covariance matrix around the optimal 
solution. This allows us to calculate a set of models that define the 
1$\sigma$ error ellipse, and then use half the range of values for any 
other property (such as $[Z/X]_{\rm s}$, $T_{\rm eff}$, or $R$) within the 
1$\sigma$ models to define an uncertainty for these non-adjustable 
parameters. For KIC\,6116048, we determined that by including the best 
$\sim$25,000 models in the likelihood-weighted mean, the resulting 
uncertainty on the surface composition $[Z/X]_{\rm s}$ (not shown in 
Figure~\ref{figA1}) was comparable to the observational error on [M/H]. 
Using this same cut, the full range of model values for $T_{\rm eff}$ 
within the 1$\sigma$ models spanned 290~K, yielding an error estimate of 
$\pm$145~K (see center panel of Figure~\ref{figA1}). We repeated the 
analysis procedure with a cut at one-half and one-quarter of this number 
of models (vertical dashed lines in the left panel of Figure~\ref{figA1}) 
until the resulting uncertainty on the model $T_{\rm eff}$ was below the 
error on the observed $T_{\rm eff}$ \citep[84~K,][]{cha14,bru12}.

By interpolating the number of models required to reproduce the observed 
$T_{\rm eff}$ error, we defined the appropriate cut that was then used to 
estimate the final uncertainties on all model properties. The center panel 
of Figure~\ref{figA1} shows the uncertainty on the model $T_{\rm eff}$ for 
the three cuts indicated in the left panel. When the 7370 best models 
sampled by the GA (vertical dotted line) were used to calculate the 
likelihood-weighted mean and standard deviation for each of the adjustable 
model parameters, the uncertainty on the model $T_{\rm eff}$ from the 
resulting 1$\sigma$ models was equal to the observational error of 84~K 
(horizontal dotted line). The range of radii within this same set of 
1$\sigma$ models define the radius uncertainty of 0.0094~$R_\odot$ (right 
panel of Figure~\ref{figA1}). As noted in section~\ref{SEC4.1}, the above 
procedure yields inherently conservative uncertainty estimates because it 
assumes that the asteroseismic data do not contribute to the determination 
of $T_{\rm eff}$ in the final solution. This assumption is certainly not 
valid for the surface composition, even ignoring the complications from 
diffusion, which explains why the final uncertainty on $[Z/X]_{\rm i}$ is 
well below the observational error on [M/H]. We repeated the above 
procedure for each star in our sample to yield the final set of 
uncertainties, which appear in Table~\ref{tab1}.\\


\newpage

  \begin{figure*}[p] 
  \centerline{\includegraphics[width=2.25in]{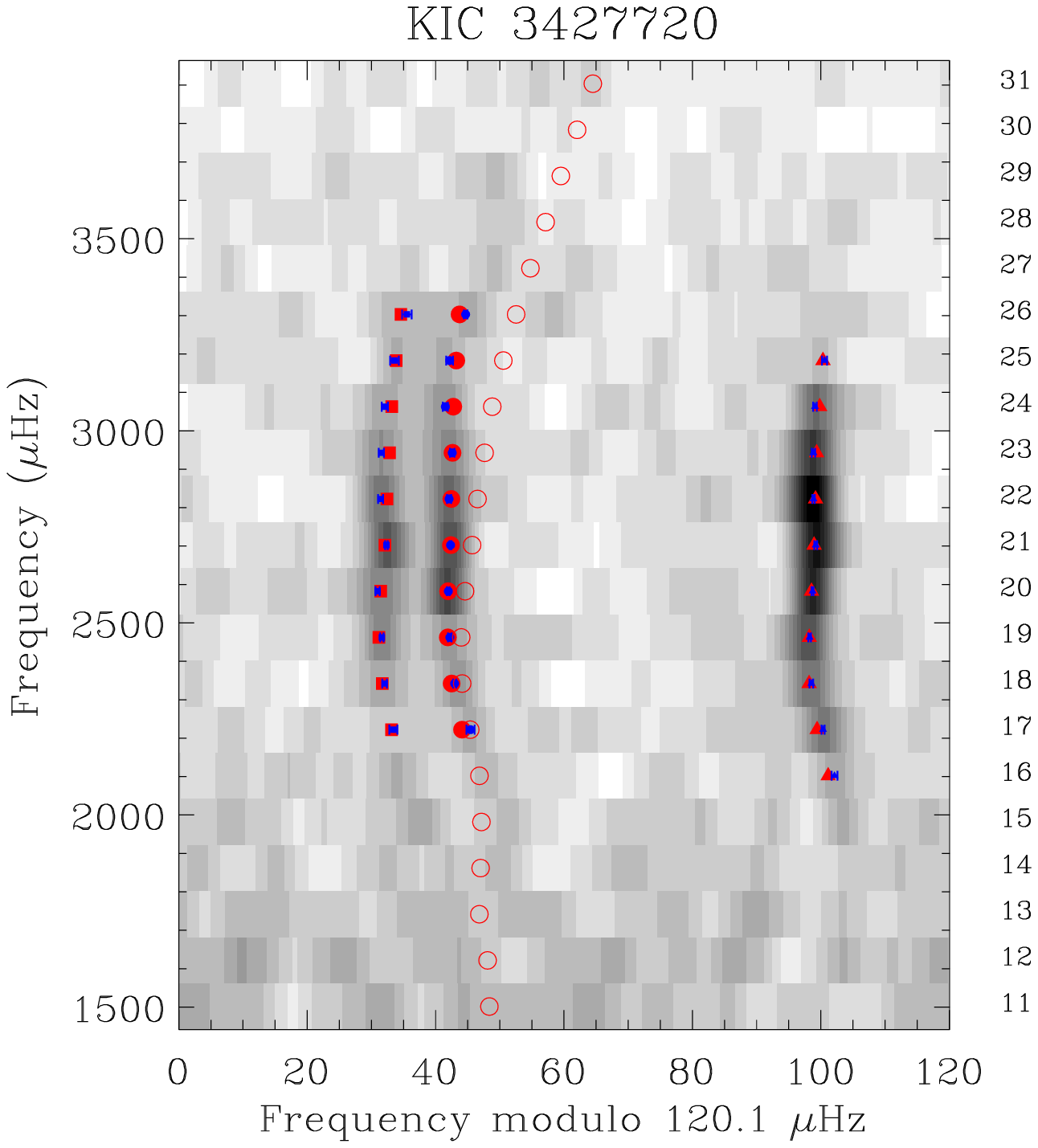}\hspace*{0.125in}\includegraphics[width=2.25in]{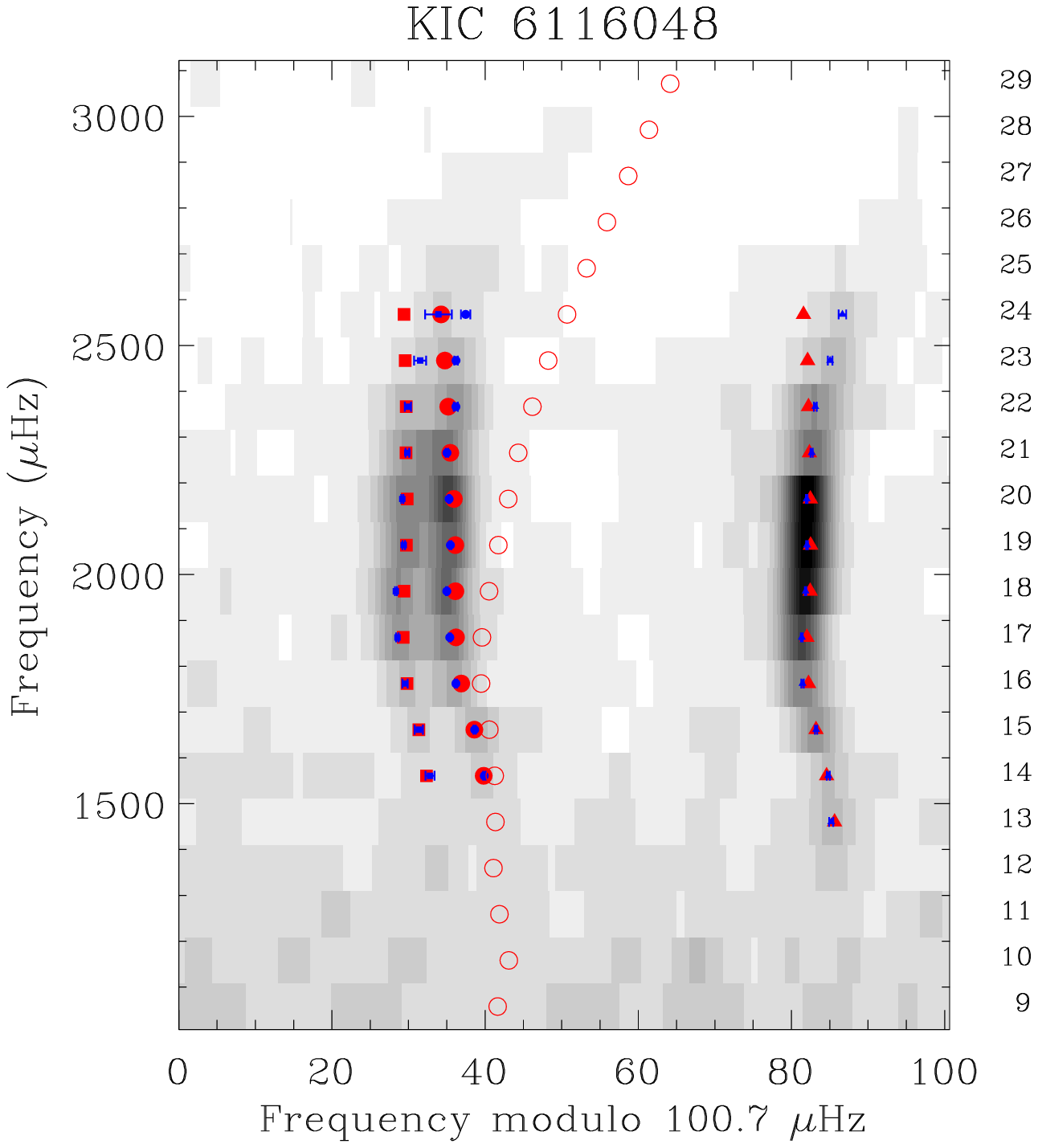}\hspace*{0.125in}\includegraphics[width=2.25in]{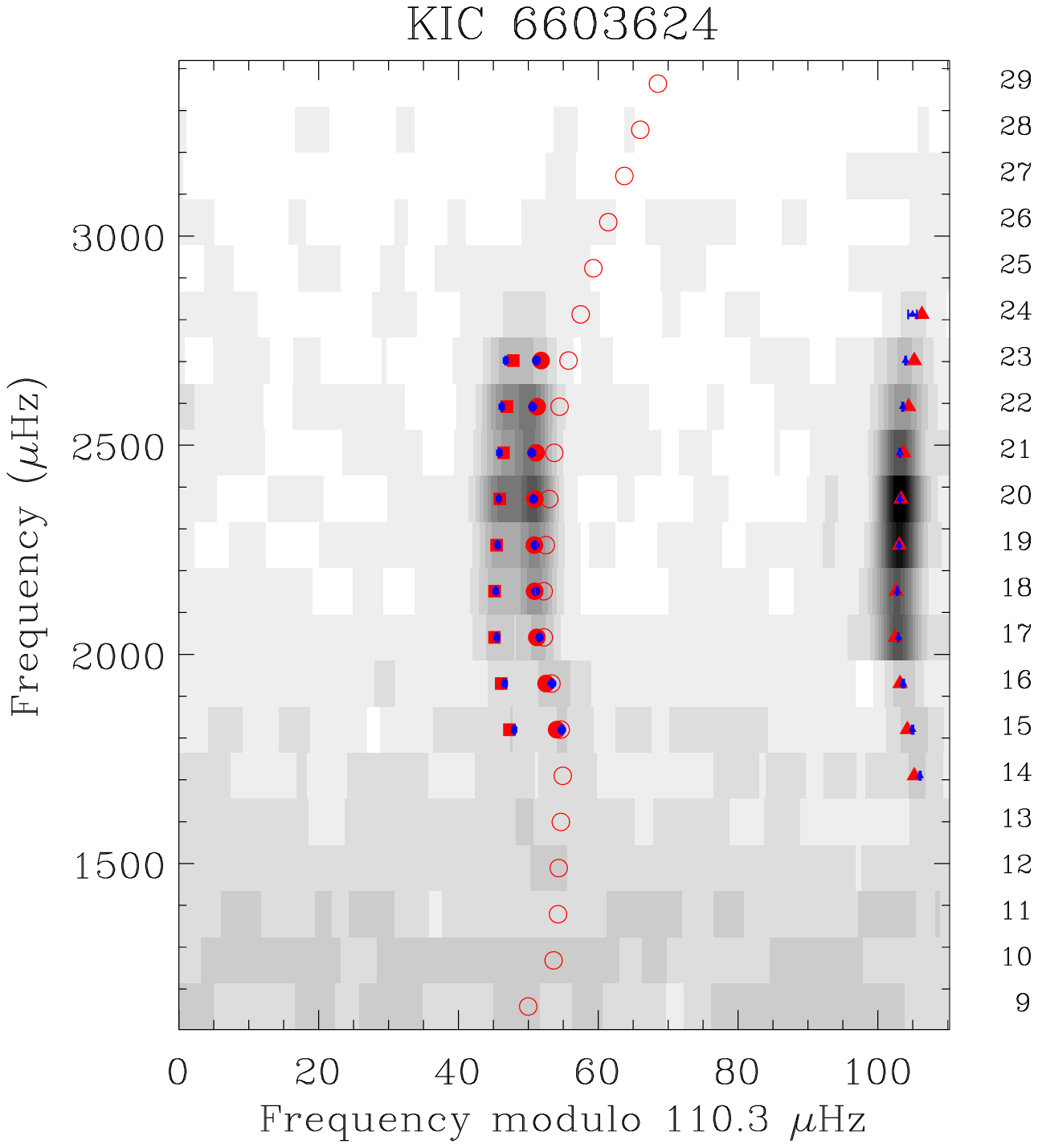}}
  \vspace*{0.125in}
  \centerline{\includegraphics[width=2.25in]{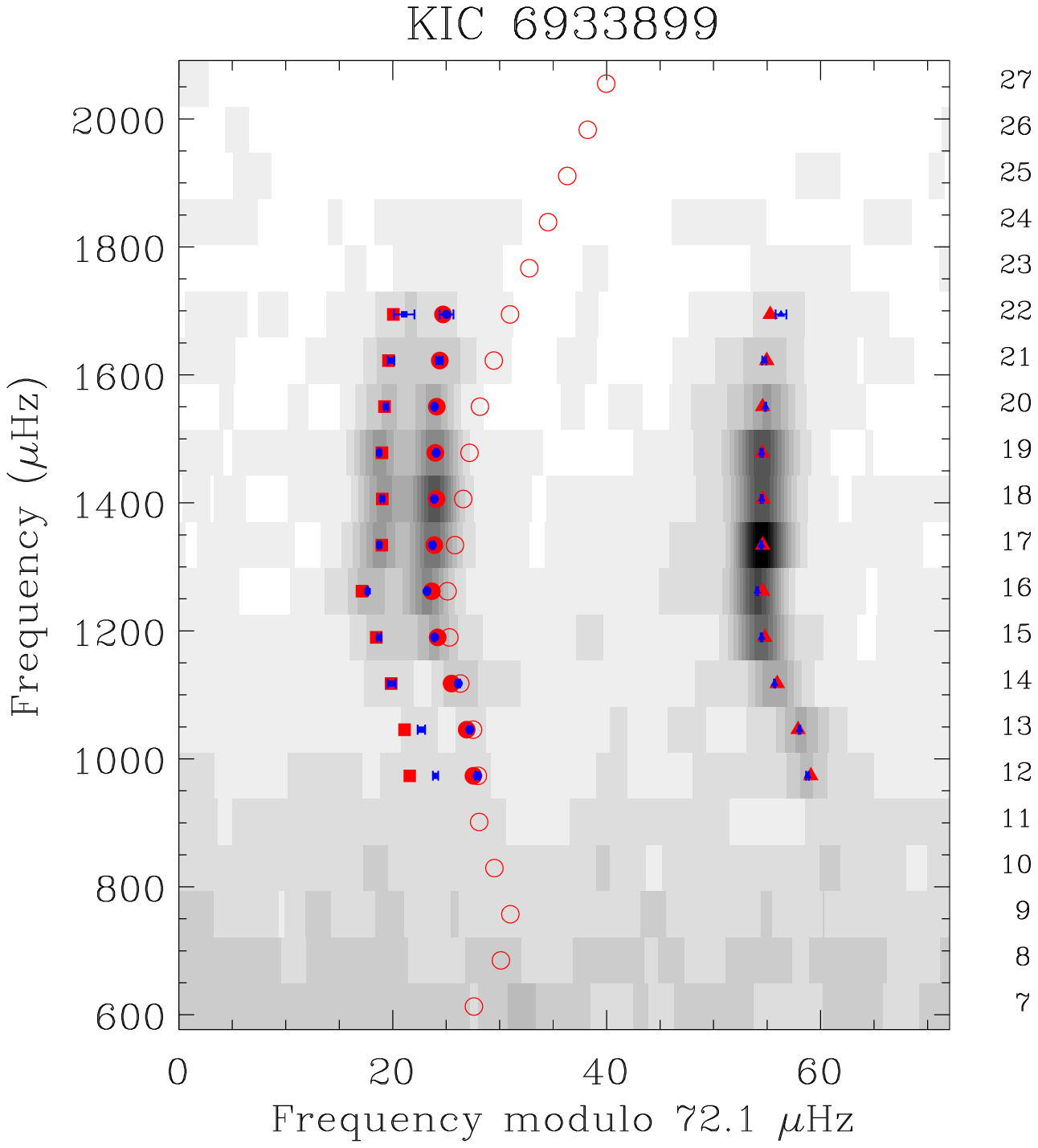}\hspace*{0.125in}\includegraphics[width=2.25in]{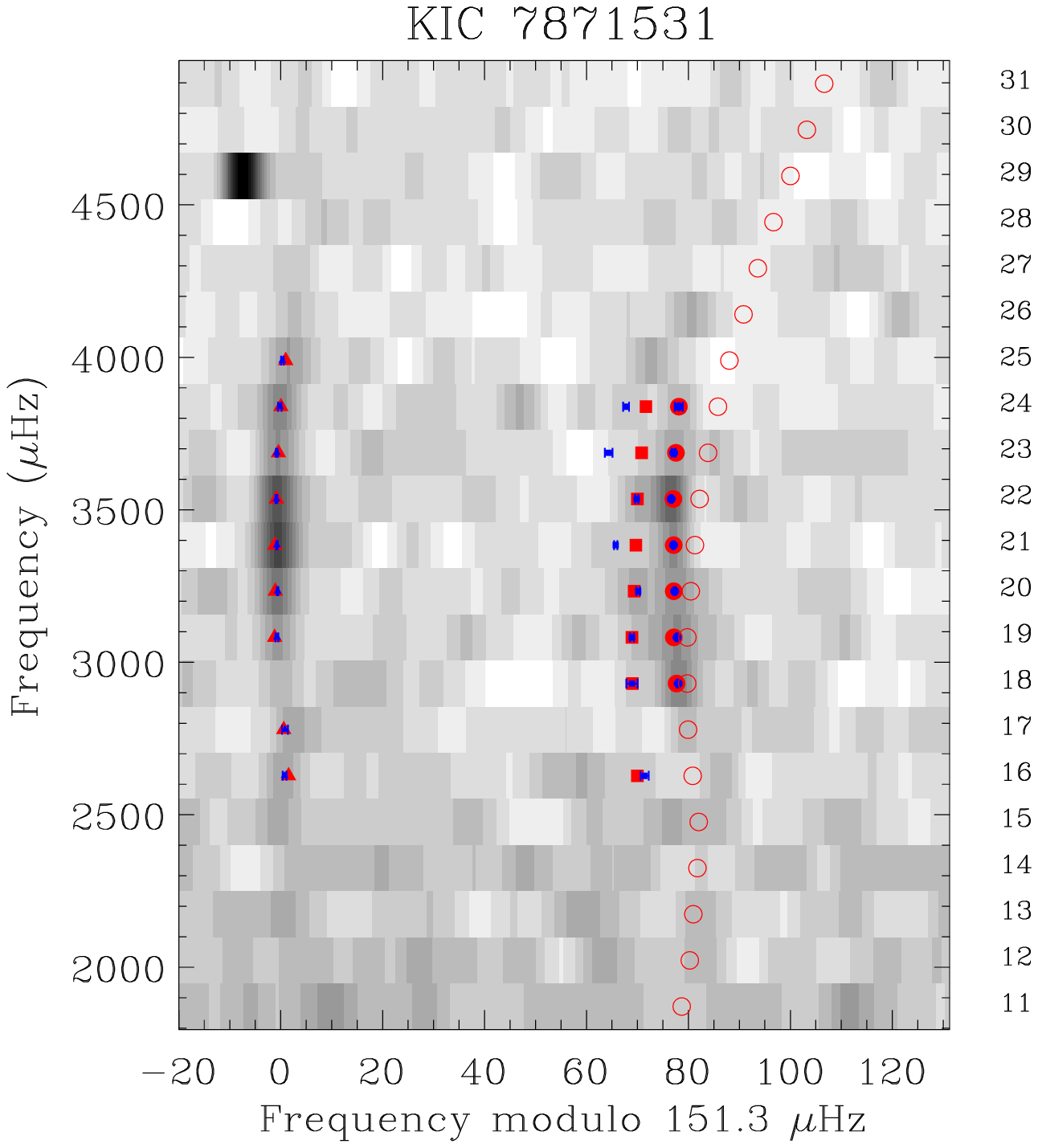}\hspace*{0.125in}\includegraphics[width=2.25in]{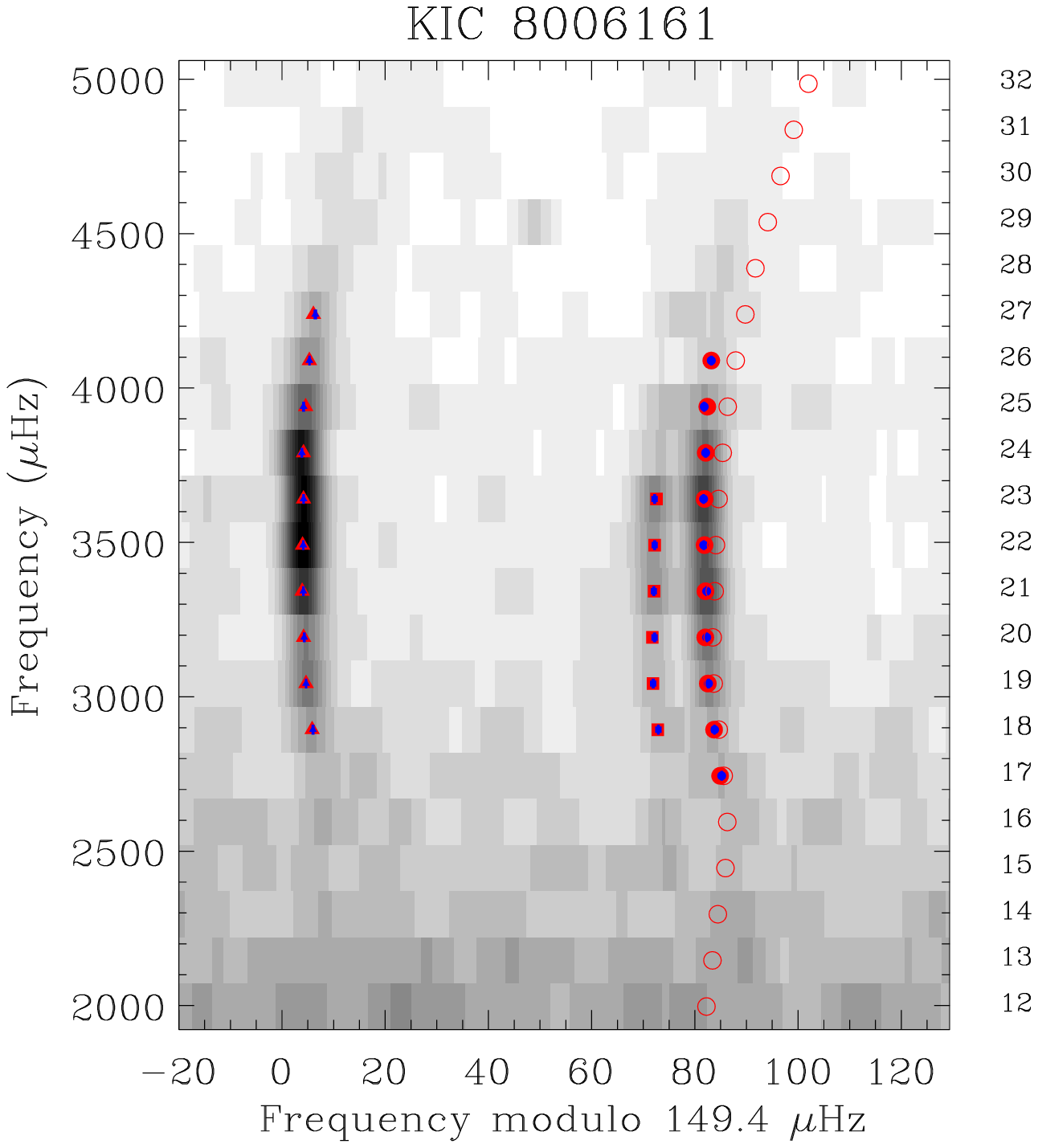}}
  \vspace*{0.125in}
  \centerline{\includegraphics[width=2.25in]{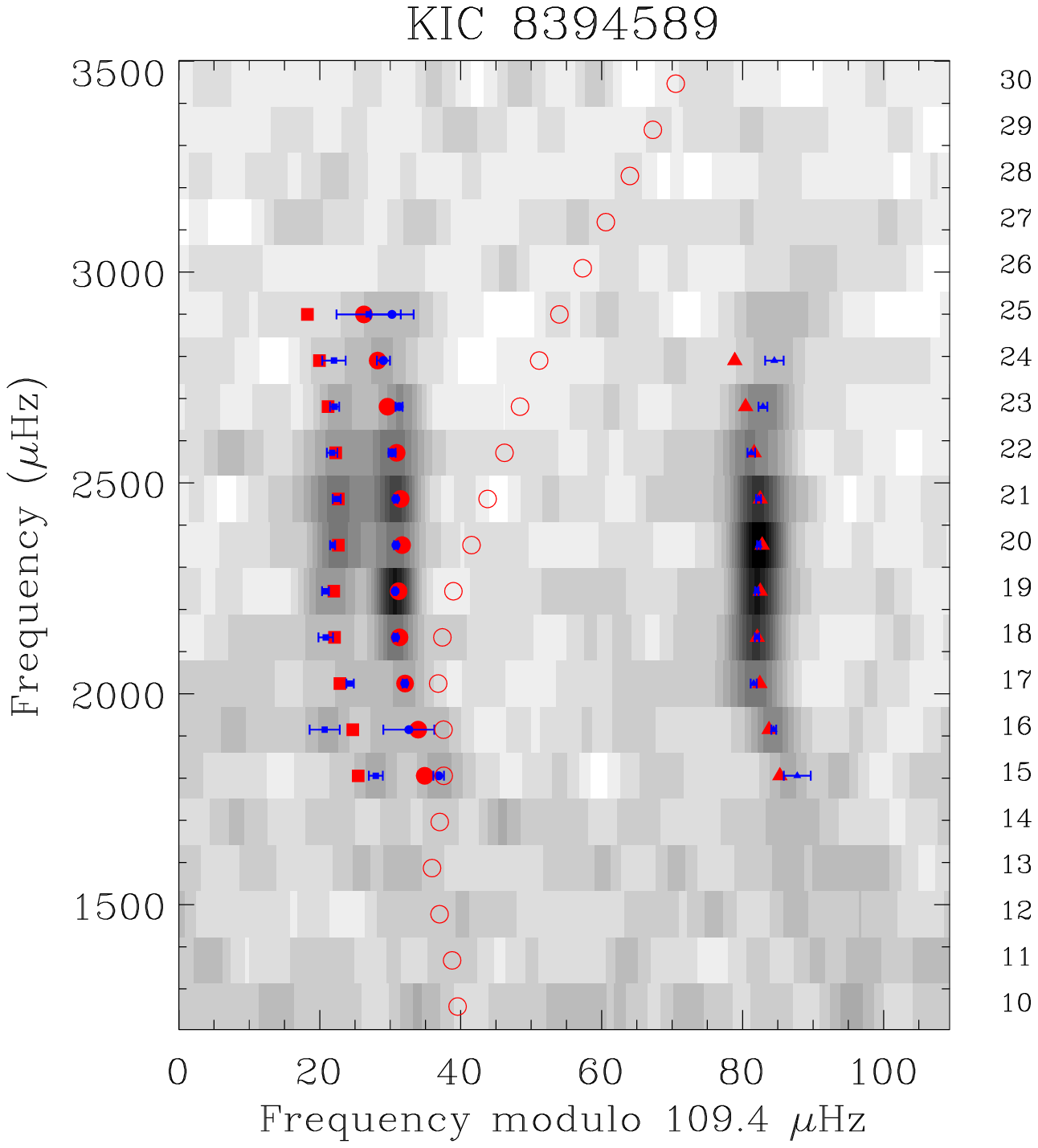}\hspace*{0.125in}\includegraphics[width=2.25in]{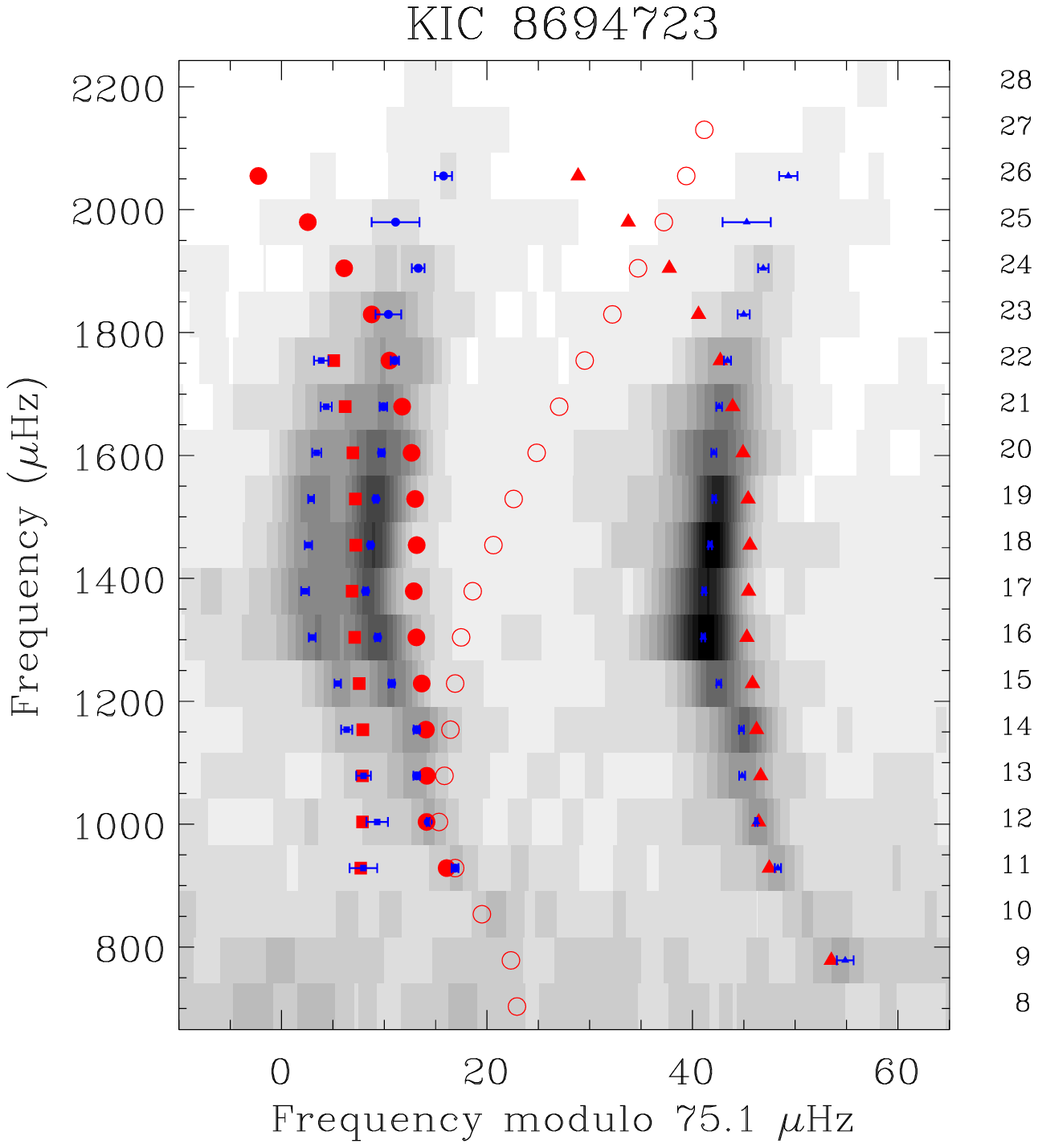}\hspace*{0.125in}\includegraphics[width=2.25in]{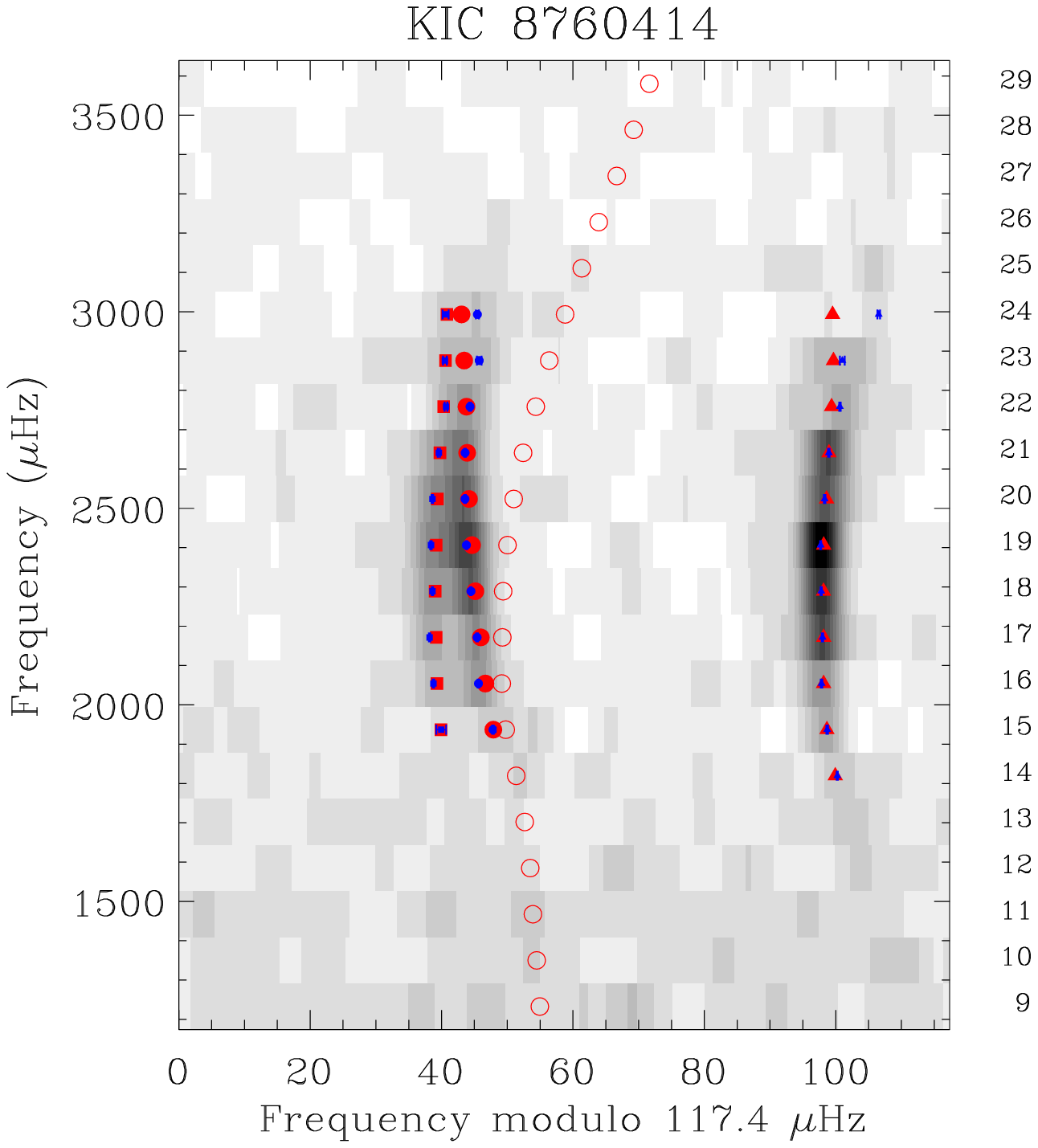}}
  \vspace*{0.5in}
  \centerline{{\bf Figure~\ref{fig3}}.~~~ ONLINE ONLY}
  \end{figure*}

  \begin{figure*}[p]
  \centerline{\includegraphics[width=2.25in]{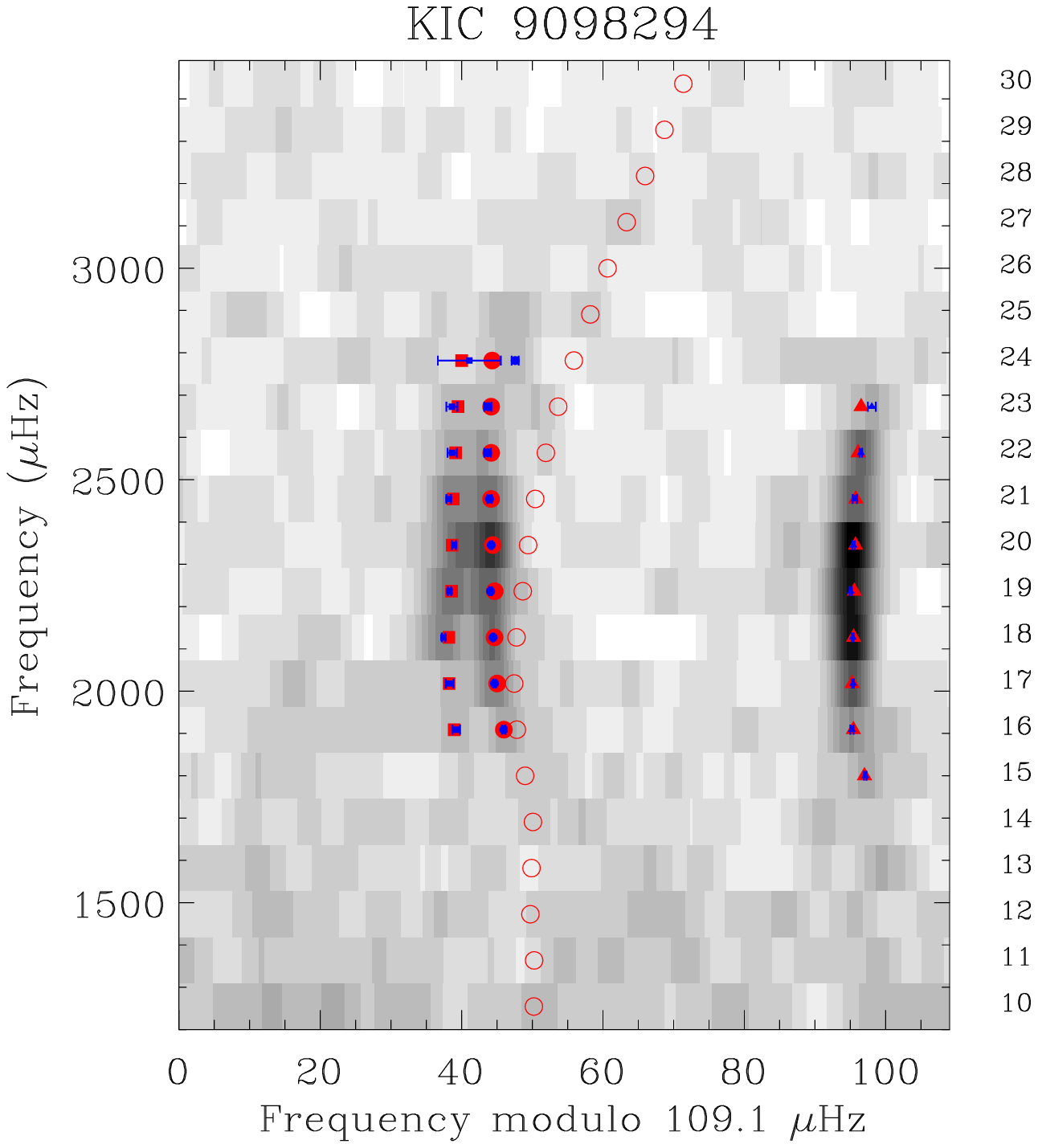}\hspace*{0.125in}\includegraphics[width=2.25in]{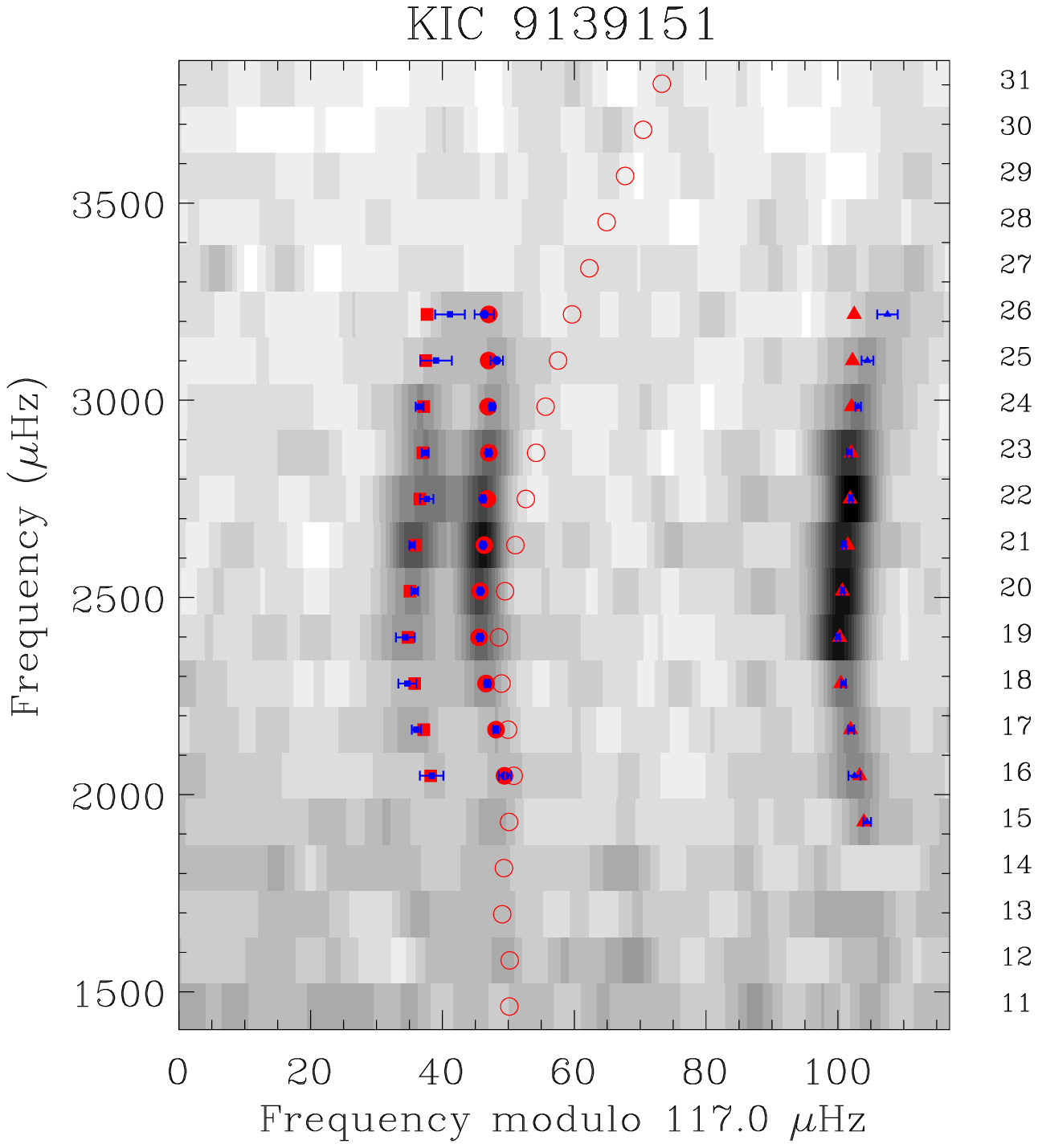}\hspace*{0.125in}\includegraphics[width=2.25in]{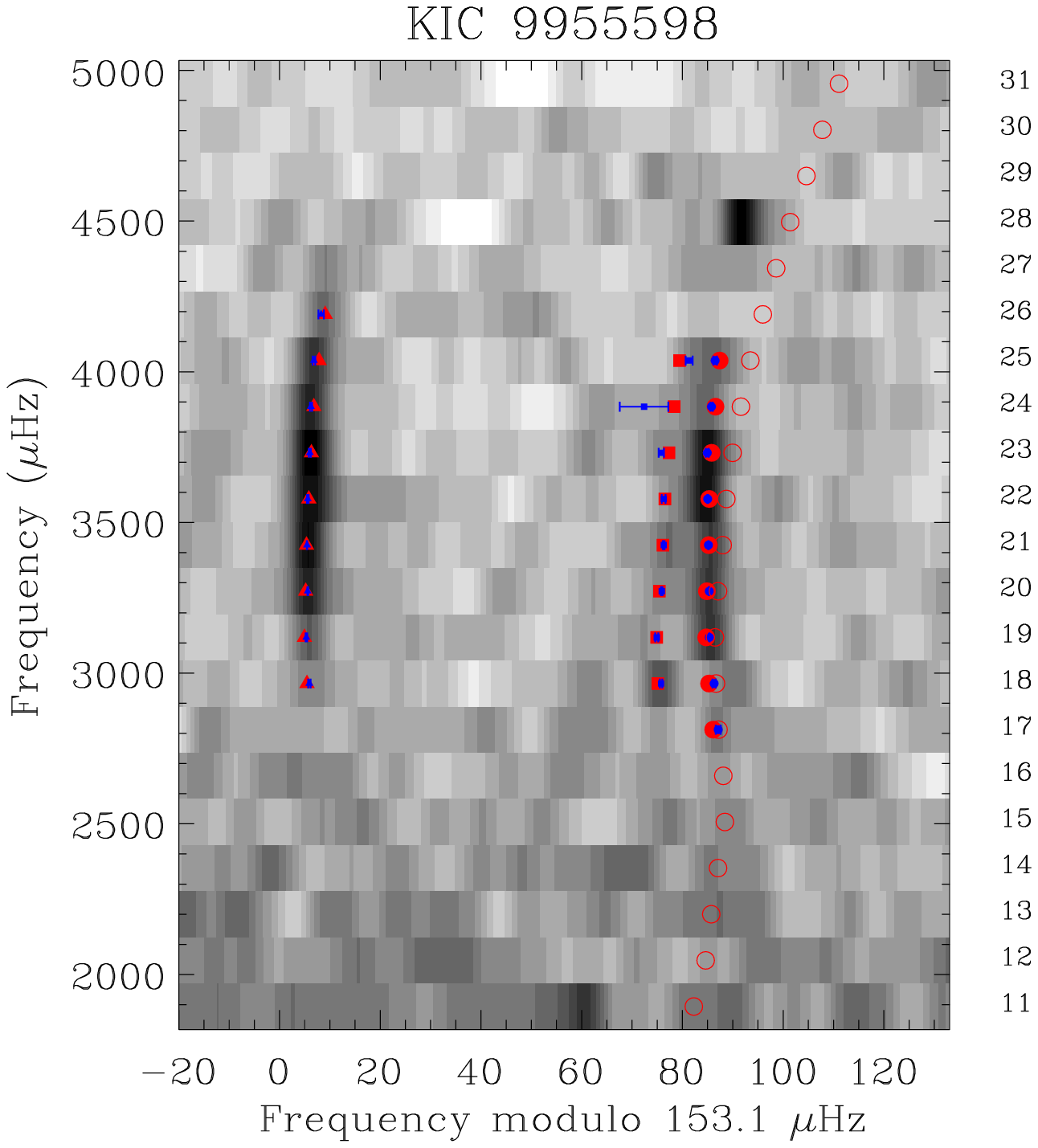}}
  \vspace*{0.125in}
  \centerline{\includegraphics[width=2.25in]{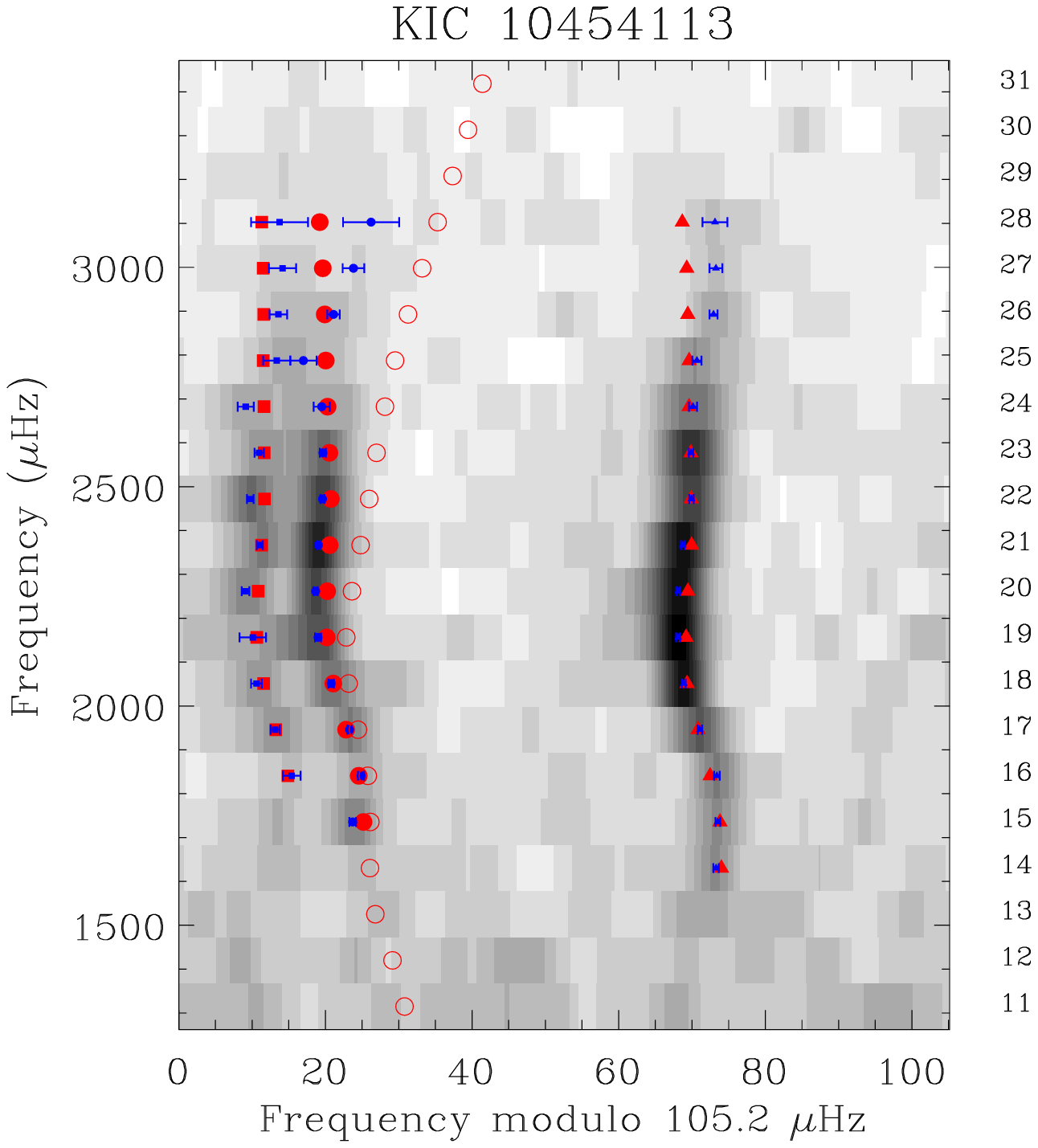}\hspace*{0.125in}\includegraphics[width=2.25in]{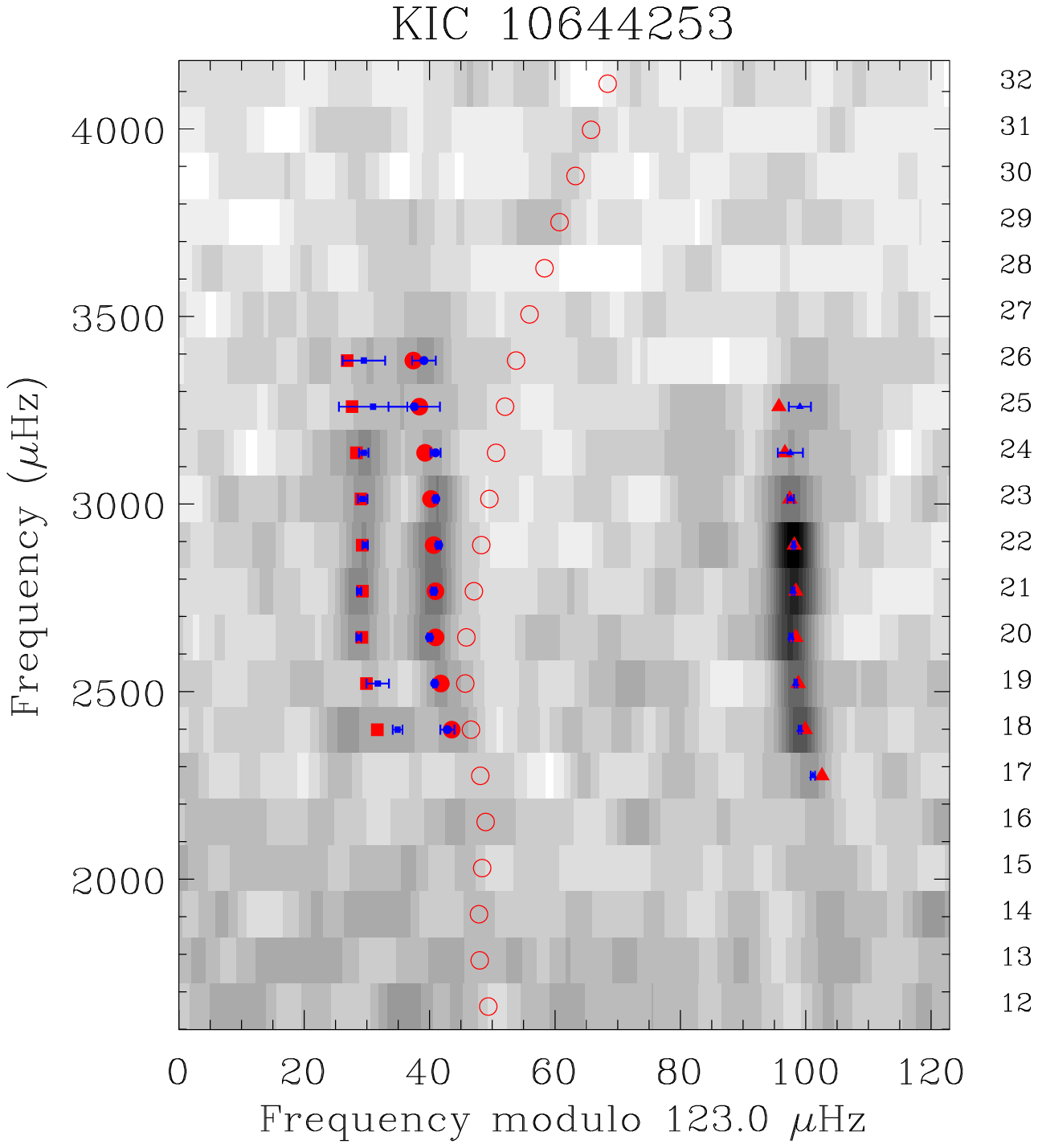}\hspace*{0.125in}\includegraphics[width=2.25in]{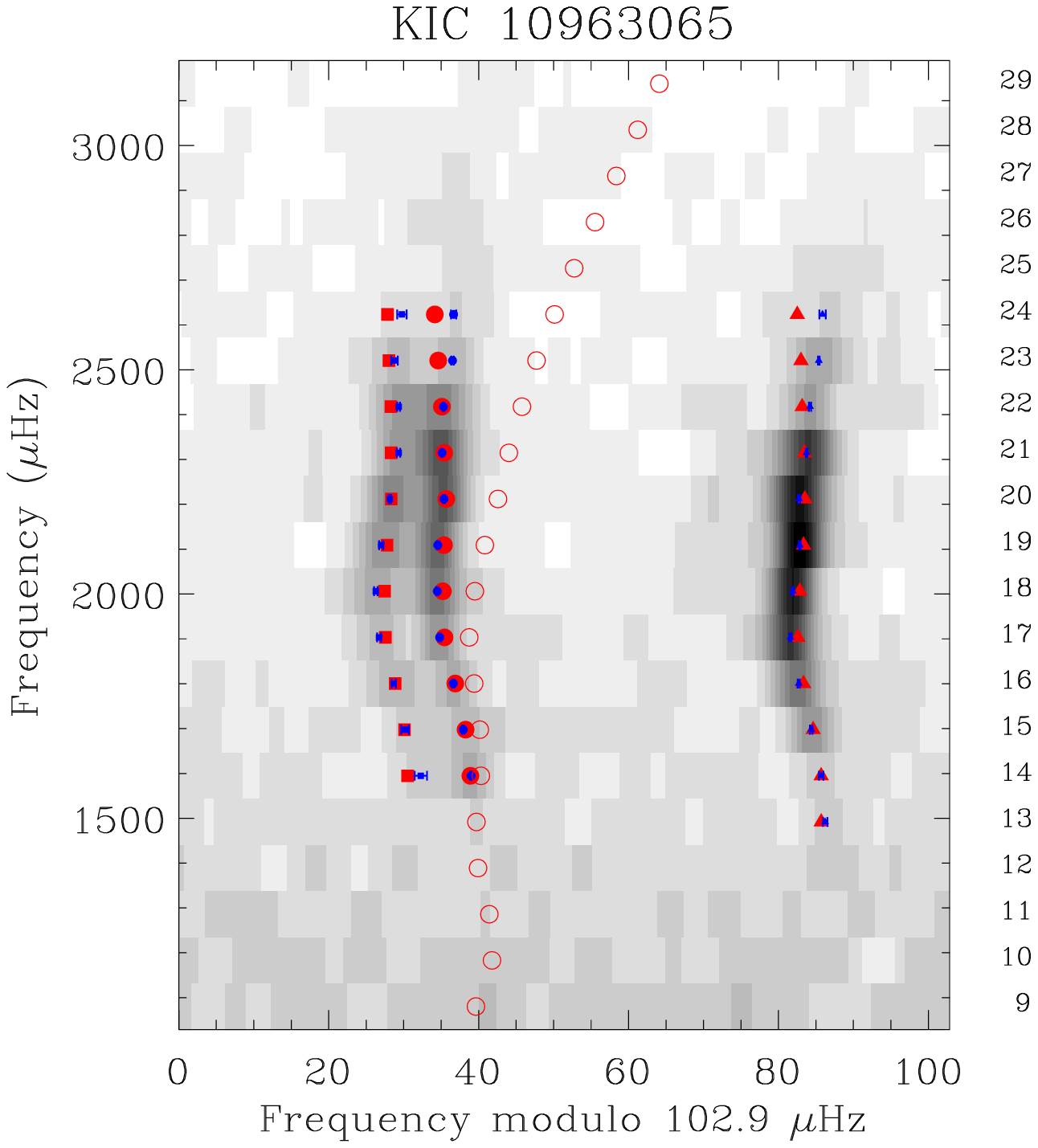}}
  \vspace*{0.125in}
  \centerline{\includegraphics[width=2.25in]{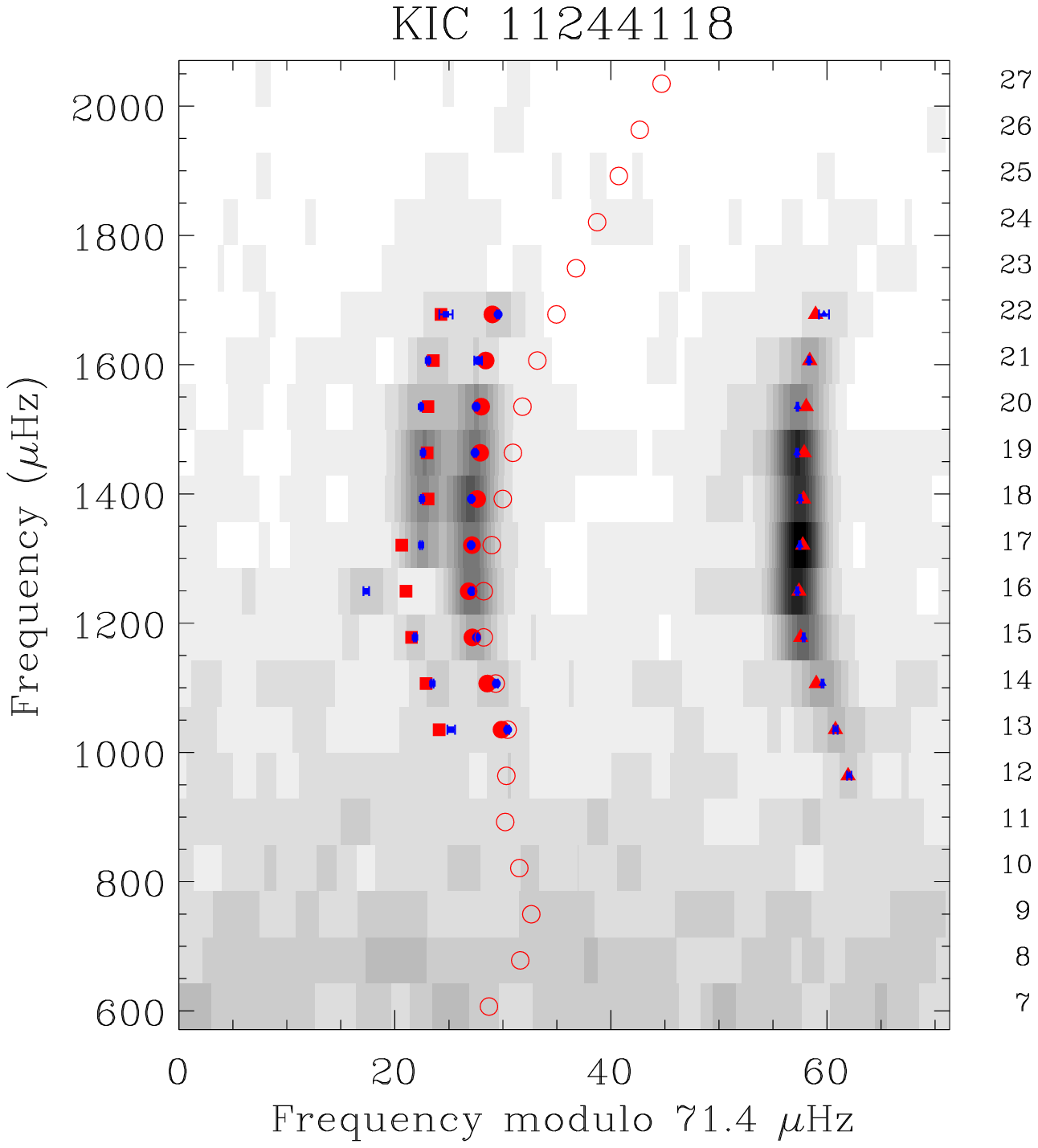}\hspace*{0.125in}\includegraphics[width=2.25in]{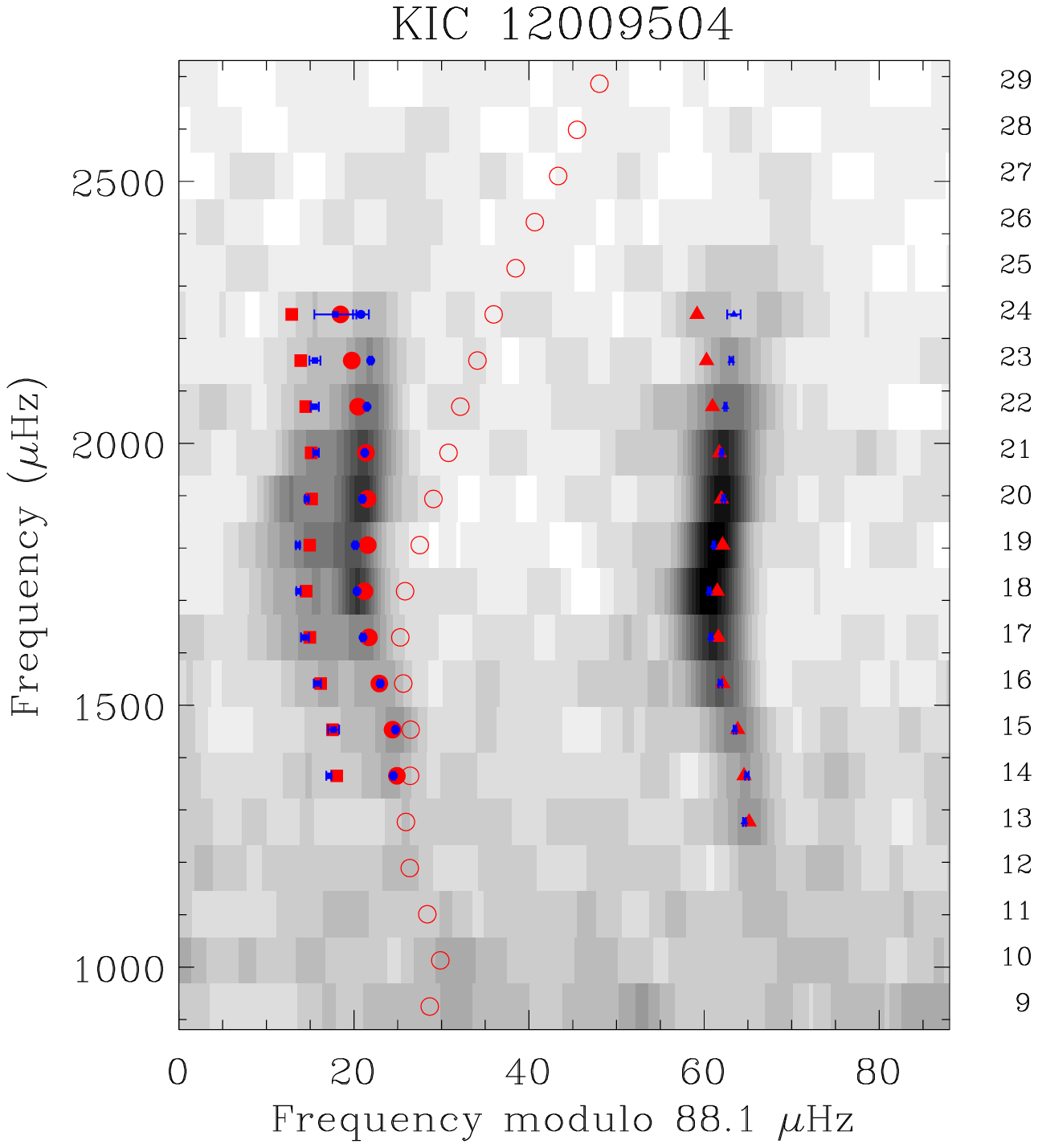}\hspace*{0.125in}\includegraphics[width=2.25in]{fig3.18.eps}}
  \vspace*{0.5in}
  \centerline{{\bf Figure~\ref{fig3}}.~~~ ONLINE ONLY (cont.)}
  \end{figure*} 

  \begin{figure*}[p] 
  \centerline{\includegraphics[width=2.25in]{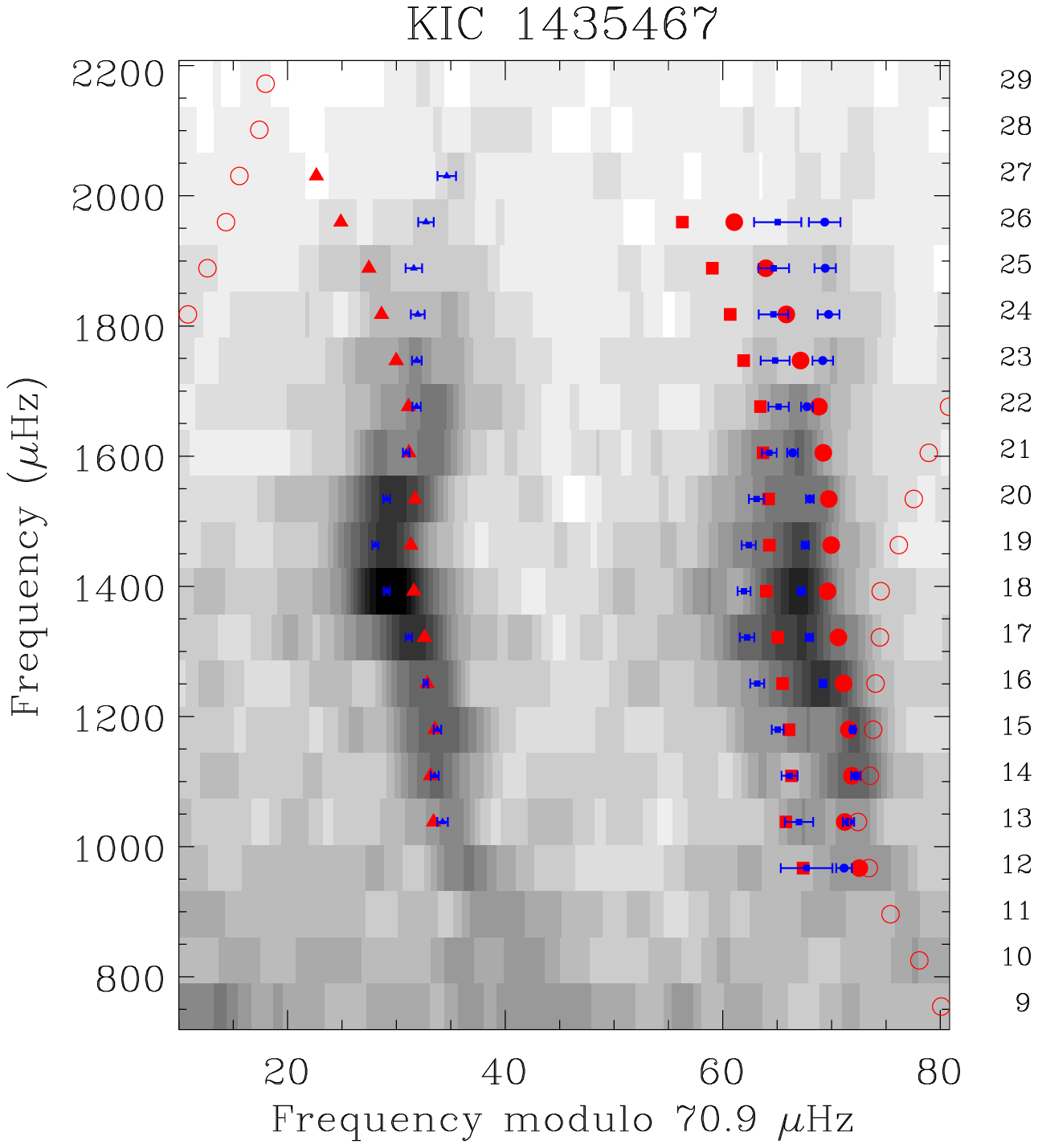}\hspace*{0.125in}\includegraphics[width=2.25in]{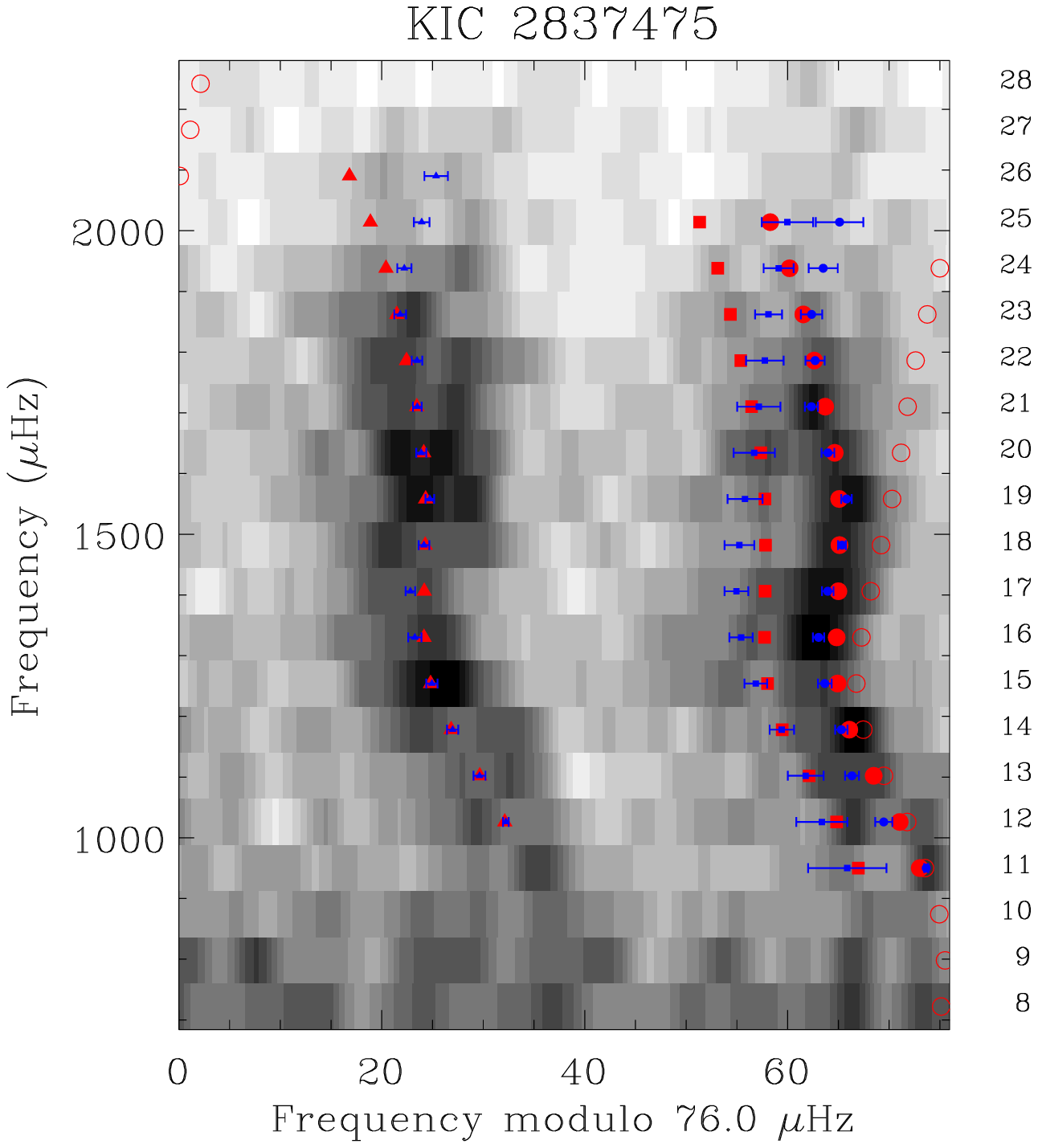}\hspace*{0.125in}\includegraphics[width=2.25in]{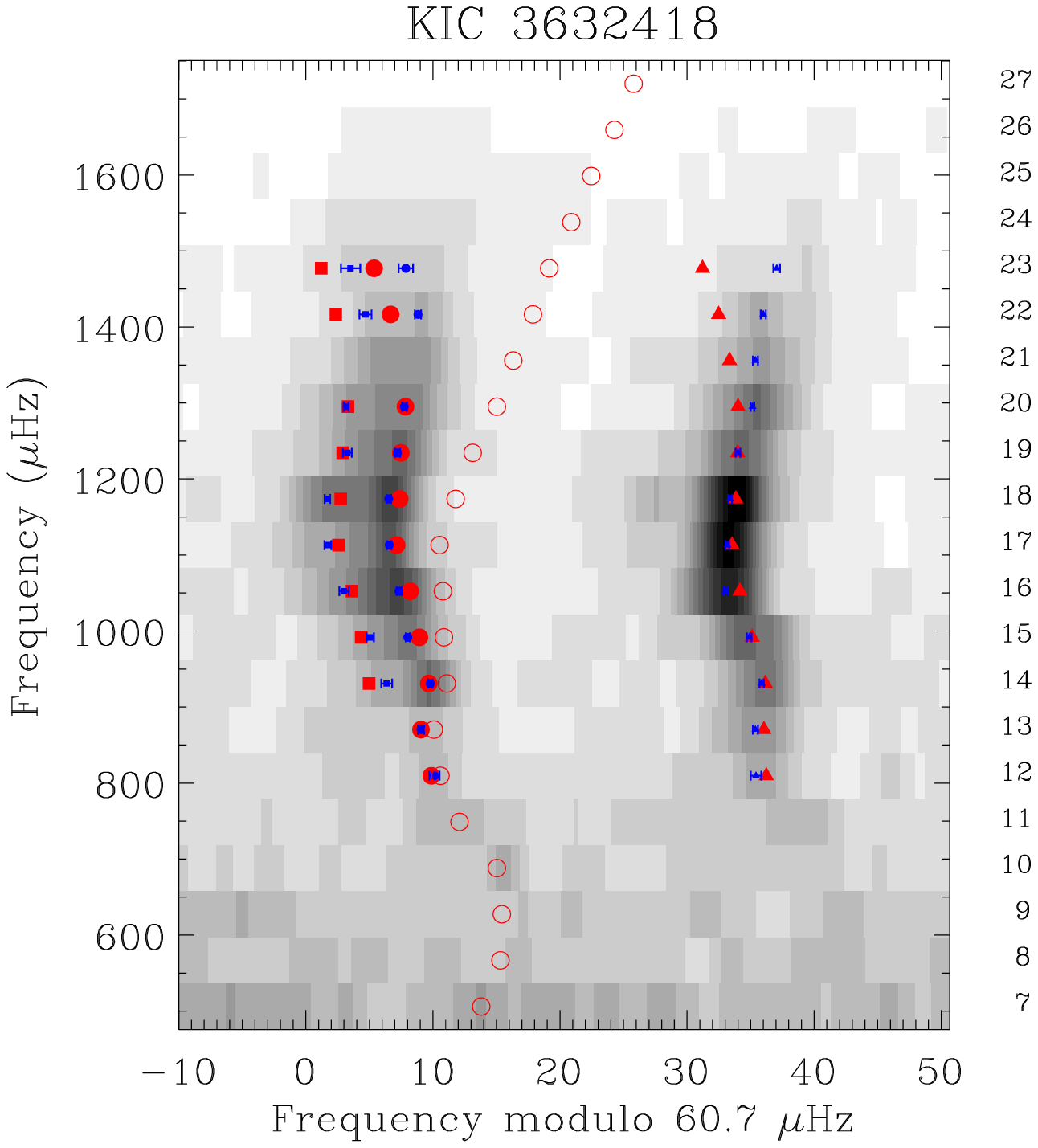}}
  \vspace*{0.125in}
  \centerline{\includegraphics[width=2.25in]{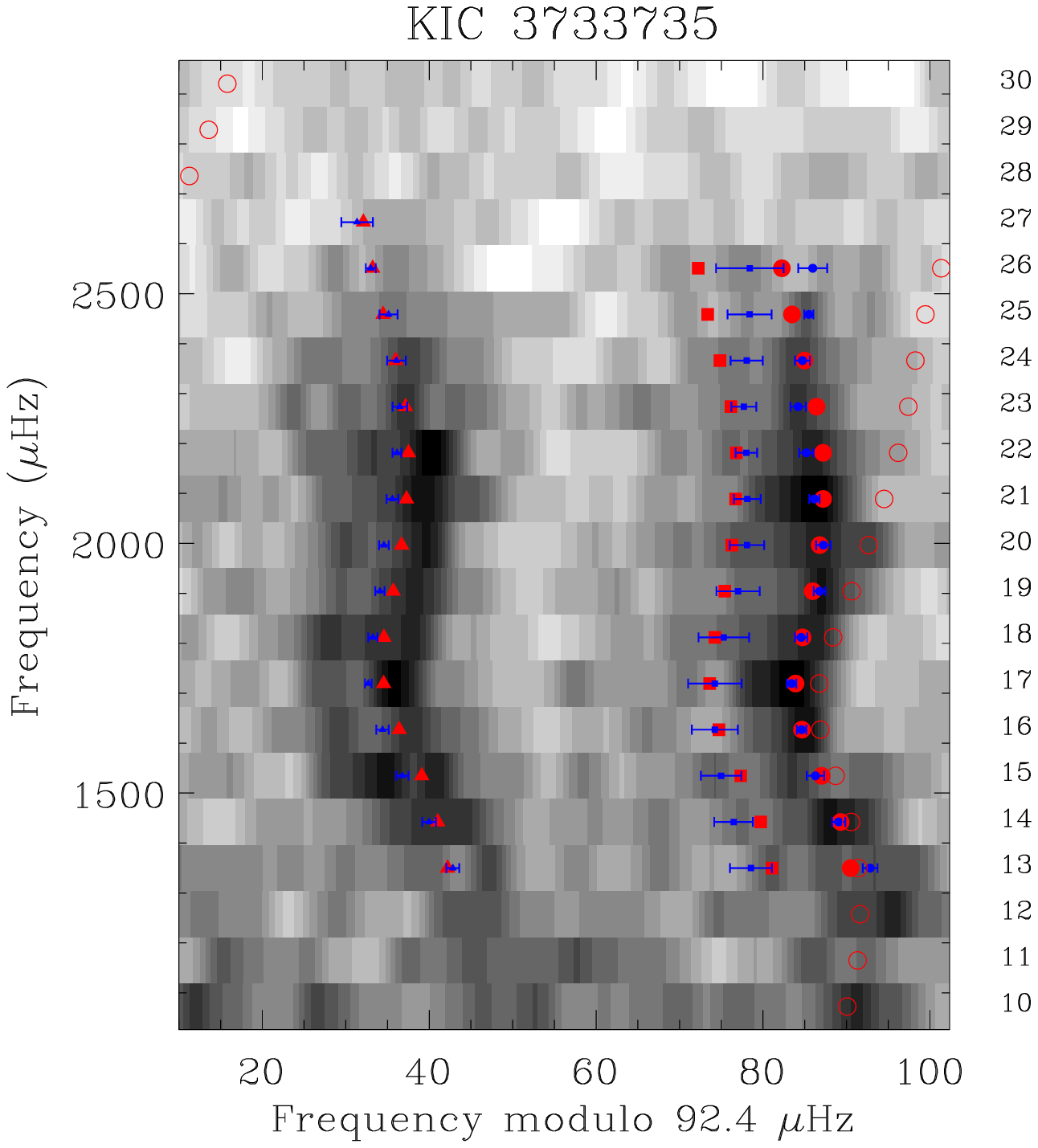}\hspace*{0.125in}\includegraphics[width=2.25in]{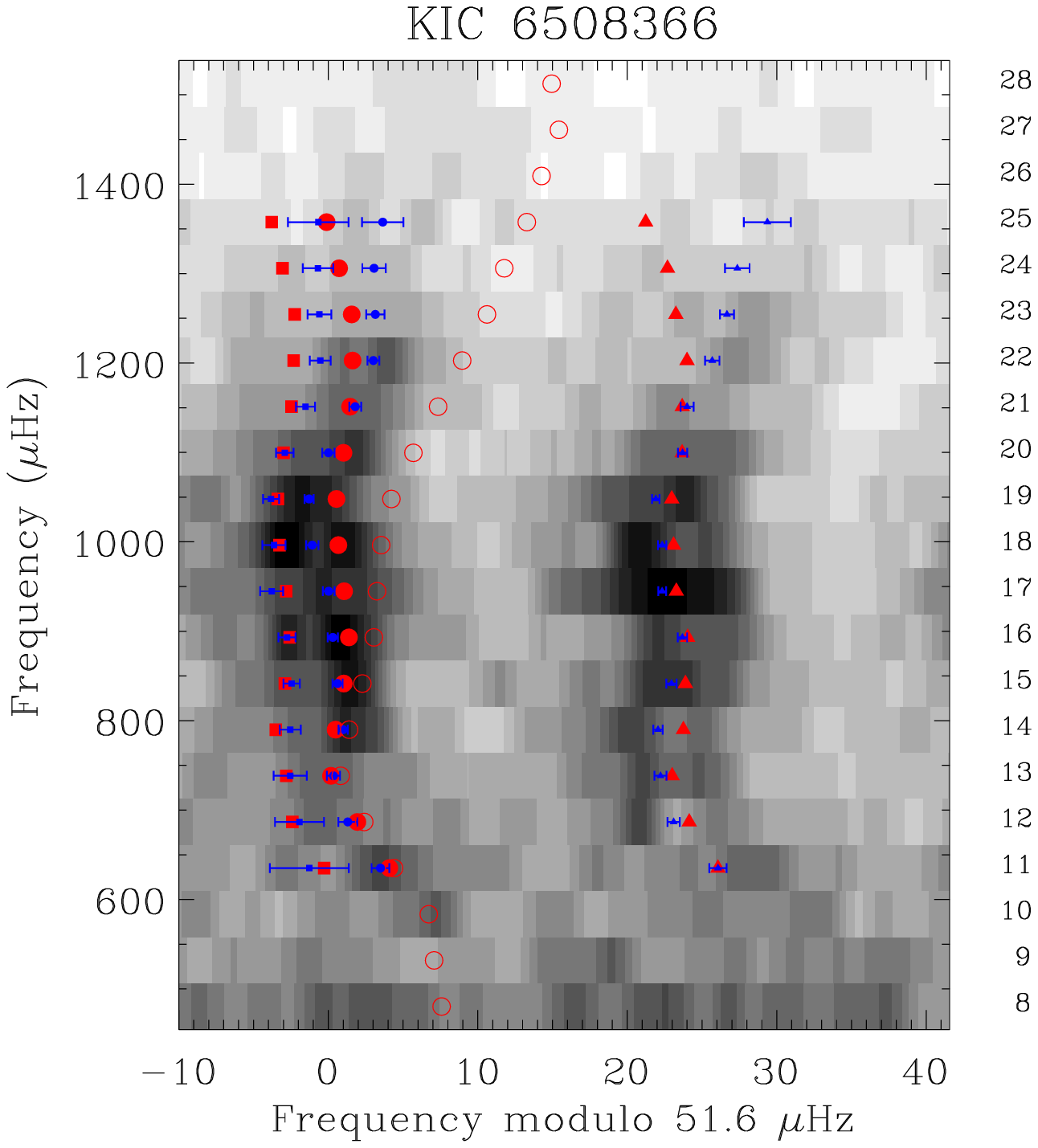}\hspace*{0.125in}\includegraphics[width=2.25in]{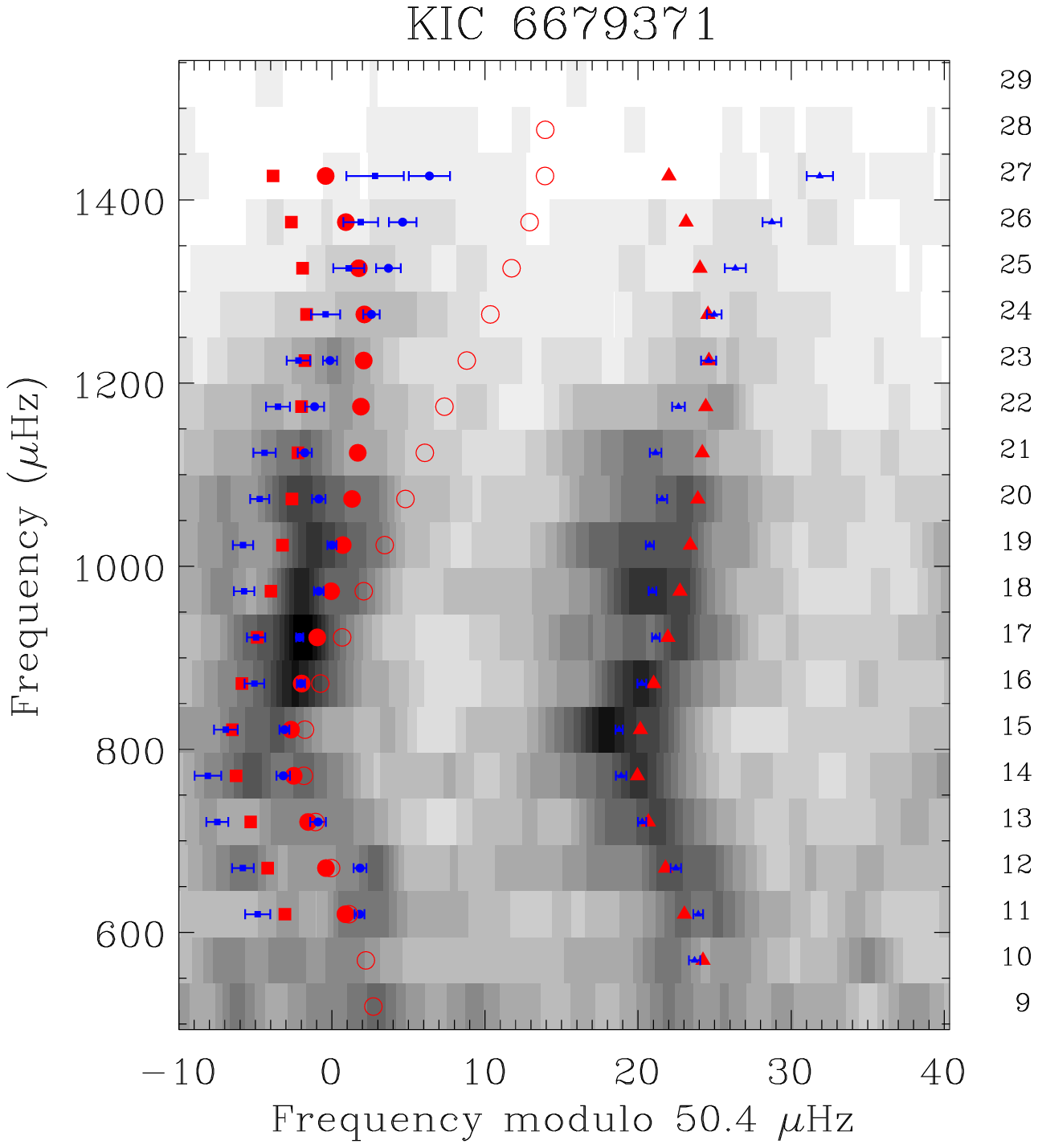}}
  \vspace*{0.125in}
  \centerline{\includegraphics[width=2.25in]{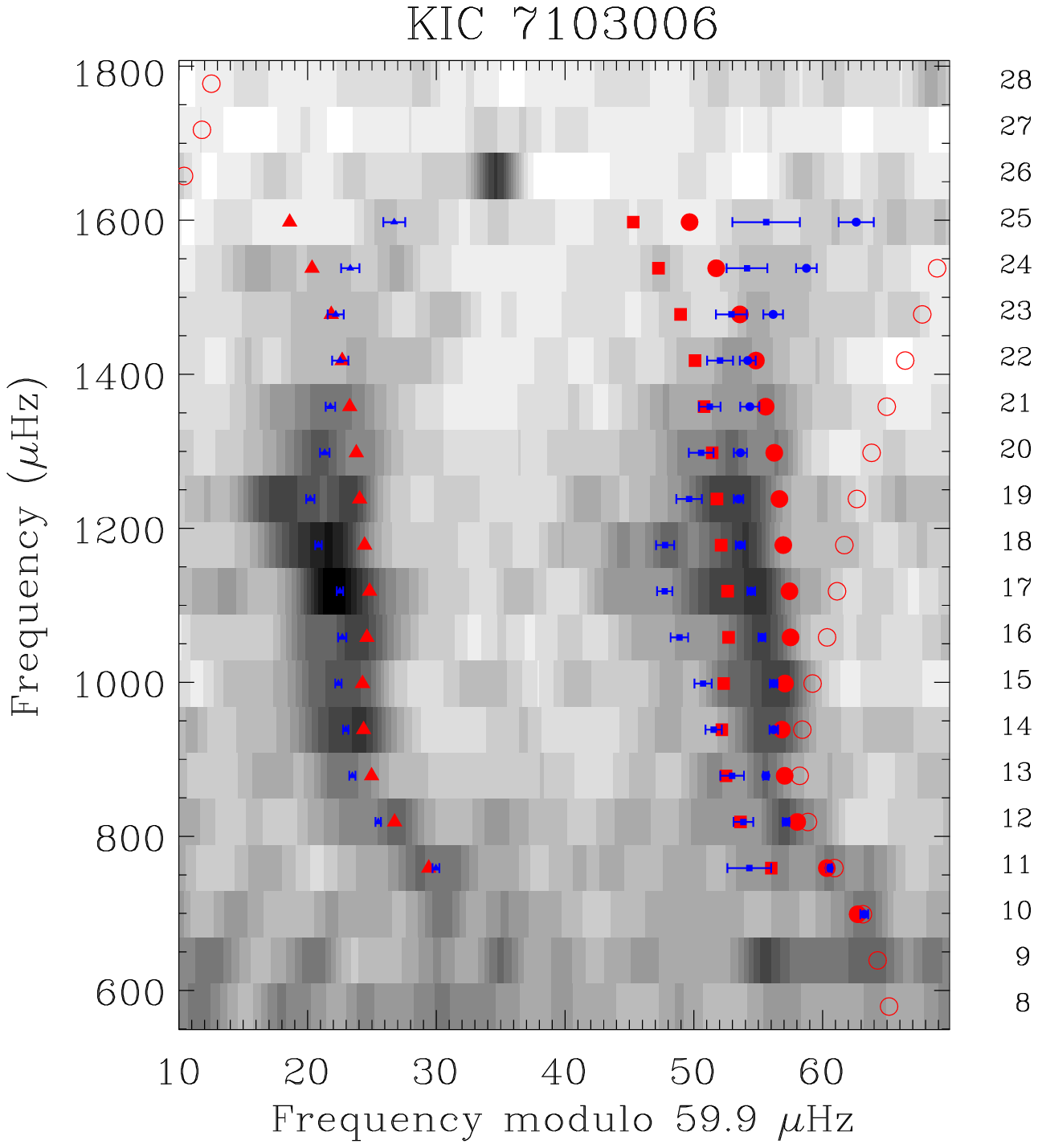}\hspace*{0.125in}\includegraphics[width=2.25in]{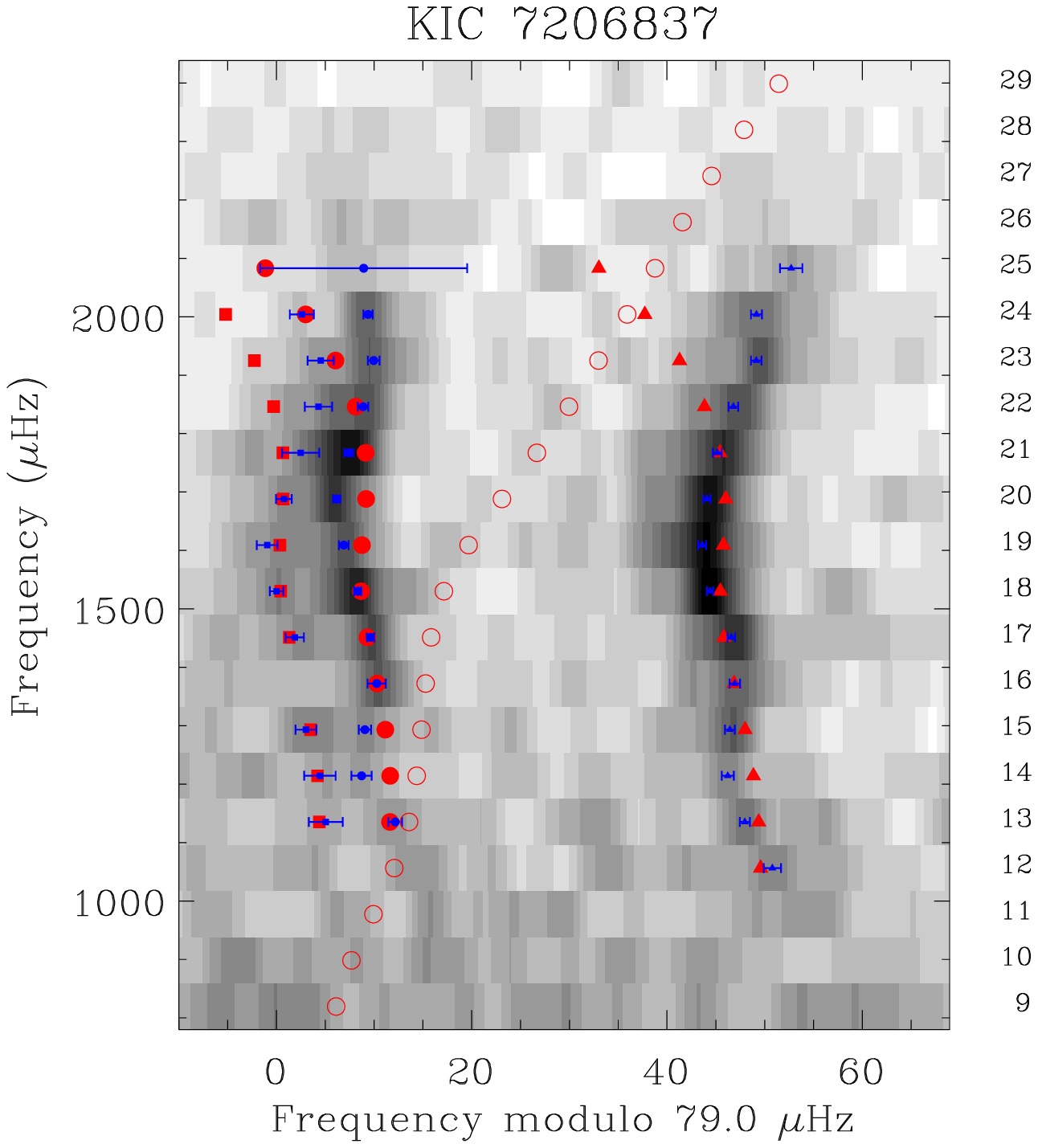}\hspace*{0.125in}\includegraphics[width=2.25in]{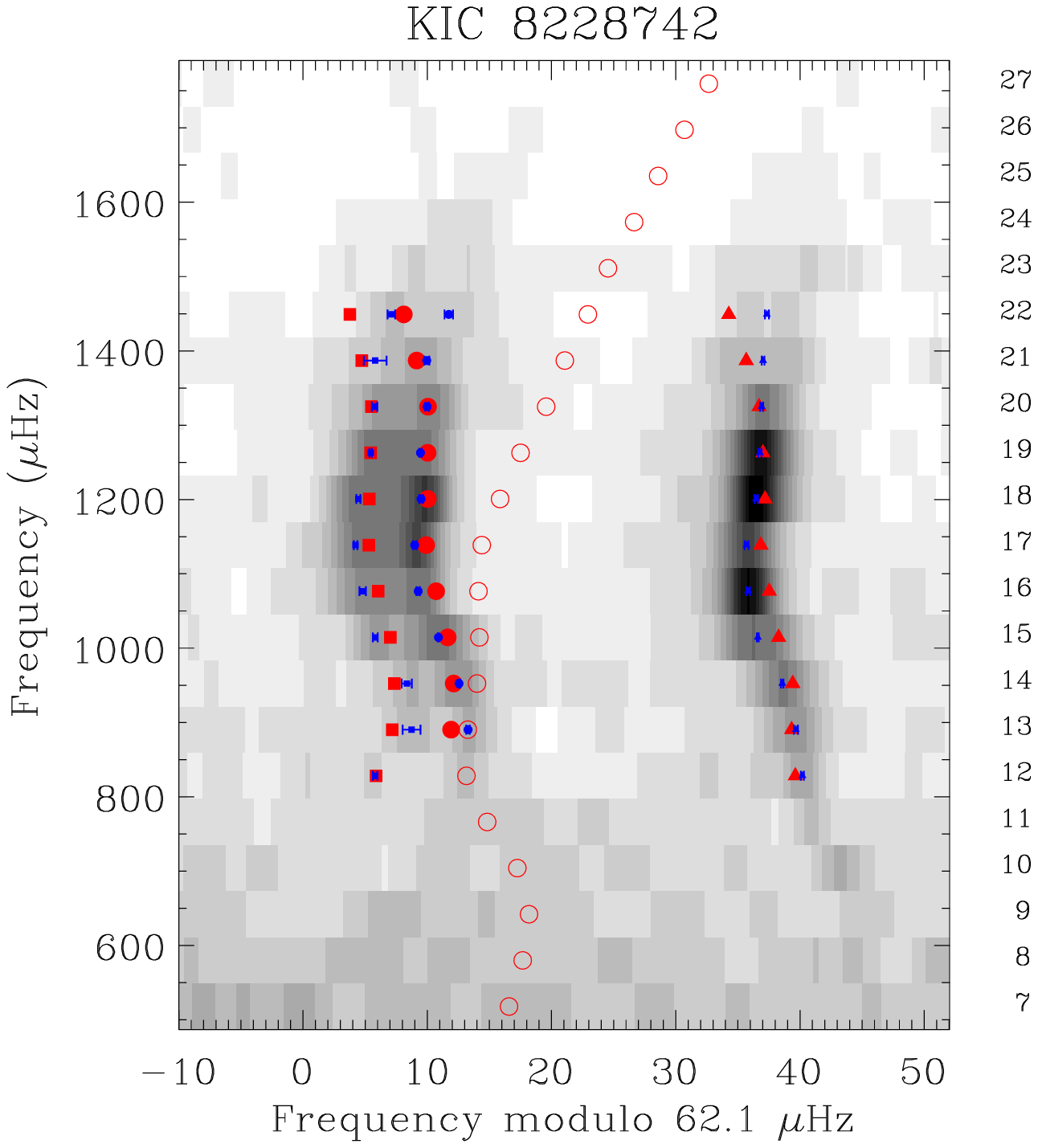}}
  \vspace*{0.5in}
  \centerline{{\bf Figure~\ref{fig3}}.~~~ ONLINE ONLY (cont.)}
  \end{figure*} 

  \begin{figure*}[p] 
  \centerline{\includegraphics[width=2.25in]{fig3.28.eps}\hspace*{0.125in}\includegraphics[width=2.25in]{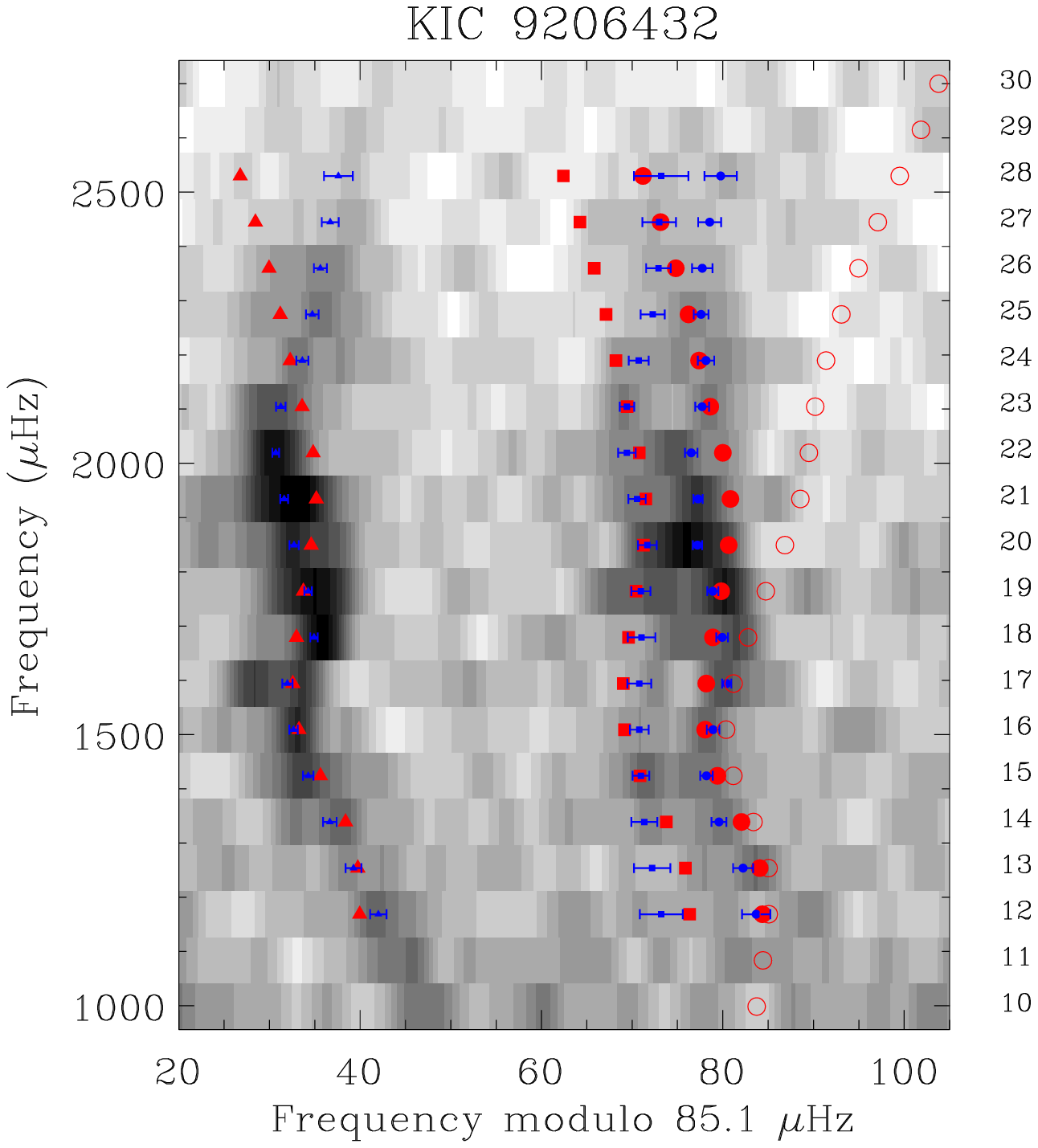}\hspace*{0.125in}\includegraphics[width=2.25in]{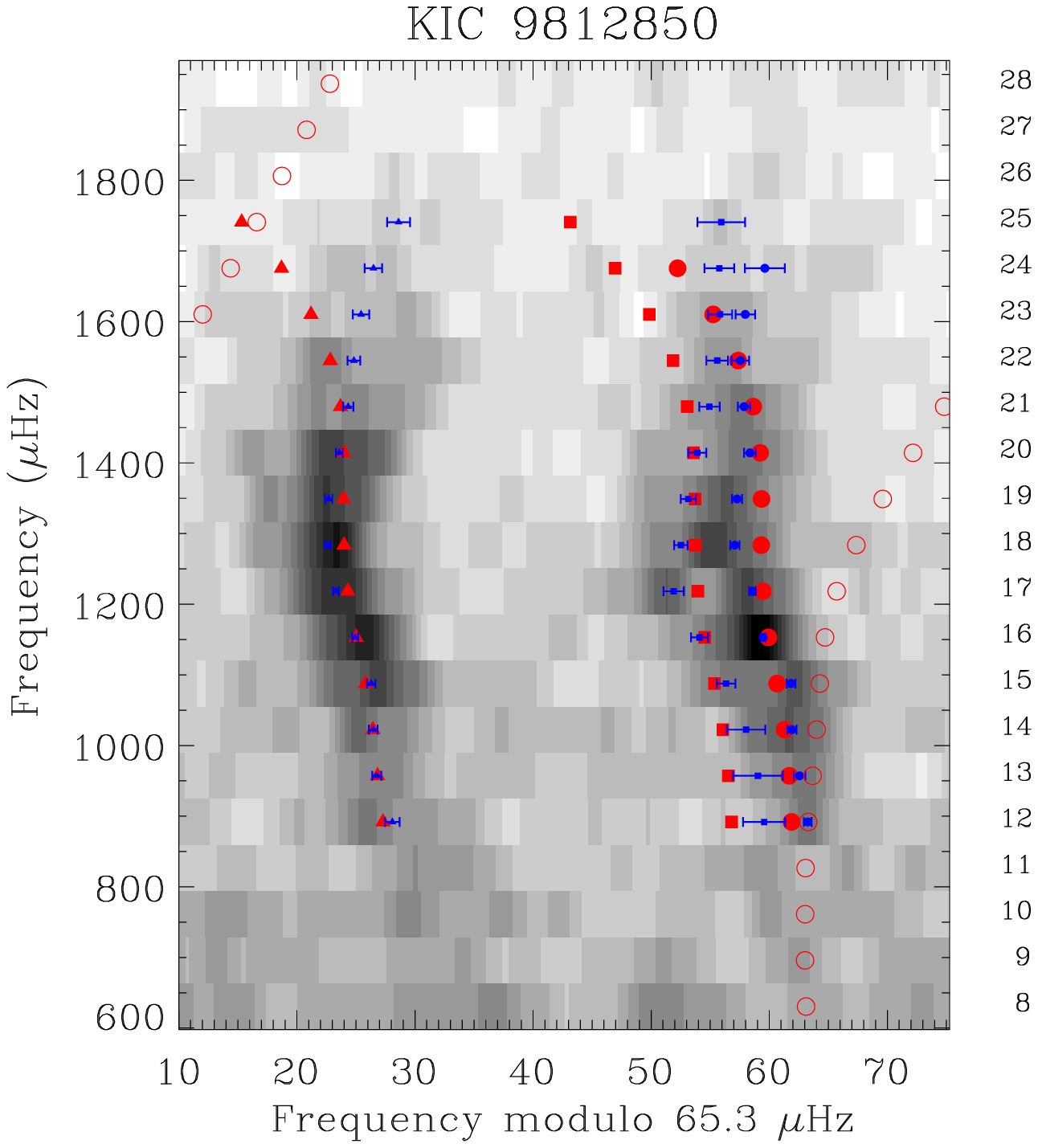}}
  \vspace*{0.125in}
  \centerline{\includegraphics[width=2.25in]{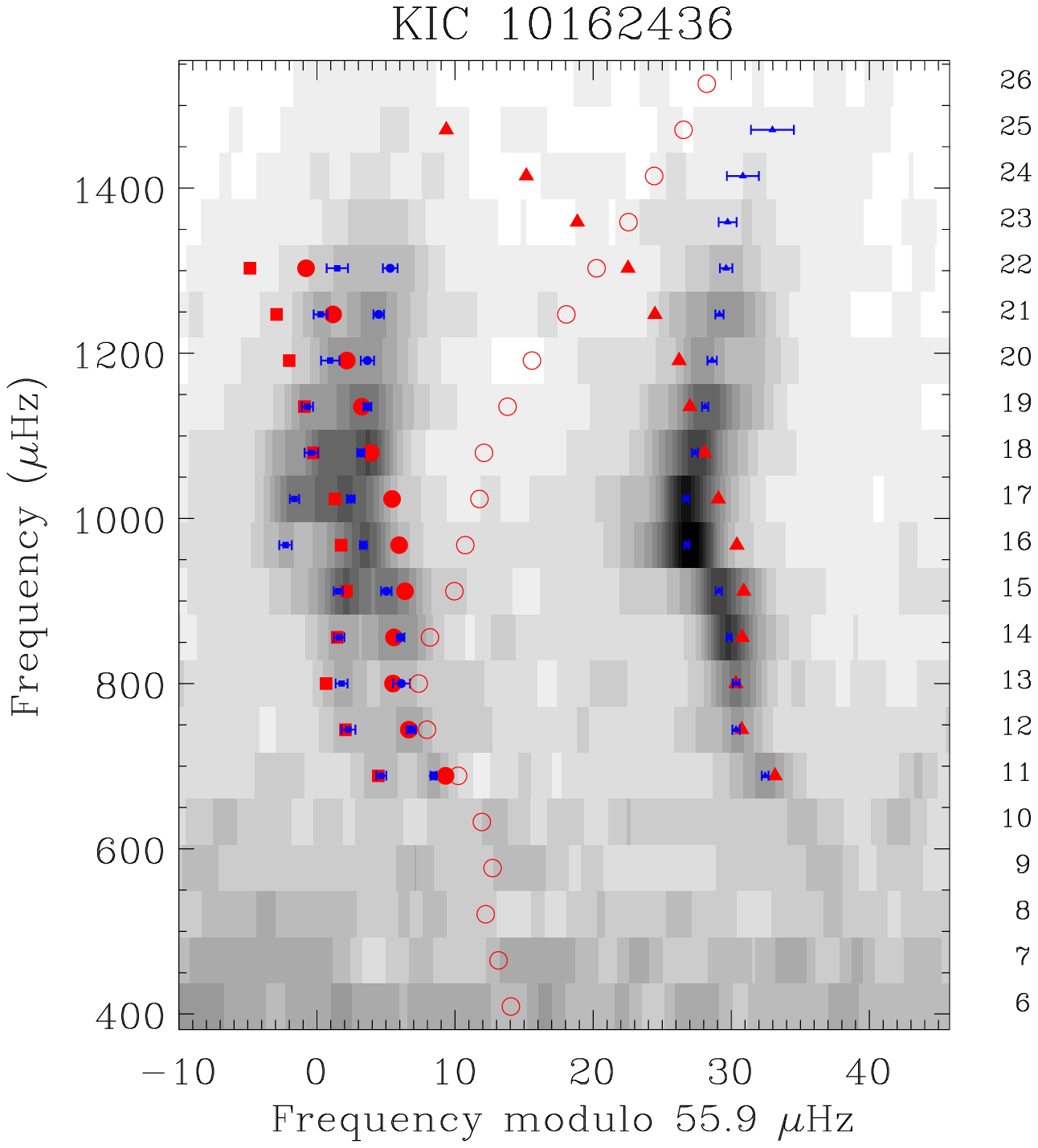}\hspace*{0.125in}\includegraphics[width=2.25in]{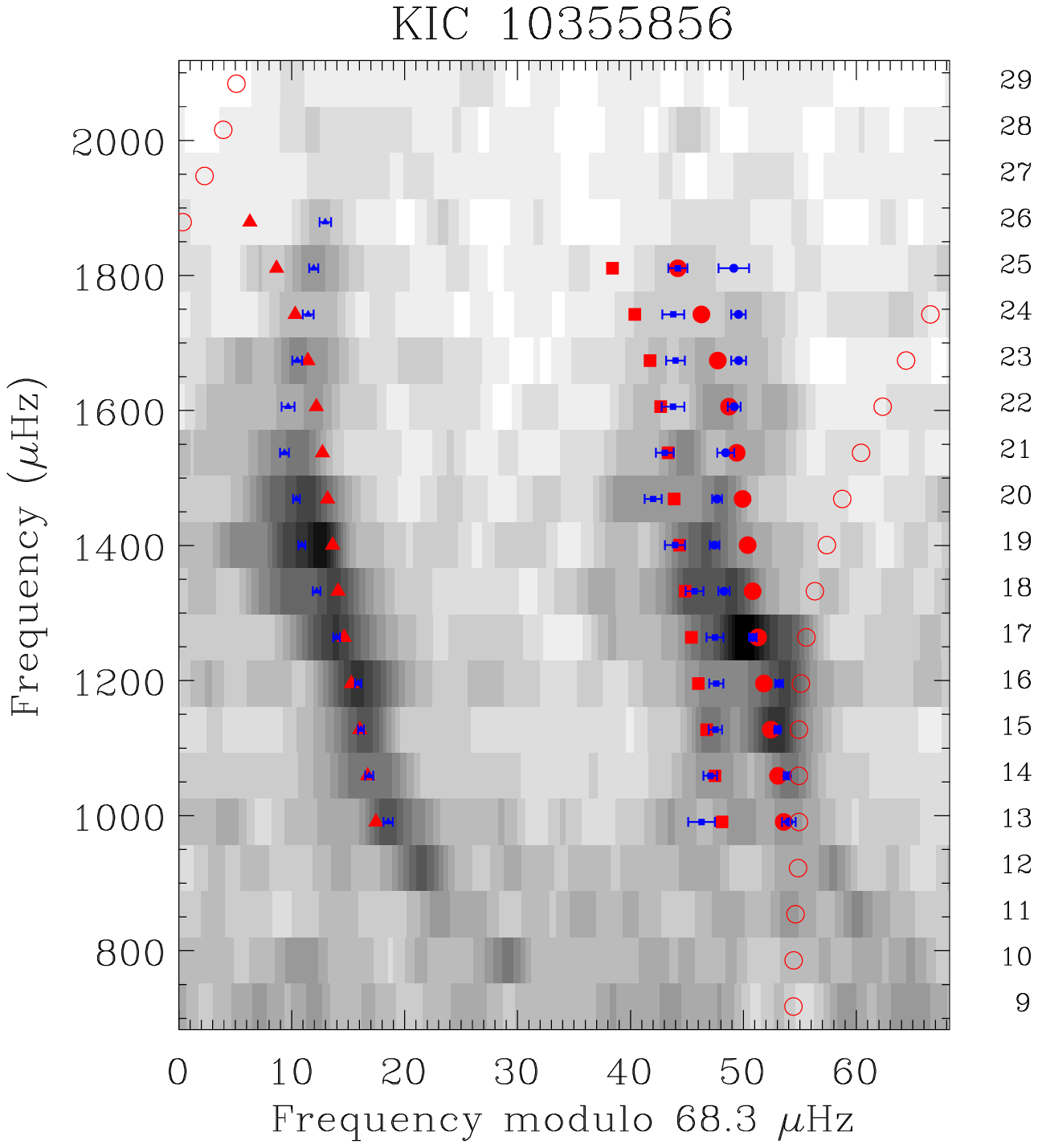}\hspace*{0.125in}\includegraphics[width=2.25in]{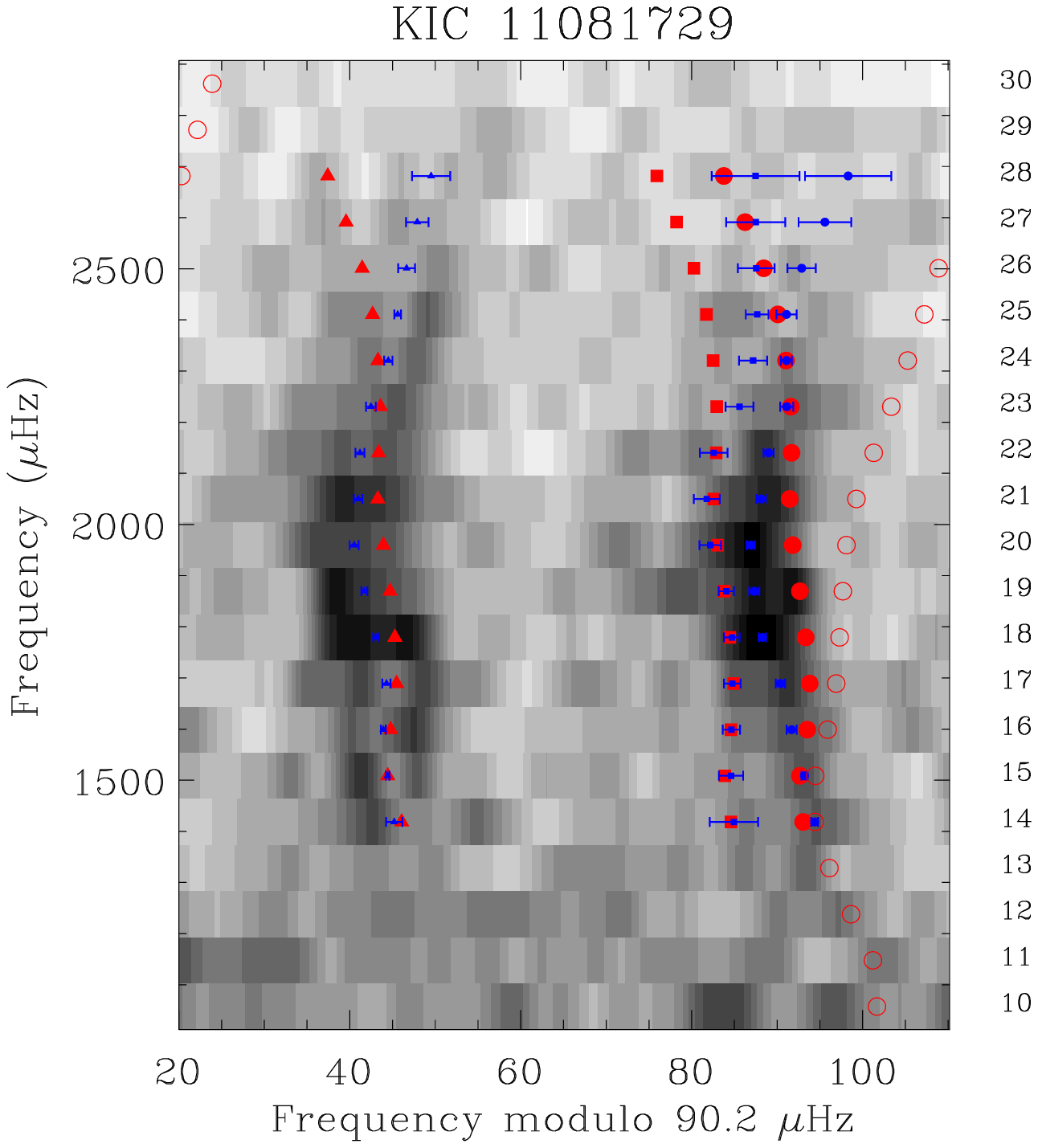}}
  \vspace*{0.125in}
  \centerline{\includegraphics[width=2.25in]{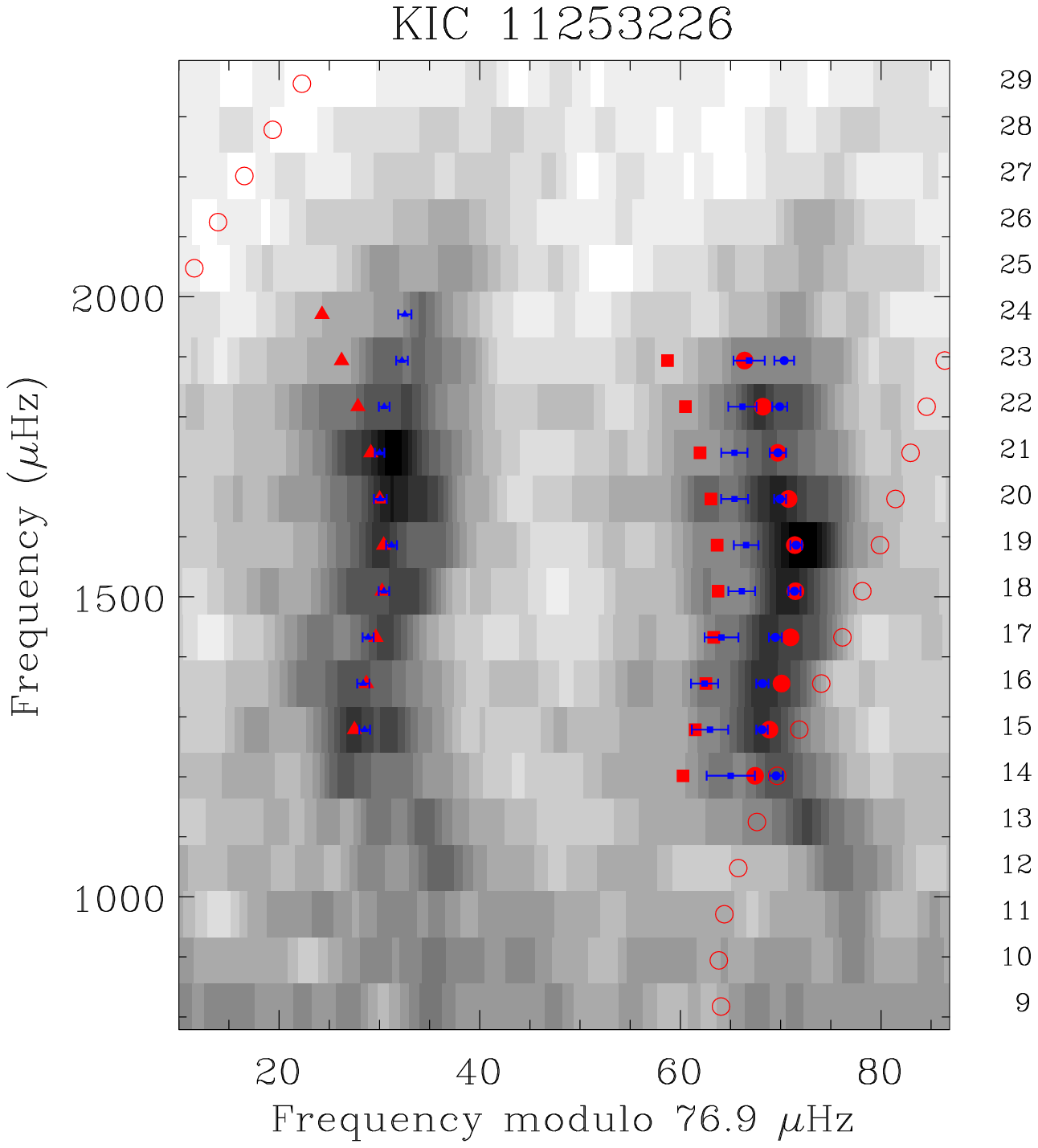}\hspace*{0.125in}\includegraphics[width=2.25in]{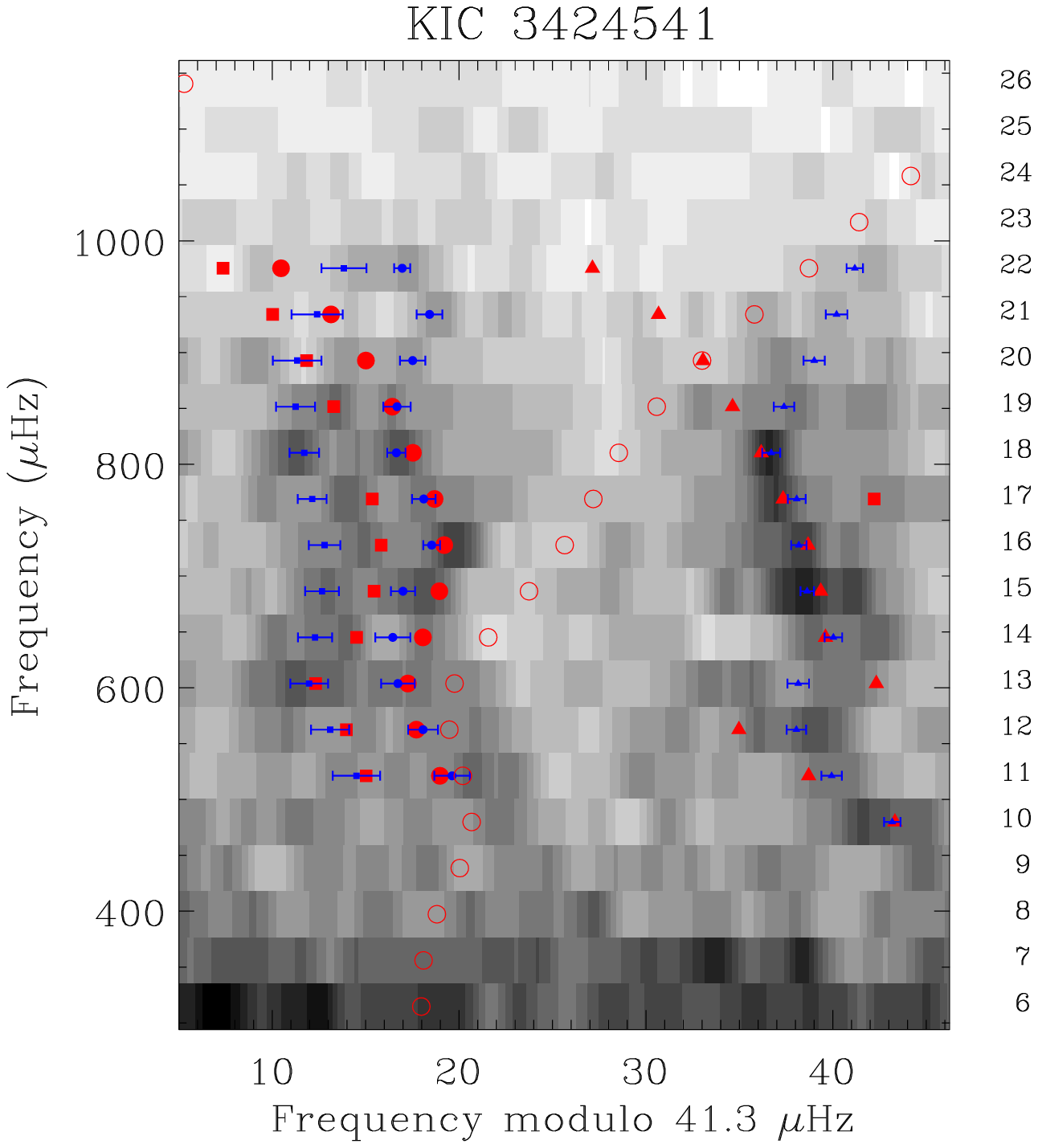}\hspace*{0.125in}\includegraphics[width=2.25in]{fig3.36.eps}}
  \vspace*{0.5in}
  \centerline{{\bf Figure~\ref{fig3}}.~~~ ONLINE ONLY (cont.)}
  \end{figure*} 

  \begin{figure*}[p] 
  \centerline{\includegraphics[width=2.25in]{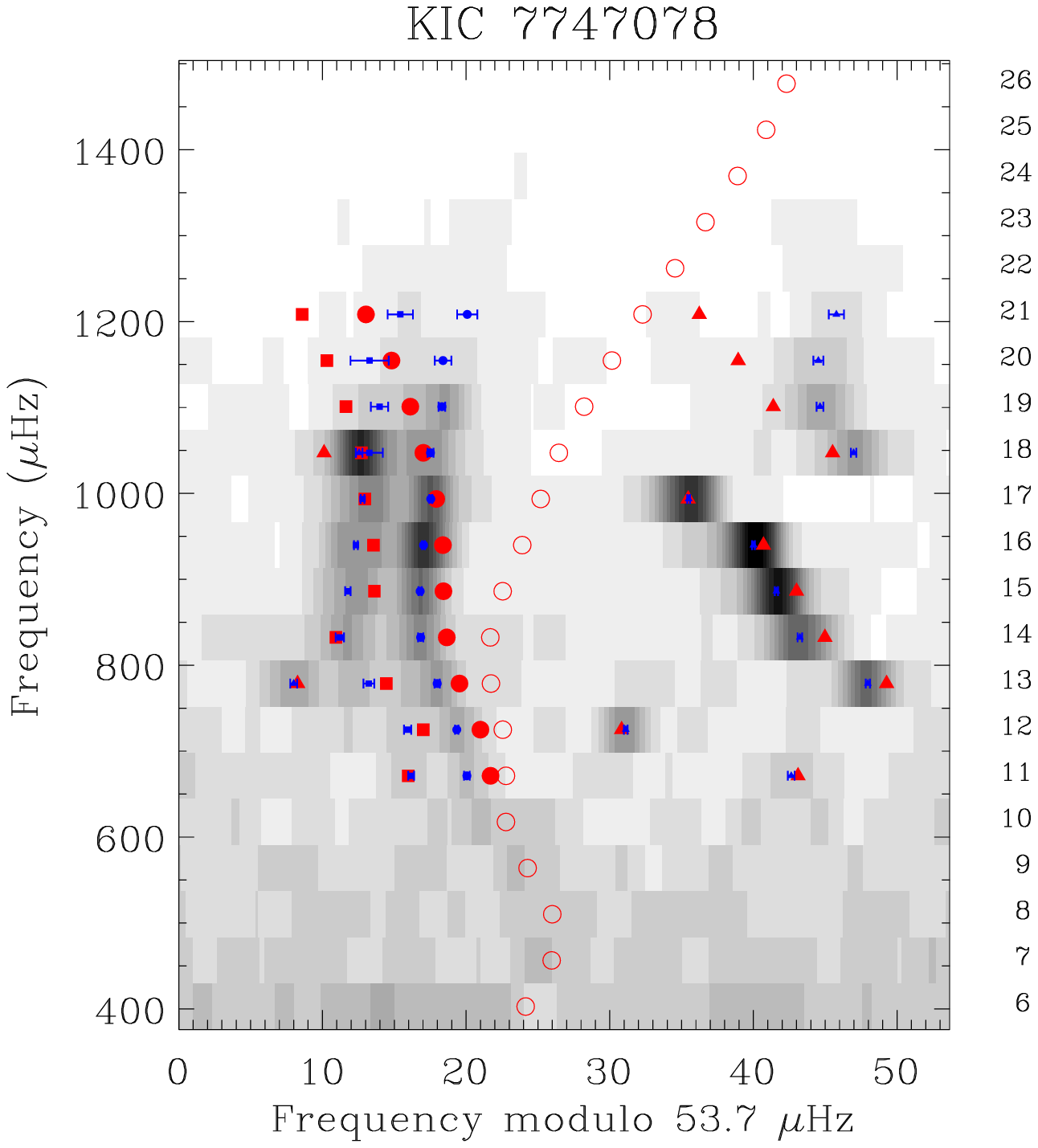}\hspace*{0.125in}\includegraphics[width=2.25in]{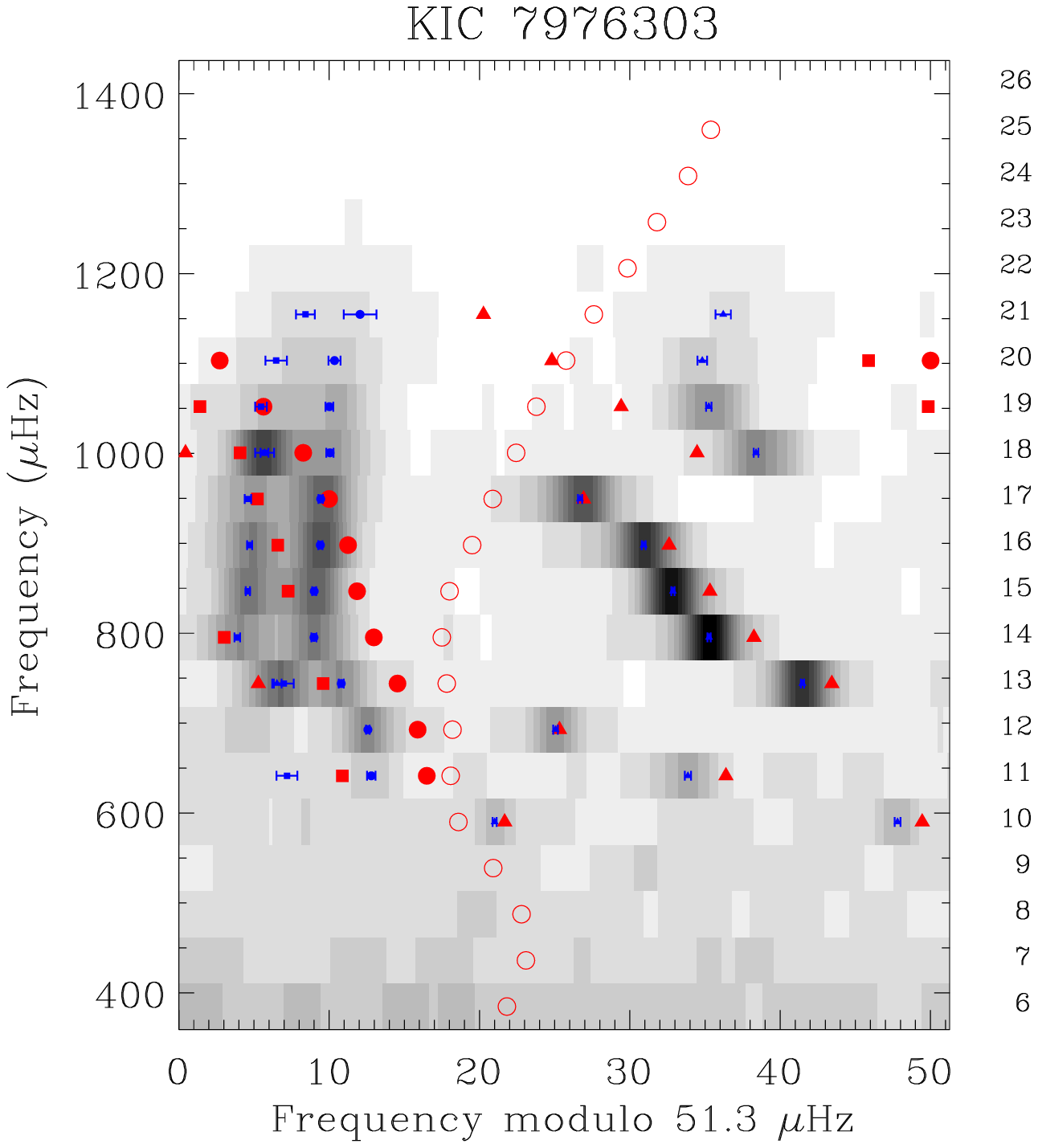}\hspace*{0.125in}\includegraphics[width=2.25in]{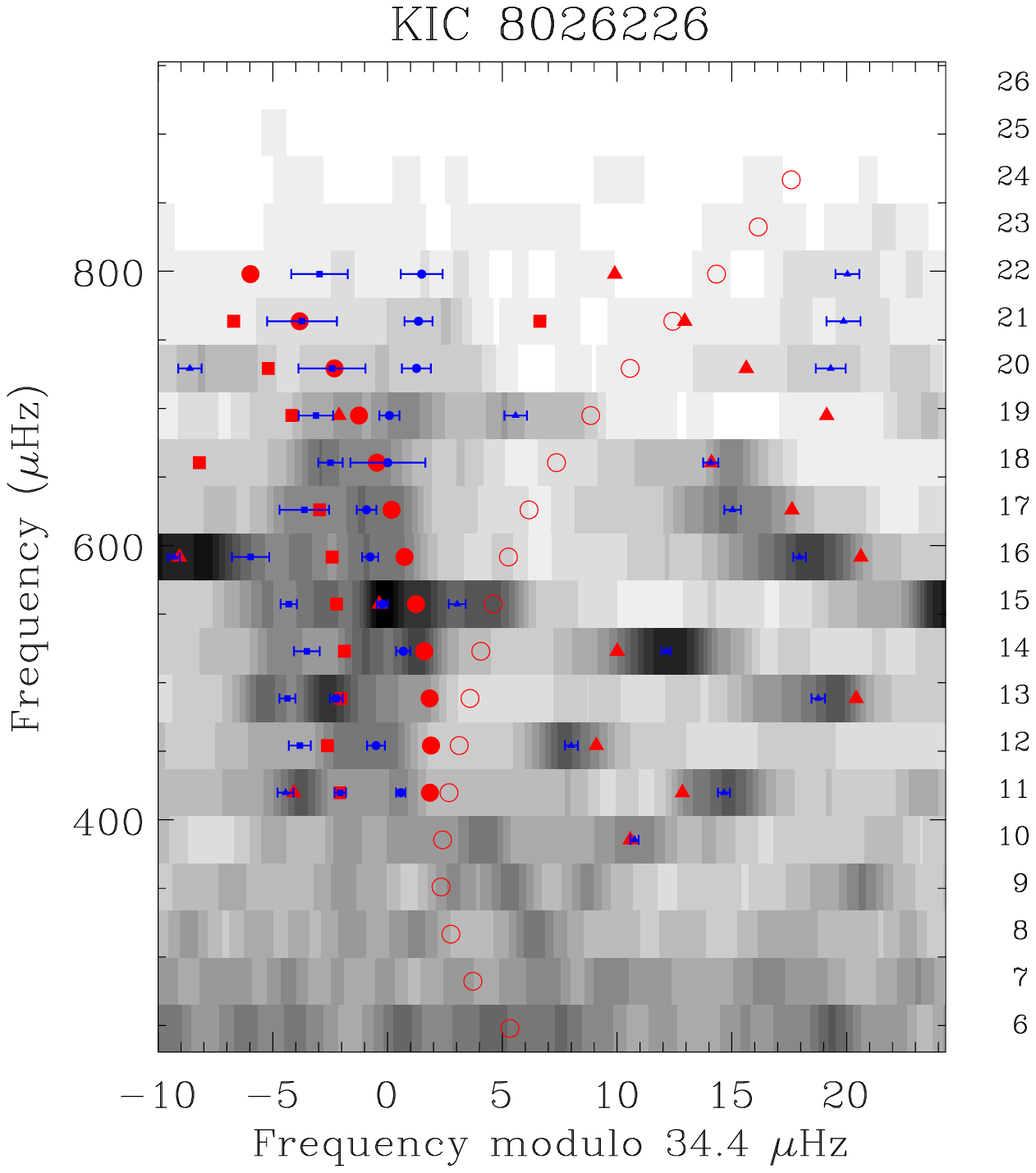}}
  \vspace*{0.125in}
  \centerline{\includegraphics[width=2.25in]{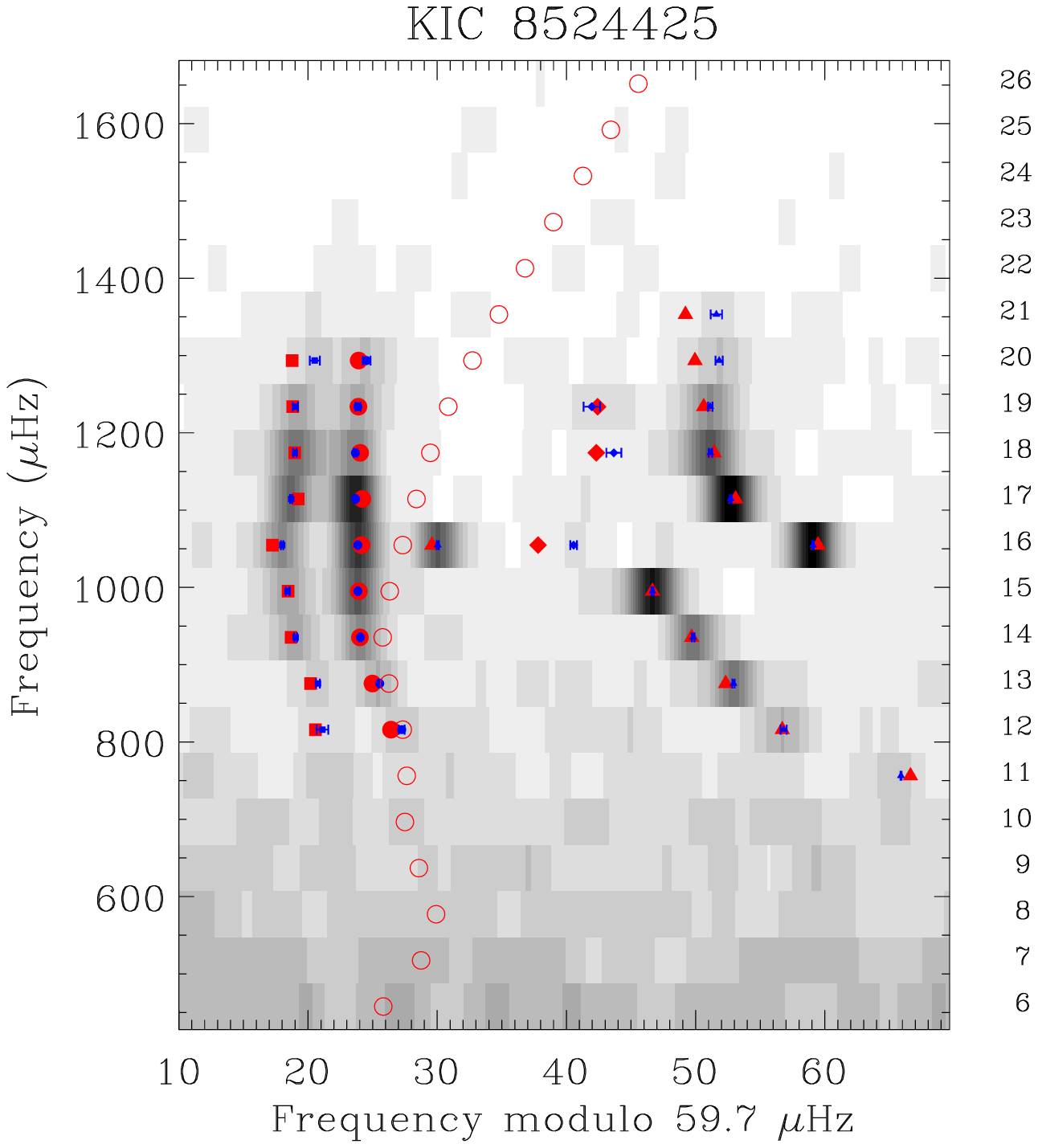}\hspace*{0.125in}\includegraphics[width=2.25in]{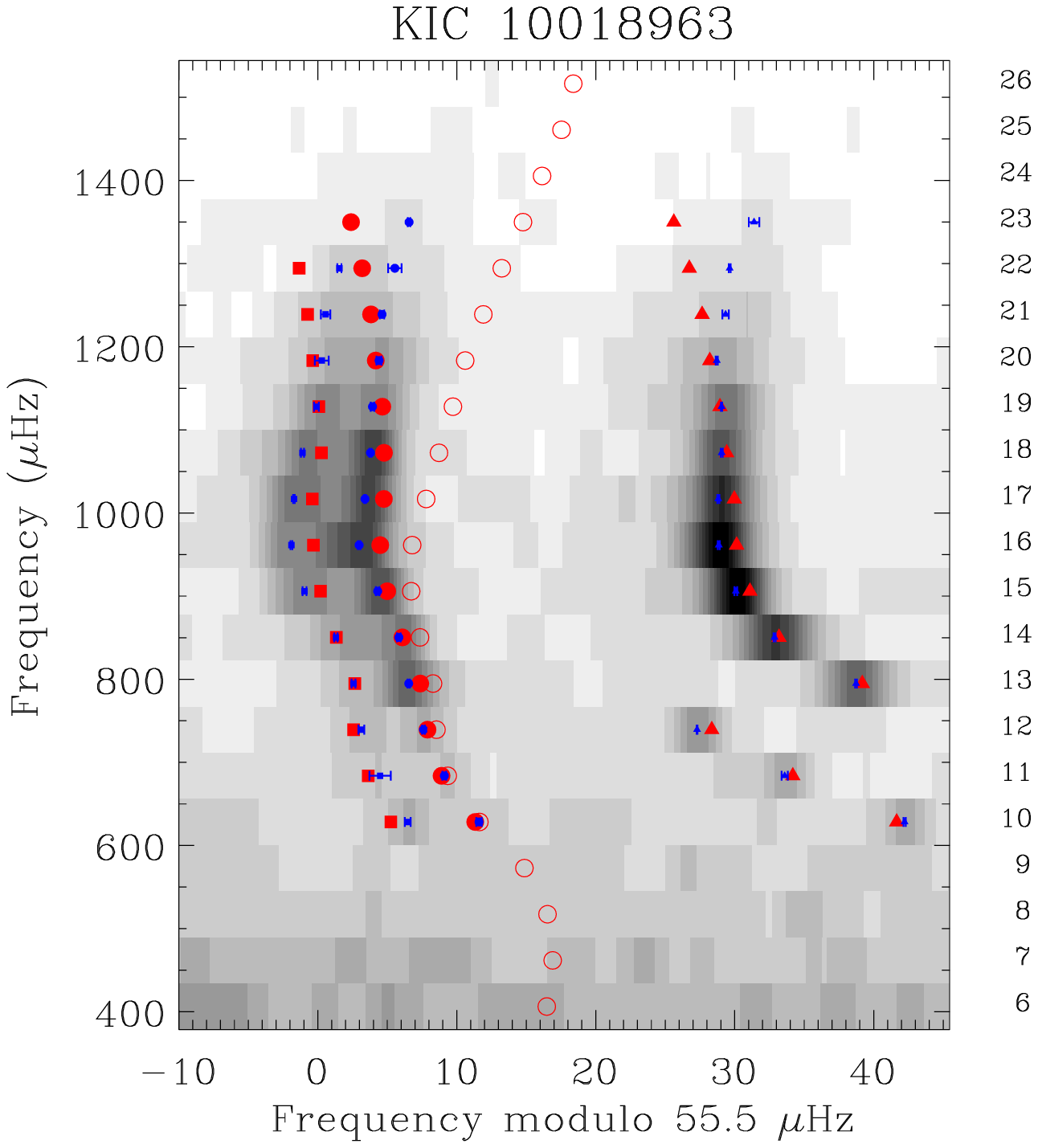}\hspace*{0.125in}\includegraphics[width=2.25in]{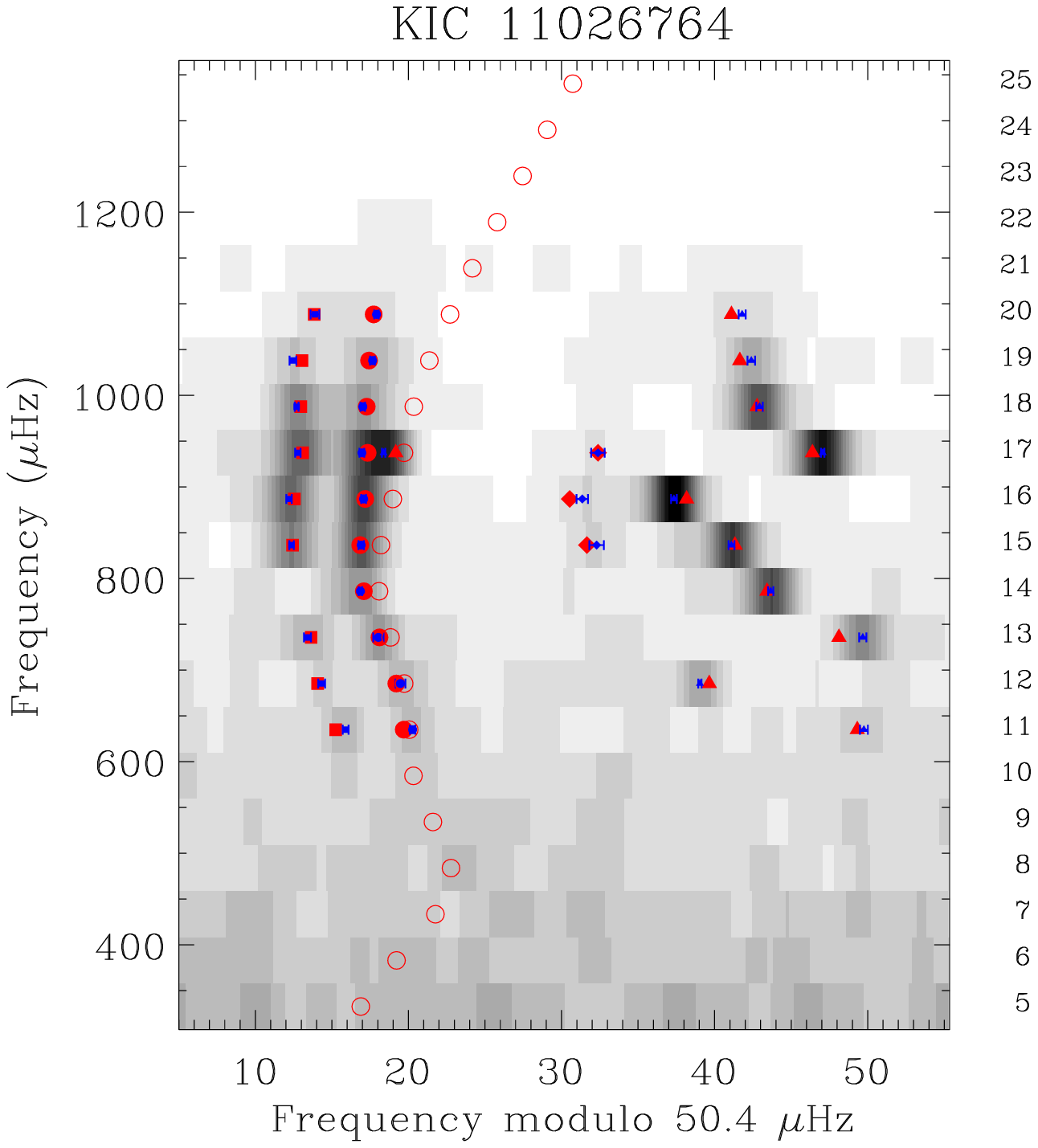}}
  \vspace*{0.5in}
  \centerline{{\bf Figure~\ref{fig3}}.~~~ ONLINE ONLY (cont.)}
  \end{figure*} 


\end{document}